\documentclass[graybox, envcountchap]{svmult}

\usepackage{mathptmx}         % selects Times Roman as basic font
\usepackage{amsmath}
\usepackage{amssymb}
\usepackage{color}
\usepackage{helvet}           % selects Helvetica as sans-serif font
\usepackage{courier}          % selects Courier as typewriter font
\usepackage{dirtree}
\usepackage{makeidx}          % allows index generation
\usepackage{graphicx}         % standard LaTeX graphics tool
\usepackage{subfig}

\usepackage{multicol}         % used for the two-column index
\usepackage[bottom]{footmisc} % places footnotes at page bottom

\usepackage{hyperref}         % for hyperlinks
\hypersetup{colorlinks=true,urlcolor=blue}

\usepackage[misc]{ifsym}

\makeindex                    % used for the subject index
\usepackage{txfonts}
\usepackage{array}
\usepackage{multirow}
\usepackage{booktabs}
\usepackage{orcidlink}

\begin{document}

%%%%%%%%%%%%%%%%%%%%%%%%%%%%%%%%%%%%%%%%%%%%%%%%%%%%%%%%%%%%%%%%%
%
%  TITLE
%
\title{Testing gravity with Extreme-Mass-Ratio Inspirals}
\titlerunning{Testing gravity with EMRIs}
% Use \titlerunning{Testing gravity with EMRIs} for an abbreviated version of
% your contribution title if the original one is too long
\author{Alejandro C\'ardenas-Avenda\~no and Carlos F. Sopuerta}
% Use \authorrunning{Short Title} for an abbreviated version of
% your contribution title if the original one is too long
\institute{Alejandro C\'ardenas-Avenda\~no\orcidlink{0000-0001-9528-1826} (\Letter) \at Princeton Gravity Initiative, Princeton University, Princeton, NJ 08544, USA. \email{cardenas-avendano@princeton.edu}
\and Carlos F. Sopuerta\orcidlink{0000-0002-1779-4447} (\Letter) \at Institut de Ci\`encies de l'Espai (ICE, CSIC), Campus UAB, Carrer de Can Magrans s/n, 08193 Cerdanyola del Vall\`es, Spain \\ Institut d'Estudis Espacials de Catalunya (IEEC), Edifici RDIT, C/ Esteve~Terradas, 1, desp.~212, Castelldefels~08860, Spain. \email{carlos.f.sopuerta@csic.es}}

%
% Use the package "url.sty" to avoid
% problems with special characters
% used in your e-mail or web address
%

\maketitle

%%%%%%%%%%%%%%%%%%%%%%%%%%%%%%%%%%%%%%%%%%%%%%%%%%%%%%%%%
%
%   ABSTRACT
%
\abstract{Extreme-mass-ratio inspirals consist of binary systems of compact objects, with orders of magnitude differences in their masses, in the regime where the dynamics are driven by gravitational wave emission. The unique nature of extreme-mass-ratio inspirals facilitates the exploration of diverse and stimulating tests related to various aspects of black holes and General Relativity. These tests encompass investigations into the spacetime geometry, the dissipation of the binary system energy and angular momentum, and the impact of the intrinsic non-linear effects of the gravitational interaction. This review accounts for the manifold opportunities for testing General Relativity, ranging from examinations of black hole properties to cosmological and multimessenger tests.}

%%%%%%%%%%  Introduction  %%%%%%%%%%

\tableofcontents

\section{Introduction}

%Introduction
%%%ACR-GW
%%%ACR-BH
Gravitational wave Astronomy, like Neutrino Astronomy~\cite{IceCube:2013cdw}, is a new form of Astronomy of the 21st century that was inaugurated with the announcement of the first-ever direct detection of gravitational waves (GWs)~\cite{LIGOScientific:2016aoc} by the Laser Interferometer Gravitational-wave Observatory (LIGO), located in two sites in US land. This first detection, made on September 14th, 2015, consisted of the coalescence and merger of a compact binary system of two Black Holes (BHs).  This event is still one of the most exciting events captured by ground-based GW detectors, in part because it can already be used for testing gravity and the nature of BHs~\cite{LIGOScientific:2016lio}, which is one of the most interesting goals of GW Astronomy, and for which this area has a vast potential of producing revolutionary discoveries. 

%Introduction Review
%%%ACR-EMRI
%%%ACR-GR
This review focuses on various aspects of beyond Einstein (or alternative) theories of gravity and how they can be tested through GW observations of compact binary systems with extreme-mass ratios. Due to the mass asymmetry of the components, the dynamics and morphology of the emitted GWs are highly complex. This makes these systems ideal laboratories for exploring gravitational theories. We present recent work dealing with multiple effects in Extreme-Mass-Ratio Inspirals (EMRIs) that can be used to test the geometry of the most compact objects in the Universe, as well as General Relativity (GR), and beyond Einstein theories of gravity, in model-dependent and model-independent ways.

%Modifications to General Relativity
%%%ACR-NS
Alternative theories of gravity modify GR in three ways: by altering the gravitational field of isolated self-gravitating objects (to which we refer sometimes as background solutions, e.g. BH or neutron star (NS) solutions); by modifying the mechanism of emission of GWs (e.g., its amplitude or frequency); or by changing the physical properties of the GWs themselves (e.g., its polarization states). The interaction between matter and new fields may lead to the emergence of ``fifth forces,'' representing energy-momentum exchanges between matter and these new fields. However, for weakly gravitating bodies, both particle physics and gravitational experiments preclude significant modifications to the dynamics.

%GW Emission in Modified Gravity
In GR, GW emission is predominantly quadrupolar~\cite{Misner:1973cw,Maggiore:2007mm}, as monopole and dipole emission are forbidden by the conservation of the matter stress-energy tensor, i.e., respectively, the total mass and the center of mass-energy of the system are conserved. In modified theories of gravity, however, the matter stress-energy tensor is not necessarily conserved, allowing for, in principle, monopole and dipole emission. In those cases, dipole radiation is the dominant effect for quasi-circular binary systems. Still, its presence and magnitude depend on the nature of the components of the binary system and of the modified theory of gravity.

\begin{figure}
\includegraphics[width=\columnwidth]{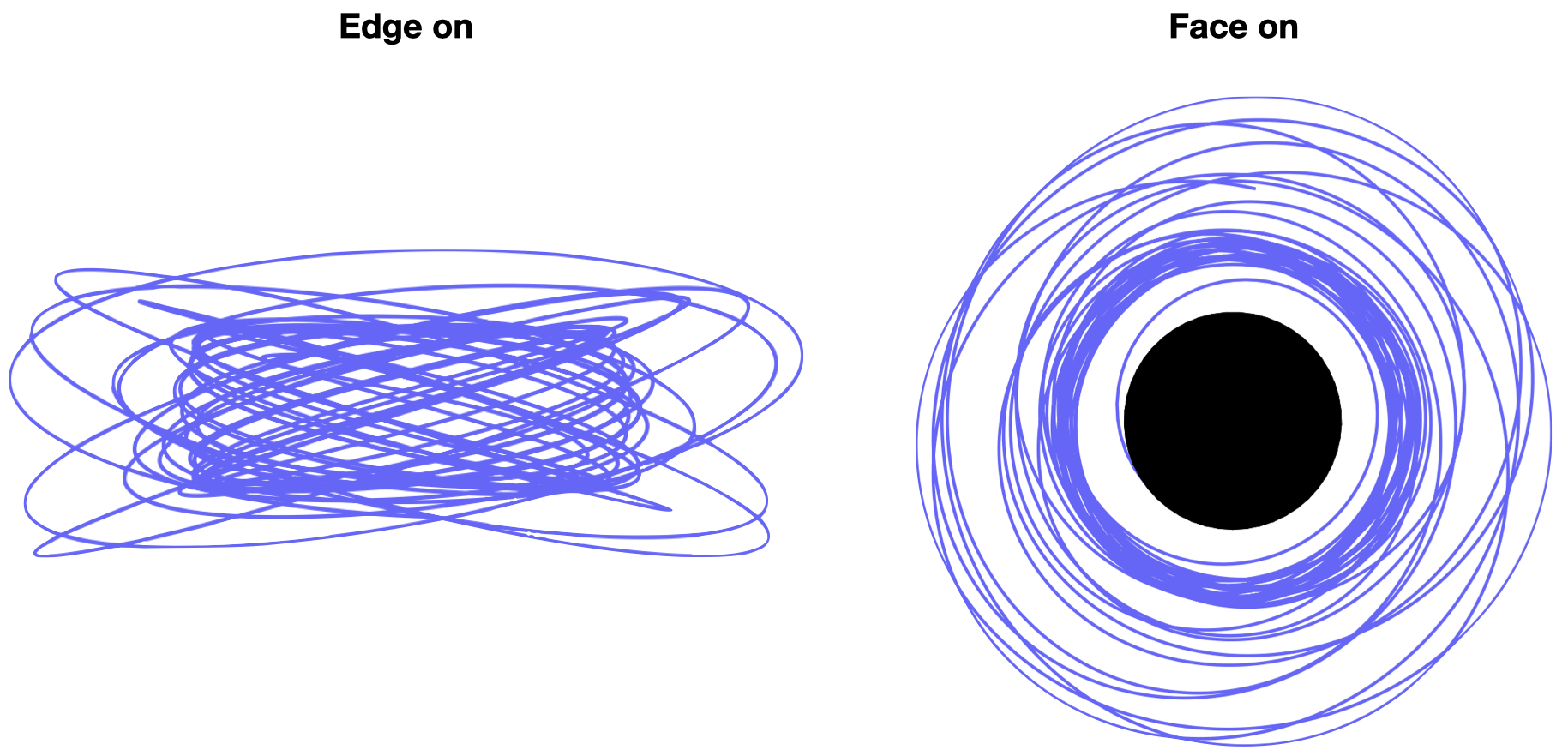}
\includegraphics[width=\columnwidth]{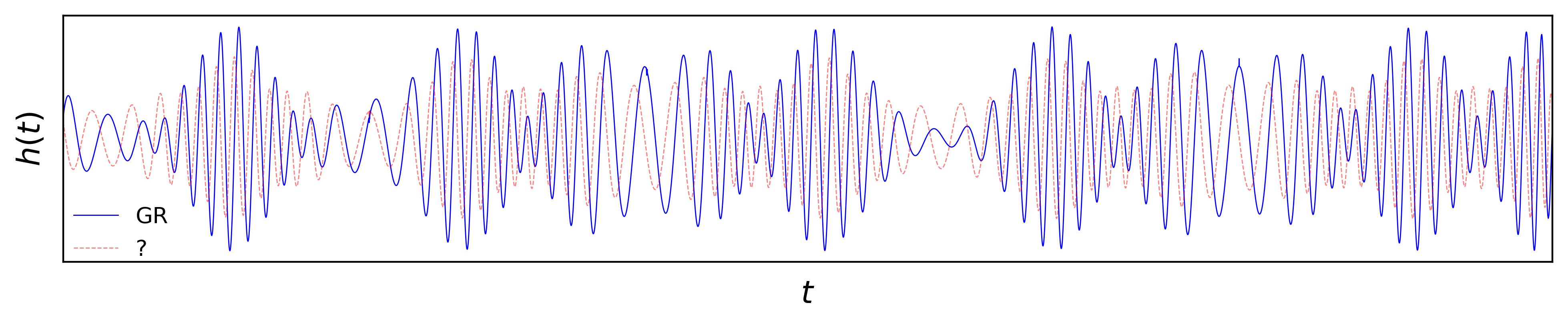}
\caption{A schematic of an EMRI trajectory (top) and its waveform (bottom). Due to the asymmetry of the components, the morphology is extremely rich. The trajectory of the secondary typically is inclined and eccentric. Even for similar initial conditions such as spin, mass ratio, eccentricity, or semilatus rectum, a waveform predicted by a beyond Einstein theory of gravity (red dashed line) can exhibit significant deviations from the prediction of GR (blue solid line).} 
\label{Fig:WFExample}
\end{figure}

%Special Cases: Extreme Mass Ratio Inspirals (EMRIs)
EMRIs are a unique class of systems with a significant difference in mass between their primary and secondary components (see Fig.~\ref{Fig:WFExample} for a schematic of the richness of the dynamics of these systems). These systems are sources in the low-frequency band of the GW spectrum, between $10^{-4}\,$Hz and $1\,$Hz\footnote{Since the mass of the primary determines the EMRI GW frequency, this frequency band translates into a range of masses of the order of $[10^4,10^8]M_\odot$. This mass range excludes, in particular, the most massive BHs that we have evidence to exist in the Universe, with masses up to $10^{10}\,M_\odot$, which fall in the frequency band of the Pulsar Timing Arrays, around the $n$Hz. Consequently, when the EMRI primary is a BH, we refer to it as a Massive BH (MBH), with masses in the mentioned mass range.}. In the {\em standard EMRI scenario}, the primary object, a (super)massive compact object, is typically assumed to be a Kerr BH, while the secondary is assumed to be a stellar-mass Kerr BH (although frequently the spin of the secondary is ignored) inspiralling into the primary object.  

%ACR-IMRI
%ACR-IMBH
%ACR-XMRI
By denoting the mass of the primary as $M$, the mass of the secondary as $m$, and the mass ratio as $q = m/M \ll 1$, the typical mass ratio range for EMRIs is: $10^{-6} < q < 10^{-4}$. There are, however, other systems very close to EMRIs and with a very similar dynamics that is worth considering. First, we have the so-called Intermediate-Mass-Ratio Inspirals (IMRIs), motivated by the possible existence of intermediate-mass black holes (IMBHs), with masses in the range $[10^2,10^5]\,M_\odot$, that is, just in between the mass ranges we have assumed for the primary and the secondary of an EMRI (see, e.g., Ref.~\cite{Giersz:2015mn}). Moreover, one usually distinguishes between IMRIs with a light primary, possibly an IMBH at the center of a globular cluster, and IMRIs with a heavy primary, possibly one of the MBH considered above. In any case, the typical mass ratio range for IMRIs is: $10^{-4} < q < 10^{-2}$.  Another category is the so-called extremely large mass-ratio inspirals (XMRIs)~\cite{Amaro-Seoane:2019umn}, with mass ratios in the range of $10^{-8} < q< 10^{-6}$, motivated by the possibility that the secondary is a brown dwarf, and consequently, with a potentially higher event rate than EMRIs. The drawback is that XMRIs can only be seen up to nearby galaxies. 

Modifications to the ``Kerrness'' of the primary, which are expected to be small, should leave an imprint on the dynamics of the secondary, which is a smaller (at least a couple of orders of magnitude less massive, i.e., $\sim[1,10^2]\,M_\odot$) compact object. Going beyond the {\em standard EMRI scenario} allows us to consider EMRIs as composed of two compact objects that are not necessarily BHs described by GR. In this way, we can talk about EMRIs containing exotic objects (such as boson stars, fuzzballs, or gravastars). 

In this generalized scenario, the nature of the secondary can also play a significant role, mainly when the modifications away from GR are controlled through curvature. Additional GW polarizations are a generic feature of alternative theories of gravity, and future space-based gravitational wave observatories open the prospect for multi-band GW Astronomy.

%ACR-LVKC
%ACR-LISA
EMRIs are one of the main sources of GWs in the low-frequency band, between $10^{-4}\,$Hz and $1\,$Hz. This band is accessible only from space due to the increase in gravity gradient noise (also known as Newtonian noise) as we go down in frequency from the high-frequency band of the ground-based detectors like the ones of the LIGO, Virgo, and KAGRA (Kamioka Gravitational Wave Detector) collaboration (LVKC). This makes it very difficult for laser-interferometric ground-based detectors, if not impossible, to be sensitive to astrophysical GW sources below $1\,$Hz, at least with the current technology. There are already plans for future detectors in space, such as the Laser Interferometer Space Antenna (LISA)~\cite{LISA:2017pwj}, DECIGO~\cite{Kawamura:2006up,Kawamura:2011zz}, Taiji~\cite{Hu:2017mde,Ruan:2018tsw}, or TianQin~\cite{TianQin:2015yph,TianQin:2020hid,Gong:2021gvw}. The proposals are currently in varying stages of development, and collectively, they will provide valuable scientific data.

%ACR-ESA
%ACR-NASA
While we will mention forecasts for most of these proposed instruments, we will mainly focus on LISA: a mission led by the European Space Agency (ESA), with the National Aeronautics and Space Administration (NASA) as a partner agency, that successfully passed the mission adoption review (which evaluates the mission definition maturity, technology readiness and implementation risks) on January 2024~\cite{Colpi:2024xhw}. The implementation phase is scheduled to begin in early 2025 with the aim of having it ready for lunch around 2035, and commencing scientific operations around 2037. The core LISA technology was demonstrated by the ESA-led mission LISA Pathfinder~\cite{Armano:2016bkm,Armano:2018kix,Armano:2019dxs,Armano:2021cwl}, which achieved a differential acceleration noise sensitivity below the required level for LISA.  

The science of LISA has been described in several documents. In particular, in the white paper {\em The Gravitational Universe}~\cite{eLISA:2013xep}, selected by ESA as the science theme of its third Large-class mission (L3) within the Cosmic Vision programme, in the LISA proposal~\cite{LISA:2017pwj}, submitted by the LISA consortium~\cite{LISAconsortium} for the ESA L3 call for missions~\cite{LISAL3call}, and more recently in the LISA Definition Study Report~\cite{Colpi:2024xhw} written for the adoption of the mission. 

%ACR-MBHB
The science of LISA is based in the following target sources of gravitational radiation~\cite{eLISA:2013xep,LISA:2017pwj,Colpi:2024xhw}: (i) coalescence and merger of binary systems of MBHBs, with masses in the range $[10^4,10^7]\,M_\odot$, in the stage where their evolution is driven by gravitational-radiation emission up to the merger and final ringdown.  The detection of these systems with LISA, up to cosmological distances with $z\sim 10$ or greater, will provide invaluable information about the process of galaxy formation and the connection with the origin and growth of the supermassive BHs lurking in their centers; (ii) the capture and subsequent inspiral of stellar-mass compact objects (mainly BHs and NSs) into (super)massive BHs, that is, EMRIs. LISA can detect these systems up to $z\sim 2$, and they can tell us about the dynamics in galactic nuclei as well as about the fundamental physics questions this review deals with; (iii) ultracompact binaries in our galaxy (and perhaps in nearby galaxies), some of them are known and hence they are guaranteed sources that can be used to calibrate the instrument (the so-called verification binaries); (iv) stellar-mass BH binaries in the inspiral phase (i.e., high-mass LIGO-Virgo-KAGRA sources with $\sim 1$ hour orbital periods); and (v) GW backgrounds, both of astrophysical and early Universe origin, the second type typically associated with physical processes in the TeV energy scale.

The detection and analysis of these sources constitutes an ambitious and revolutionary science programme~\cite{eLISA:2013xep,LISA:2017pwj}, and the details can be found in the series of white papers released by the LISA Consortium~\cite{LISA:2022kgy,LISACosmologyWorkingGroup:2022jok,LISA:2022yao,Seoane:2021kkk,Savalle:2022xpv}). The science case of EMRIs is unique to the low-frequency GW band, and there are several articles and reviews where it has been presented and analyzed (see, e.g., Refs.~\cite{Gair:2012nm,Amaro-Seoane:2007osp,Amaro-Seoane:2012lgq,Sopuerta:2012hg,Amaro-Seoane:2014ela,Babak:2017tow,Gair:2017ynp,Berti:2019xgr,Berry:2019wgg,2020arXiv201103059A,Barausse:2020rsu} and references therein). For the TianQin space mission, prospects for EMRIs have been analyzed in Ref.~\cite{Fan:2020zhy}. For the future, as possible follow-up missions to LISA, the AMIGO concept~\cite{Baibhav:2019rsa}, proposed as a paper for the Voyage 2050 Science programme of ESA, will be an ideal configuration of a space-based detector to carry out the science described in this review. Additionally, there are proposals for atom-interferometric GW detectors, both ground- and space-based detectors, that aim at reaching the deciHertz on ground~\cite{Badurina:2019hst,Canuel:2019abg}, and hence to be able to detect light IMRIs, and also the milliHertz in space~\cite{AEDGE:2019nxb}, thus reaching the EMRI territory and the possibility of detecting heavy IMRIs.

EMRIs have a great potential for revolutionary science discoveries in Astrophysics~\cite{LISA:2022yao}, Cosmology~\cite{Laghi:2021pqk,LISACosmologyWorkingGroup:2022jok}, and Fundamental Physics~\cite{Barausse:2020rsu,LISA:2022kgy}, which is the target of this review. Our general approach to EMRIs provides the right setup to analyze the potential of EMRIs for fundamental physics, in particular for testing the nature of BHs, including their existence, and to test GR and alternative theories of gravity. 

\begin{table}
\centering
\begin{tabular}{>{\centering}m{4cm}>{\centering}m{5cm}>{\centering}m{2cm}}
\toprule 
\textbf{Type of test} & \textbf{Examples} & \textbf{Section} \tabularnewline
\midrule 
Cosmological tests & Hubble constant & \ref{sec:propagation} \tabularnewline
\midrule 
Dipole radiation & Detection of scalar fields & \ref{sec:secondary}\tabularnewline
\midrule 
Dispersion relation & Mass of the graviton & \ref{sec:propagation} \tabularnewline
\midrule 
Enviromental effects & A dark matter distribution & \ref{Sec:Enviroment} \tabularnewline
\midrule 
Gravitational wave polarization & Vector modes & \ref{sec:polarization}
\tabularnewline
\midrule 
The nature of compact objects & Quadrupolar structure &  \ref{Sec:NatureofBHs} \tabularnewline
\bottomrule
\end{tabular}
\caption{\label{tab:tests}Some examples of the various effects and tests discussed throughout this review.}
\end{table}

In what follows, we will present a brief overview focused only on the recent (vigorous and fast!) progress made in the context of EMRIs. See Table~\ref{tab:tests} for a quick roadmap of some of the contents of this review. While acknowledging the limitations of our scope and the inherent biases towards the literature we are familiar with, our conscientious effort to provide a panoramic perspective on the recent advancements ensures that this review encapsulates a compelling glimpse into the expansive tapestry of progress. 

\subsection*{Notation and conventions}
In this review, otherwise stated, we use geometric units in which $c=G=1\,$.

\subsubsection*{List of acronyms:}
\begin{itemize}
    \item[] AAK: Augmented Analytic Kludge
    \item[] AGN: Active Galactic Nuclei.
    \item[] AK: Analytical Kludge.
    \item[] BBH: Binary Black Hole.
    \item[] BH: Black Hole.
    \item[] BHPT: Black Hole Perturbation Theory.
    \item[] b-EMRI: binary Extreme Mass-Ratio Inspiral.
    \item[] DM: Dark Matter.
    \item[] ECO: Exotic Compact Object. 
    \item[] EFT: Effective Field Theory. 
    \item[] EMRI: Extreme Mass-Ratio Inspiral.
    \item[] EOB: Effective One Body.
    \item[] ESA: European Space Agency.
    \item[] FF: Fitting Factor. 
    \item[] GR: General Relativity.
    \item[] GW: Gravitational Wave.
    \item[] IMBH: Intermediate-Mass Black Hole.
    \item[] IMRI: Intermediate Mass-Ratio Inspiral.
    \item[] ISCO: Innermost Stable Circular Orbit. 
    \item[] KAGRA: Kamioka Gravitational Wave Detector. 
    \item[] KAM: Kolmogorov--Arnold--Moser.
    \item[] LIGO: Laser Interferometer Gravitational-wave Observatory. 
    \item[] LVKC: LIGO-Virgo-KAGRA Collaboration.
    \item[] LISA: Laser Interferometer Space Antenna.
    \item[] MBH: Massive Black Hole.
    \item[] MBHB: Massive Black Hole Binary. 
    \item[] NASA: National Aeronautics and Space Administration. 
    \item[] NK: Numerical Kludge.
    \item[] ODE: Ordinary Differential Equation.
%    \item[] PTA: Pulsar Timing Array.
    \item[] PBH: Primordial Black Hole.
    \item[] PDE: Partial Differential Equation. 
    \item[] PPN: Parametrized Post Newtonian.
    \item[] PM: Post Minkowskian approximation.
    \item[] PN: Post Newtonian approximation.
    \item[] SNR: Signal-to-Noise Ratio.
    \item[] SPA: Stationary Phase Approximation.
    \item[] TDI: Time-Delay Interferometry.
    \item[] XMRI: Extremely large Mass-Ratio Inspiral.
\end{itemize}

\subsubsection*{The stationary phase approximation:}
\label{App:SPA}

%ACR-SPA
The stationary phase approximation (SPA) is a generic technique used to approximate the value of an integral, particularly when there are oscillatory functions involved. It is based on the idea that the integral is dominated by the contributions from regions where the phase of the integrand is stationary or nearly stationary. Here we just provide a simple example. The interested reader can read, for example, Chapter 6 of Ref.~\cite{bender1999advanced}. In the GW context, we want to compute, for example the Fourier transform of the waveform~\cite{Cutler:1994ys}
\begin{equation}
    \tilde{h}\left(f\right)=\int_{-\infty}^{\infty}e^{2\pi ift}h\left(t\right)dt \,.
\end{equation}
For example, for the function $B(t) = A(t) \cos(\phi t)$, with $d \ln A/dt \ll d\phi(t)/dt$ and $d^2\phi/dt^2 \ll (d\phi/dt)^2$, the above integral in the SPA is
\begin{equation}
    \tilde{B}\left(f\right)\approx \frac{1}{2} A\left( t\right) \left(\frac{df}{dt}\right)^{-1/2} e^{i \left\{ 2 \pi f t - \phi \left[t\left( f \right) \right] -\pi/4\right\} } \,,
\end{equation}
where $t$ is the time at which $d\phi(t)/dt = 2\pi f$. The SPA is considered to be efficient for tests of gravity with EMRIs in the frequency domain~\cite{Hughes:2021exa}. 

\subsubsection*{Comparing waveforms: The fitting factor}
\label{App:FF}

Once a waveform is generated in a modified theory, it is typically compared to a GR solution. If the waveform in the modified theory can be captured with a GR solution, then the match between the two waveforms would be high (or, equivalently, the mismatch is low). On the other hand, if a feature cannot be captured with a GR template, the match between the two waveforms would be low, and one can distinguish the solutions.  

Given two waveforms $h_1$ and $h_2$, which depend on the source and observation parameters, the fitting factor (FF) for match is defined as~\cite{Cutler:1994ys}:
\begin{align}
FF(h^{}_1, h^{}_2):=\frac{\left<h^{}_1 \left|\right. h^{}_2\right>}{\sqrt{\left<h^{}_1 \left|\right. h^{}_1\right> \left<h^{}_2 \left|\right. h^{}_2\right>}}
    \,,\label{eq:fitting-fact}
\end{align}
where the inner product is given by
\begin{align}
    \left<h^{}_1 \left|\right. h^{}_2\right> := 4R\int\frac{\Tilde{h}^{}_1(f)\Tilde{h}_2^*(f)}{S^{}_n(f)}df
    \,. \label{eq:inner-prod}
\end{align}
In the previous expression, the asterisk denotes a complex conjugate, $\Tilde h$ is the Fourier transform of $h$, and $S_n$ is the noise power spectral density of the detector, typically taken to be the sky-averaged of the detector noise. Since each waveform depends on several parameters, the above fitting factor implicitly should be understood as a maximization process over all the relevant quantities. 

This measure is useful because two waveforms are indistinguishable if the mismatch $\mathcal{M}\equiv1-FF$ satisfies
\begin{equation}
    \mathcal{M}\lesssim \frac{1}{2 \rho^2}\,,
\end{equation}
where $\rho^2$ denotes the signal-to-noise (SNR) of the observed signal.  

\subsubsection*{Phase corrections:}

In this review, as is customary, we report the relative correction of the full Fourier-phase in the SPA as
\begin{equation}
    \Psi\left( f\right) = \Psi_{\rm{GR}}^{(0)}\left[1 + \left( \rm{PN}\, {\rm{corrections}}\right) + \delta\psi \right]\,,
\end{equation}
where $\Psi_{\rm{GR}}^{(0)}$ denotes the leading term of the phase's post-Newtonian (PN) expansion. Given a correction/modification, one can construct the GW template in the frequency domain simply as 
\begin{equation}
\tilde{h} (f)= A e^{i \Psi\left(f\right) } \,. 
\end{equation}
%

%%%%%%%%%%  Astrophysical Channels to EMRI Formation  %%%%%%%%%%
\subsection{Astrophysical Channels to EMRI Formation}
%{\color{red}{Carlos}}

A crucial question to understand what kind of science we can do with EMRIs is what astrophysical mechanisms can produce EMRIs and how many are expected for each mechanism. Moreover, an additional and more detailed question is what is the distribution of the physical parameters that characterize them (e.g., masses, spins, or luminosity distance) for each mechanism? Several studies have tried to answer these questions and produce some quantitative estimates of the event rates and the parameter distribution. We will not enter into details here, but we will give a glimpse of the state of affairs in these questions. Similar questions have been asked about closely connected events to EMRI, in particular for IMRIs~\cite{Miller:2003sc,Miller:2008fi,Vazquez-Aceves:2022wmv} and XMRIs~\cite{Amaro-Seoane:2019umn,Vazquez-Aceves:2021xwl}. In particular, IMRIs are potentially excellent sources for doing fundamental physics with LISA~\cite{Miller:2008fi}. Details and abundant literature on these subjects can be found in the recent review produced by Astrophysics Working Group of the LISA Consortium~\cite{LISA:2022yao}. We can already anticipate that the event rates are quite uncertain due to a poor observational knowledge of many of the ingredients necessary to answer these questions, especially in the IMRI case. In particular, the relevant dynamical processes are quite complex as they involve very different spatial and temporal scales (see, e.g., Ref.~\cite{Amaro-Seoane:2012lgq} for details).  Then, it is difficult to know what to expect, but it also gives even more relevance to the space-based GW observatories like LISA as they will tell us more precisely about these questions. Depending on the actual EMRI/IMRI event rates, if they are at the high-end of the predictions, there is the possibility of having a stochastic background of gravitational radiation produced by these systems (see, e.g., Refs.~\cite{Bonetti:2020jku,Pozzoli:2023kxy} for details).

The most important EMRI channel, and the best studied, is the formation in gas-poor galactic nuclei due to various relaxation processes. It is common to call this type of scenario a \emph{dry}\footnote{The distinction between \emph{wet} (which is less understood) and \emph{dry} channels arises from the presence or absence, respectively, of a dense environment or accretion disk, respectively, influencing the dynamics of the inspiral process~\cite{Amaro-Seoane:2012lgq,Babak:2017tow}.} formation channel, in the sense that the galactic nuclei needs to be relaxed in order to have time to form around it a highly dense (core or cusp) distribution of compact objects that can, at some point, get gravitationally bounded to the central compact object (the primary), so that with time they will emit GWs in the low-frequency band. The more compact objects there are in this distribution the more likely will be to form an EMRI. Typically, the EMRI system formed in this way has a large eccentricity, but little is known about the distribution of the orbital inclination. As we have mentioned above, the dynamics is quite complex, and many factors count (see Ref.~\cite{Amaro-Seoane:2012lgq} for a detailed discussion), including the spin of the primary~\cite{Amaro-Seoane:2012jcd}. A different formation mechanism proposed is the formation and tidal disruption of binaries around a MBH~\cite{ColemanMiller:2005rm}. The idea is that one of the components of the binary, once tidally disrupted, gets gravitationally bounded to the MBH, while the other one is ejected and can become a hypervelocity star. A remarkable feature of these EMRIs is that, in contrast to the aforementioned ones, they have a quite low eccentricity, although in certain scenarios we may also expect high-eccentricity EMRIs~\cite{Raveh:2020jxg}. There are other proposals for EMRI formation but their rates are even more uncertain. Some of them are, for example: capture of cores of stars in disks~\cite{DiStefano:2001ci,Davies:2005tc}; formation of EMRIs in Active Galactic Nuclei (AGNs)~\cite{Levin:2003ej,Levin:2006uc,Secunda:2020cdw,Pan:2021ksp,Pan:2021oob}; or EMRIs triggered by supernova kicks in galactic centers~\cite{Bortolas:2019sif}.  

In the case of IMRIs, as we have mentioned above, it is important to distinguish between light IMRIs, where the primary, in the standard scenario, is an IMBH; and heavy IMRIs, where the primary is a MBH. There is a variety of scenarios for producing them, but the event rates are very uncertain. This is due in part to the very limited observational evidence about the existence of IMBHs (see, e.g., Ref.~\cite{Mezcua:2017mm}). In any case, let us emphasize again, that it is important to keep these systems in mind as they can be very important for fundamental physics and their description and dynamics is intimately related to the one of EMRIs.

Until now, we have just mentioned general EMRI/IMRI formation mechanisms. But plenty of special cases deserve special attention due to their relevance for fundamental physics and cosmology. For example, it would be exciting to detect EMRIs or IMRIs with third-generation ground-based or space-based GW observatories, where the secondary is a subsolar-mass BH of primordial origin~\cite{Guo:2017njn,Barsanti:2021ydd}. These detections can be the key to prove the existence of BHs of primordial origin (PBHs), which may be the constitutive element of dark matter (DM)~\cite{Carr:2016drx,Villanueva-Domingo:2021spv}. Of similar interest would be to detect EMRIs/IMRIs with a secondary in the mass gap~\cite{Pan:2021lyw}.

%%%%%%%%%%  EMRI Detectability with LISA  %%%%%%%%%%
\subsection{EMRI Detectability}\label{emri-detectability}

There are several factors that one has to consider to assess the detectability of EMRIs and the precision with which we can extract the physical parameters from the GW detector data stream. This has been studied in detail for the case of the future LISA detector and can be found in many different papers and reports on the mission. Here, we only mention the most important aspects relevant to EMRIs.

First of all, we can look at the case in which we consider individual EMRI signals in the data stream of a detector like LISA, where we take into account a model for the instrumental noise and the confusion noise created by the population of galactic binaries, which is dominated by white dwarf binaries. In the case of the LISA mission, in Ref.~\cite{Babak:2017tow}, using several astrophysical models for EMRI populations, estimations of the event rates, SNRs, and parameter-estimation precision were given. Let us just mention three important conclusions from that study. First, all the astrophysical models considered predict that LISA should be able to detect a few EMRIs per year (with the most optimistic projections, the event rate could reach a few thousand per year). Second, in the case of the LISA mission, we should be able to make very precise estimations of the EMRI physical parameters. In particular, we should recover intrinsic parameters like the (redshifted) masses, the spin of the primary and orbital eccentricity with a precision of $\sim 10^{-6}
- {10}^{-4}$. The precision for the determination of the luminosity distance should be around $\sim 10\%$ and for the sky localization, around a few square degrees. Third, and of special interest for this review, tests of the multipolar structure of the primary (see SubSec.~\ref{multipole-moments-of-the-primary} for a detailed discussion) should be possible at a percent-level or even better. This summarizes the situation in the case of the \emph{standard scenario} (neglecting the spin of the secondary). 

There are a number of factors that can make the detection of EMRIs either better or worse. For instance, the standard scenario ignores the possible astrophysical environmental effects, which can affect the EMRI dynamics and hence, their detectability and the precision to extract physical parameters (see Sec.~\ref{Sec:Enviroment} for a detailed discussion).  This means that in order to be able to make tests of gravity with EMRIs, we must have an excellent understanding of the astrophysical environmental effects that can be confused with deviations either from the BH paradigm or even from GR itself. Another potential danger for precise parameter estimation is the appearance of degeneracies in the parameter space of EMRIs.  This question has been recently studied in Ref.~\cite{Chua:2021aah}, where the authors introduce new tools for this study and propose strategies to mitigate the possible appearance of degeneracies. It may also be possible that, as we get closer to more precise waveform EMRI models (see Sec.~\ref{Sec:WaveformGeneration} for details), degeneracies in the parameters become less and less important. For instance, it has been recently shown in Ref.~\cite{Burke:2020vvk}, for the case of LISA, that in the case of near-extremal BHs (with a dimensionless spin parameter $a\sim 0.9999$ or even closer to unity), we could measure the spin of the primary with extraordinary precision, of the order of 3-4 orders of magnitude better as compared with the case of moderate spins (with $a\sim 0.9$).

Still within the standard scenario, one of the most common questions about EMRIs is about the effect and detectability of the spin of the secondary, the determination of which may provide invaluable information about the nature of the secondary. This fascinating topic has been receiving increasing attention in recent years. On the other hand, detectability has been studied using Numerical Relativity (NR) waveforms~\cite{Piovano:2021iwv}. There are also studies of the detectability of the spin of the secondary using the induced quadrupole deformation on the primary~\cite{Rahman:2021eay}. 

Another challenge in doing fundamental physics with GW detections of EMRIs is the systematic errors one can make in the data analysis process, which is particularly relevant for high SNR sources, as some of the EMRI sources can be. These systematic errors have to do with the overlapping of signals from different sources and also inaccuracies in the waveform models used for parameter estimation (there are studies for the case of MBH mergers, see, e.g., Ref.~\cite{Cutler:2007mi}, and Ref.~\cite{Hu:2022bji} for a study considering  3G detectors).

In Ref.~\cite{Hu:2022bji}, it is argued that waveform inaccuracies contribute most to the systematic errors, but multiple overlapping signals could magnify the effects of systematics owing to the incorrect removal of signals. They also point out that using selected golden binaries\footnote{Golden binaries here refer to those low-frequency binary GW sources that are bright enough to provide a very good parameter estimation, in particular of the sky localization (see, e.g., Refs.~\cite{Sberna:2020ycl,Hughes:2004vw}). In the case of LISA, they are usually considered in the context of tests of the nature of BHs and of the gravitational interaction. For the case of MBHBs, they have SNR$>100$ in the post-merger phase, and in the case of EMRIs, they have an accumulated SNR$>50$ (see, e.g., Ref.~\cite{LISA:2017pwj}).} is even more vulnerable to false deviations from GR, pointing to the need to develop adequate techniques and analyses to prevent these problems. Mitigation strategies have been recently proposed in Ref.~\cite{Owen:2023mid} for the case of current ground-based detectors.

Regarding data analysis strategies for EMRIs, they have to be part of what is known as the \textit{global fit}, which is probably the best data analysis strategy for the low-frequency band. The idea consists in fitting simultaneously the different GW sources and also the different noises in the data (see, e.g., Ref.~\cite{Littenberg:2023xpl} for a prototype of what a \textit{global fit} is expected to look like for the case of LISA). Nevertheless, most strategies to date were developed for the search of single EMRI events. Some techniques/strategies for detecting individual EMRIs and extracting their physical parameter have been proposed~\cite{Cornish:2008zd,Gair:2008zc,Babak:2009ua,Wang:2012xh,Babak:2014kqa,Zhang:2022xuq}. Also, new tools for EMRI data analysis have been recently developed~\cite{Chua:2019wgs,Chua:2021aah,Chua:2022ssg,Saltas:2023qec}. A useful tool based on machine learning to predict the SNR of an EMRI from its parameters have been presented in Ref.~\cite{Chapman-Bird:2022tvu}.

%%%%%%%%%%  EMRI modelling  %%%%%%%%%%
\section{EMRI modelling}
%\section{EMRI modelling and waveform families}

The goal of EMRI modelling is to ultimately produce waveform template banks that can be used for GW data analysis, that is, to detect the source and estimate the corresponding physical parameters (together with their posterior probability distributions). To achieve this objective, we need to have a deep understanding of the dynamics of EMRIs. The main particular aspect of EMRIs is the presence of very different scales, both spatial and temporal, due to the extreme mass ratios involved. It is precisely the smallness of the mass ratio $q$ that allows us to make the first and most important approximation to EMRI modelling: to ignore the structure of the secondary and treat it as a \emph{particle} (an energy-momentum distribution supported on a timelike curve) endowed with the relevant physical properties (mass, spin, and any other physical properties~\cite{Dixon:1970zza,Dixon:1970zz,Dixon:1974xoz}, the so-called \emph{hair} in the case of BHs). It is known that in full GR the presence of particles introduces some difficulties~\cite{Ehlers:2003tv,Bezares:2015eg,Geroch:2017hdb}, but in a context where perturbative techniques can be used, as in the case of EMRIs, the point-mass approximation provides simplifications to the treatment of the dynamics. The alternative approach, considering an extended energy-momentum distribution for the secondary, leads to a much more complex description in which the matter equations of motion may be even more challenging than the gravitational sector.  In any case, the main challenge in describing the dynamics of EMRIs is understanding how the secondary trajectory around the primary interacts with its own gravitational field.

In practice, the modelling of EMRIs involves the integration of different sets of equations: the equations describing the dynamics of the gravitational field (Einstein's equation in GR); the equations describing the orbital motion of the secondary around the primary (the equation of motion for the particle worldline); and the equations for matter fields that may affect the geometry of the primary, secondary or their dynamics. Without imposing any approximations, this represents a system of non-linear coupled partial differential equations (PDEs) and ordinary differential equations (ODEs) which, in general, requires the use of numerical methods. NR is the field that was developed precisely with this aim. Given the success of NR to solve the binary black hole (BBH) problem in GR~\cite{Pretorius:2004jg,Pretorius:2005gq,Baker:2005vv,Campanelli:2005dd}, one may think of using it to model EMRIs. However, given the current state of affairs, it is not an option because of the presence of at least two very different scales in the EMRI problem. In terms of lengths we have the primary versus the secondary sizes, which scales with the mass ratio. In terms of times, we have the orbital timescale versus the radiation reactions timescale, which also scales with the mass ratio. Therefore, using NR would require the use of sophisticated techniques of refinement of space and time discretizations to capture these two different scales. But even in the best possible scenario, the problem will be still not feasible from the point of view of numerical resources. Despite these challenges, there has been some proof-of-principle studies~\cite{Lousto:2010ut,Lousto:2020tnb,Lousto:2022hoq} to push NR to intermediate mass ratios ($q\sim 10^{-3} - 10^{-2}$) for a few orbital cycles.

In view of the situation with NR, most efforts have been focused to a semianalytic approach based on BH perturbation theory (BHPT; see Ref.~\cite{BHPToolkit} for a website with useful tools to make computations within this framework), the \textit{self-force} programme~\cite{Mino:1996nk,Quinn:1996am} (see, for instance, Refs.~\cite{Barack:2009ux,Poisson:2011nh,Pound:2015tma,vandeMeent:2017zgy,Barack:2018yvs} for reviews). This programme, which is nowadays considered the best approach to model EMRIs, has seen very important advances in the last years which we briefly describe below. The fact that the self-force programme is based on BHPT means that it has to be generalized for cases in which the primary is not a BH. In any case, any approach to EMRIs dynamics can be seen as a simplification/approximation to the self-force approach.

The basic idea behind the self-force approach is that the EMRI spacetime metric can be described as the metric associated only to the primary plus perturbations generated by the presence of the secondary, which is described as a particle. The interaction of these gravitational perturbations with the secondary itself (gravitational backreaction) is what drives the inspiral. This backreaction, which appears as the action of a \emph{force} vector (the self-force) that governs the deviation of the secondary from geodesic motion around the primary, has to be computed at different perturbative orders. And then, the effect of the corrected motion on the gravitational waveform has to be estimated (it enters at the next perturbative order). 

In analogy with the case of electrodynamics~\cite{Barut:1980aj,Jackson:1999jd,Rohrlich:2000fr}, where the self-force approach was already introduced (see Refs.~\cite{Abraham:1904ma,Lorentz:1936ha,Dirac:1938pd} for the case of special relativity and Refs.~\cite{DeWitt:1960fc,Gralla:2009md} for the extension to curved spacetimes), the self-force has two different pieces, a regular and a singular one (with the singularity located at the position of the secondary). This decomposition was presented by Detweiler and Whiting~\cite{Detweiler:2002mi} by studying the Green function associated with the perturbative Einstein equations. The singular part (which satisfies the inhomogeneous equations with the particle as the source) is inversely proportional to the proper distance from the point at which it is evaluated to the particle, and it corresponds to the field tidally distorted by the local (around the secondary) Riemann tensor, and it exerts no force on the secondary. The regular part (which satisfies the homogeneous perturbative equations) is what determines the actual self-force (the interaction of the particle with its own gravitational field). In this way, Detweiler and Whiting showed that the motion of the secondary around the primary can be seen as geodesic motion in the perturbed spacetime, but in terms only of the regular piece of the metric perturbations~\cite{Detweiler:2002mi}. This is a remarkable result that gives a more intuitive perspective to the self-force formalism.

This analysis of the structure of the self-force also indicates an important difference between the case of flat and curved spacetimes. In contrast with the case of flat spacetime, where the effect of the self-force is purely dissipative, in curved spacetime the self-force also has also a conservative piece. In the case of EMRIs, the action of this conservative piece induces effects as, for instance, decreasing the rate of periastron advance~\cite{Osburn:2015duj}.

Currently, the self-force programme has been able to produce accurate first-order perturbative self-force computations for generic orbits both around a Schwarzschild primary~\cite{Barack:2010tm} and a Kerr primary~\cite{vandeMeent:2017bcc}. In the case of Schwarzschild, a whole inspiral using the first-order self-force has been produced~\cite{Warburton:2011fk}. This has recently been extended to eccentric inspirals into a rotating BH in Ref.~\cite{Lynch:2021ogr}. The computation uses an action angle formulation of the method of osculating geodesics for eccentric, equatorial (i.e., spin-aligned) motion in Kerr spacetime. The forcing terms are provided by an efficient spectral interpolation of the first-order gravitational self-force in the outgoing radiation gauge. On the other hand, there is currently significant progress towards second-order self-force computations in the standard scenario:~\cite{Pound:2019lzj,Upton:2021oxf,Wardell:2021fyy,Spiers:2023mor}, which are required to reach the precision needed for the detection and parameter estimation. There are also some self-force calculations for the case of a spinning secondary~\cite{Mathews:2021rod}, and, beyond GR, some progress has been made towards introducing scalar charge for the secondary~\cite{Spiers:2023cva}.  

Looking at the signal of a typical EMRI in a detector like LISA, we know that it can spend of the order of $10^4-10^6$ ($= N_{\rm cycles}$) GW cycles in band. This  means that EMRIs emit long-lived GW signals that accumulate their SNR over the observable lifetime of the inspiral.  Actually, one year before the plunge of the secondary into the primary, an EMRI system can spend up to $\sim 10^5$ GW cycles in band, most of them of a highly relativistic nature. This number of cycles scales with the inverse of the mass ratio, $N_{\rm cycles} \propto q^{-1}$, although in the relatively short transition from inspiral to plunge it scales as $q^{-3/5}$~\cite{Ori:2000zn}. Furthermore, since typical space-borne instruments operating in the low-frequency band consists of a constellation of satellites, one expects to get a good localization, within $10\; \rm{deg}^2$~\cite{Berry:2019wgg}, due to the frequency modulation induced by the motion of the constellation. To ensure meticulous measurements of a system's features, it's crucial to possess a precise depiction of its frequency development. 

In the limit of zero mass ratio, $q\rightarrow 0$, we are in regime where the secondary is a test mass (with zero spin) and hence, it follows a geodesic of the primary, a Kerr BH in the standard scenario.  An important characteristic of the geodesic motion around a Kerr spacetime is that it is completely integrable, which is a consequence of the existence of hidden symmetries of the metric, in particular of a second-rank Killing tensor that leads to a new conserved quantity along the geodesics, the Carter constant~\cite{Carter:1968rr,Walker:1970un} (see Subsec.~\ref{Sec:Conse} below for details). 
As a consequence, geodesics of the Kerr spacetime are characterized by three fundamental frequencies/periods that set the orbital time scale, which is determined by the mass of the primary, $M$.  On the other hand, we have the inspiral time scale, the typical time for the orbit to shrink, which is determined by the mass of the secondary, or equivalently, by $q M$. 

Up to here, we have described the fundamental ingredients of EMRI dynamics following the standard scenario. This gives us the perspective to face deviations either due to a modification of the primary (non-Kerr spacetime) or to a modification of GR, or both.
The evolution of the frequency of the GWs emitted depends \emph{both} on the conservative (time-symmetric) dynamics, as well as on the dissipative (time-asymmetric) dynamics, i.e., 
\begin{equation}
\label{eq:freqevol}
\frac{df}{dt}=\underbrace{\left(\frac{dE}{df}\right)^{-1}}_{\textrm{conservative}}\overbrace{\frac{dE}{dt}}^{\textrm{dissipative}}\,.
\end{equation}
When discussing scenarios beyond the standard one, and in particular alternative theories of gravity, it is possible to modify one or both contributions. Before discussing the various types of tests, let us take a quick look at how these modifications are typically introduced in both sectors.  To that end, we are going to first differentiate modifications that directly affect the conservative dynamics of EMRIs from modifications that affect the dissipative dynamics.  In this sense, it is important to realize that this distinction is only about the nature of the modifications with respect to the standard scenario.  It is clear that changes of a conservative nature affect the dissipative dynamics and the converse.  Then, the conservative effects that we describe in what follows refer to changes that affect the background geometry of the primary.

%%%%%%%%%%
%
\subsection{Conservative changes to the EMRI evolution}
\label{Sec:Conse}
We are now going to look at changes of the standard scenario that alter the background spacetime. Of particular relevance are changes of the form:
\begin{equation}
\label{eq:modmetric}
g^{\rm primary}_{\mu\nu} = g_{\mu\nu}^{\textrm{GR}} + \delta g^{}_{\mu\nu} \,,
\end{equation}
where the first term typically denotes the metric of the primary in GR, the Kerr metric in the standard scenario, and the second term is a generic modification (perturbation of the background). For instance, the perturbation encapsulated in the second term in this equation can be induced by a beyond GR modification, but it can also represent, for example, an environmental effect or a tidal deformation from a third body within GR, or even a combination of them. However, as we will see in Sec.~\ref{Sec:Enviroment}, a modification of the form~\eqref{eq:modmetric} is very unlikely to capture the \emph{astrophysical bumpiness} expected to be around the primary. Thus, for the purposes of this review, we will mostly assume it arises from a modification of GR. 

As mentioned earlier, the EMRI dynamics can be modelled, locally in time, by geodesics of the background spacetime. Under that assumption, we can write the relativistic Hamiltonian as
\begin{equation}
\label{eqn:Ham}
H = \frac{1}{2 m}g^{\mu\nu}p^{}_{\mu}p^{}_{\nu}\,,
\end{equation}
where here $g_{\mu\nu}$ denotes the metric of the primary [see Eq.~\eqref{eq:modmetric}], $m$ is the rest mass of the secondary, and its corresponding (conjugate) four-momenta is $p_{\mu} \equiv m \; u_{\mu}$, with $u^{\mu}= dx^\mu/d\tau$, where $\tau$ is the proper time, its normalized four-velocity (satisfying $g_{\mu\nu}u^\mu u^\nu=-1$). At this level of approximation, the orbital motion of the EMRI is determined by solving Hamilton's equations
\begin{equation}
\label{eq:EOMH}
\dot{q}^{\mu} = \frac{\partial H}{\partial p^{}_{\mu}}\,, \qquad 
\dot{p}^{}_{\mu} = -\frac{\partial H}{\partial q^{\mu}}\,,
\end{equation} 
where the overhead dot denotes a derivative with respect to the affine parameter of the geodesic, and since we are considering massive test particles, it can be taken to be proper time. Note that in this relativistic Hamiltonian, $g^{\mu\nu}$ includes a modification in the form shown in Eq.~\eqref{eq:modmetric}.

Analogously, we can describe the motion from a Lagrangian perspective, and write the Lagrangian as
\begin{equation}
{\mathcal L} = \frac{m  }{2} g^{}_{\mu\nu} \frac{d x^\mu}{d\tau}  \frac{d x^\nu}{d\tau}\,,
\end{equation}
where $\tau$ is proper time. As the rest mass of the orbiting test particle is constant along the trajectory, we have that 
\begin{equation} 
\label{4vel}
p^\mu p^{}_\mu = -m^2 \,.
\end{equation}
Thus, the value of the Lagrangian is a conserved quantity, i.e., ${\mathcal L} =-m/2$.

\begin{figure}
\includegraphics[width=\columnwidth]{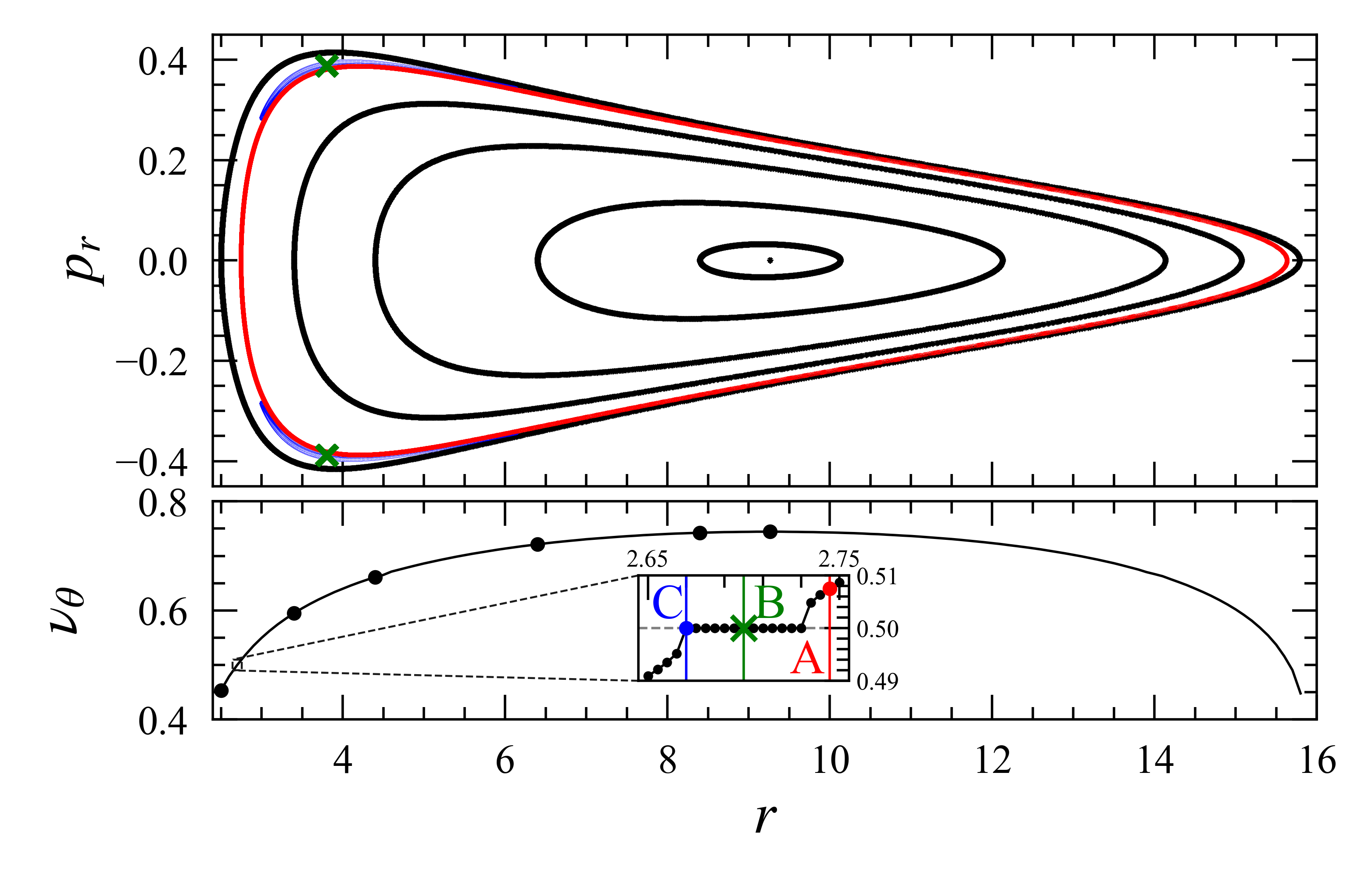}
\caption{Poincaré surfaces of section (top) and the rotation curves (bottom), for orbits with $E = 0.95$, $L_z = 2.5\,M$, and $a=0.8$ in a slowly-rotating Kerr BH (i.e., the Kerr solution to $\mathcal{O}\left( a^2\right)$, which is \emph{not} integrable~\cite{Cardenas-Avendano:2018ocb}). In the lower panel,  particular initial conditions are marked with black dots corresponding to the depicted surfaces of the section shown above. The inset shows the details around a plateau, corresponding to the resonant KAM tori at rotation number $\nu_\theta=0.5$. Trajectories with initial conditions $r=2.745\,M$ (denoted with the letter A and plotted in red), $r=2.700\,M$ (denoted with the letter B and plotted in green), and $r=2.670\,M$ (denoted with the letter C and plotted in blue) are also highlighted in the top panel with corresponding colors. Under radiation reaction, a particle starting at the initial ``regular" condition A should evolve towards the chaotic region. Conditions B and C are part of the same family of Birkhoff islands and therefore share the same rotation number value (a plateau in the rotation curve $\nu_\theta \left( r \right)$).} 
\label{Fig:PSS}
\end{figure}

Assuming that the primary is stationary and axisymmetric, the metric is then independent of the coordinates $t$ and $\phi$, which provides two constants of motion, namely the specific energy at infinity $E$ and the axial component of the specific angular momentum at infinity $L_z$:
\begin{eqnarray}
p^{}_t & \equiv & \frac{\partial {\mathcal L}}{\partial \dot{t}}
= g^{}_{tt} \dot{t} + g^{}_{t\phi} \dot{\phi} = - E \,,\label{Eq:E}  \\
 p^{}_\phi &\equiv& \frac{\partial {\mathcal L}}{\partial \dot{\phi}}
= g^{}_{t\phi} \dot{t} + g^{}_{\phi\phi} \dot{\phi} = L_z  \,, \label{Eq:L} 
\end{eqnarray}
where, again, the dot denotes a derivative with respect to the proper time $\tau$. The term ``specific'' is used to indicate that $E$ and $L_z$ are, respectively, the energy and the angular momentum (along the symmetry/spin axis) per unit rest-mass. These two constants of the motion, $E$ and $L_z$, are already enough to reduce the system to only two degrees of freedom and its associate phase space to four dimensions. This means, that one can easily study the geodesic motion in phase space, by solving the geodesic equations
\begin{equation} 
\label{geodesicNRR}
\frac{d^2 x^\mu}{d\tau^2} =- \Gamma^\mu_{\nu\rho} 
\frac{d x^\nu}{d\tau}
\frac{d x^\rho}{d\tau}  \,,
\end{equation}
where $\Gamma^\mu_{\nu\rho}$ denotes the Christoffel symbols, by performing a cut of the four dimensional torus where each trajectory lives. The resulting 2D curve (in phase space) of a trajectory is known as a Poincar\'e surface or section, which we show in Fig.~\ref{Fig:PSS}. In practice, it is constructed by recording the values of a coordinate and its conjugate momentum, every time it crosses a plane in physical (configuration) space. For the example shown in Fig.~\ref{Fig:PSS}, we have chosen the equatorial plane $\theta=\pi/2$. 

As we have mentioned above, geodesic motion in GR has a remarkable property that may not necessarily hold in a modified theory: it is integrable~\cite{Carter:1968rr}. The (Liouville) integrabilty~\cite{Contopoulos_2002} means that if there are $n$ linearly independent integrals of motion in a system of $n$ degrees of freedom, then there exists a coordinate transformation to angle-action variables such that the equations of motion can be written in quadrature form. Operationally, it means that one can easily integrate the equations of motion in terms of action-angle variables. In GR, the motion is integrable because of the presence of an additional constant of motion (the others are $m$, $E$ and $L_z$), known as the Carter constant~\cite{Carter:1968rr} (which is not uniquely defined and in the literature one can find different definitions):
\begin{equation} 
\label{Eq:CarterConstant}
Q = u_\theta^2 + \cos^2 \theta \left[ a^2 \left( m^2 - E^2 \right) + \left(\frac{L_z}{\sin \theta}\right)^2 \right]\,.
\end{equation}
This constant of the motion is a consequence of the existence of a second-rank Killing tensor~\cite{Walker:1970un}, $\xi_{\mu\nu}$ ($\nabla_{(\mu}\xi_{\rho\sigma)}=0$), so that the Carter constant is obtained as: $Q = \xi_{\mu\nu} u^\mu u^\nu$, with $u^\mu$ being the secondary four-velocity. 
Thus, all the trajectories, that describe bound motion, will form in phase space smooth closed curves (black lines in Fig.~\ref{Fig:PSS}). 

Without gravitational backreaction (i.e., the test mass approximation), all initial conditions for bound orbits generate periodic or quasi-periodic trajectories. Consequently, different initial conditions correspond to different values of the ratio of the characteristic frequencies of the system. For geodesics in the Kerr background geometry, the ratio of the polar to the radial frequency of the motion, $\omega_\theta/\omega_r$ is the relevant quantity to study the motion. Periodic orbits happen when this ratio is a rational number $k/j$, with $(j, k)\in \mathbb{N}$, and consequently, the phase-orbit repeats itself after $j$ windings. On the other hand, quasi-periodic orbits happen when the ratio of $\omega_\theta/\omega_r$ is irrational, and the phase-orbit is densely covered.

When two or more characteristic frequencies of the system are in rational proportion the system is said to be in \emph{resonance}, the phase-orbit is not densely covered, and a pattern is created. In the presence of modification that is not integrable, the Kolmogorov--Arnold--Moser (KAM) theorem implies that some of the invariant tori will be deformed and survive, while others are destroyed. Which ones? The theorem does not tell! For sufficiently small perturbations, such as the ones expected from a modification of GR that is still consistent with current tests, almost all non-resonant tori will be deformed~\cite{Cardenas-Avendano:2018ocb}. For quasi-periodic orbits, the resulting deformed phase-orbits are called KAM curves. Close to the resonant tori, however, it is expected that nonlinear resonances and chaos to develop, signaling a deviation from vacuum GR~\cite{LukesGerakopoulos:2010rc}. To summarize, around a resonance, a modification of GR that is not integrable (e.g., due to the absence of a fourth constant of the motion), will show a difference in the EMRI evolution. Note that resonances are very likely to occur during an EMRI evolution, and even expected to last for a few hundred orbital cycles at mass ratios of the order of~$10^{-6}$~\cite{Ruangsri:2013hra}.  

Therefore, even at the purely conservative level, one can already see differences and interesting features in the evolution with respect to Kerr geodesics. These features, however, do not necessarily have to come from wild solutions or modifications to GR. If one, for instance, considers a slowly rotating Kerr metric, i.e., the Kerr solution up to a particular order in the spin, the resulting trajectories already show up these features~\cite{Cardenas-Avendano:2018ocb}. For example, the top panel in Fig.~\ref{Fig:PSS} shows some trajectories in phase space for different initial conditions with $a=0.8$, $E=0.95$ and $L_z = 2.5\,M$, while the bottom panel shows the rotation curve for all the initial conditions that provide bounded motion for those values of spin, energy, and angular momentum. The rotation curve $\nu_\theta(r)$ offers information on the localization and emergence of Birkhoff islands (a set of small KAM curves appear), and abrupt changes signal the presence of chaotic orbits~\cite{Contopoulos_2002}. Embedded in the lower panel is a zoom of the rotation curve around a regime that presents a chaotic feature: a plateau in the rotation number. The rotation number is defined as~\cite{Contopoulos_2002} 
\begin{equation}
\label{Eq:rotnum}
\nu^{}_\theta(r) = 
\lim_{N \rightarrow \infty} \frac{1}{2\pi N}\sum_{i=1}^{N}\theta^{}_i\,,
\end{equation}
where $\theta_i$ is the clockwise angle subtended by two vectors, defined from the invariant point, to two consecutive successive piercings of the Poincare's surface of section, i.e., $\theta_i = \measuredangle (\vec{v}_{i+1},\vec{v}_{i})$. The invariant point is the fixed point corresponding to the periodic orbit, which crosses the two-dimensional slice defining the Poincaré surface, which here was taken to be the equatorial plane, at only one point with $P_r = 0$. The rotation curve of the system is obtained by evaluating the rotation number as a function of the location of the Poincar\'e surface of section in phase space, which for this example was defined as the location of the surface by the minimum value of the radial coordinate sampled by that surface. 

The rotation number serves at least two purposes. Firstly, it identifies the system dynamics. Secondly, it detects a constant ratio of orbital frequencies that should translate into a pattern of frequencies in the GWs emitted. If there is a plateau in the rotation number, it indicates the presence of chaos. This can be used as a test of GR and the Kerr hypothesis~\cite{Gair:2007kr,LukesGerakopoulos:2010rc}. That is why these type of features have been studied for different metric solutions (see, for example, Refs.~\cite{Ryan:1995wh,Collins:2004na,Gair:2007kr,LukesGerakopoulos:2010rc,Cardenas-Avendano:2018ocb,Destounis:2020kss,Destounis:2021mqv,Deich:2022vna} for both theory-specific and theory-agnostic examples). The notion is, in a nutshell, that if the motion associated with $g+\delta g$ is not integrable (chaotic), then features in the GWs emitted by EMRIs could signal either a departure from the strong-equivalence principle or a violation of the Kerr hypothesis~\cite{LukesGerakopoulos:2010rc,Destounis:2021mqv}.  

Until now, we have not considered the radiation effects generated by an EMRI as the particle moves along a geodesic. To get an inspiral, one needs to include the effect of losses (fluxes) of at least energy and angular momentum. This can be done by using, for example, the ~\emph{adiabatic} approximation, where by means of balance laws one can correct the orbital parameters defining the geodesic (energy, angular momentum, and Carter constant in the case of a Kerr BH primary) according to the fluxes associated to each orbital parameter (see Eq.~\eqref{linearizedfluxes} in next subsection for details on the dissipative effects of the self-force). 

Studying the differences with respect to GR by changing the metric requires to build the trajectory from Eq.~\eqref{eq:modmetric}, compute an inspiral using some fluxes (e.g., the linearized ones according to the prescription shown in Eq.~\eqref{linearizedfluxes}, compute the waveform (using any of the methods we describe below in Sec.~\ref{Sec:WaveformGeneration}), and compare the resulting non-Kerr waveforms to Kerr waveforms, via, for example, fitting factor (see~\ref{App:FF}). Typically, what is found is that, for the major part of the parameter space, and away from resonances, there could be a \textit{confusion} problem~\cite{Babak:2006uv}: while the approximate GWs emitted in a perturbed/modified spacetime are different from those emitted by the same compact object moving along the same orbit in a pure Kerr spacetime, with the same mass and spin as the perturbed one, it is possible to still find the conditions where their orbital trajectories will have the same $\omega_r$ and $\omega_\phi$ frequencies, and therefore be indistinguishable. 

Studying the dephasing under the adiabatic approximation has been pursued more than a decade ago~\cite{Gair:2007kr}. However, around resonances, the dynamics is thought to be where a ``smoking gun'' of departures from Kerr motion and observable signatures can take place~\cite{Apostolatos:2009vu,LukesGerakopoulos:2010rc}. Several studies have shown the presence of features in the trajectories. However, trajectories are not an observable, so one needs to compute gravitational waveforms to properly study these effects. 

Using a metric \emph{designed} to break integrability and enhance the difference from GR, derived in Ref.~\cite{Destounis:2020kss}, Ref.~\cite{Destounis:2021mqv} computed approximate gravitational waveforms, by employing a hybrid \emph{kludge} scheme~\cite{Barack:2003fp,Gair:2005ih} to evolve EMRIs with a non-Kerr primary, and the Einstein-quadrupole approximation to model the GW emission. They showed that non-integrability displays a \textit{glitch} phenomena, where the frequencies of GWs increase abruptly, as expected from previous works that studied only the trajectories, when the orbit crosses the Birkhoff islands. The presence or absence of these features in future data may therefore allow not only for tests of GR but also of fundamental spacetime symmetries~\cite{Cardenas-Avendano:2018ocb,Destounis:2021rko,Deich:2022vna}. 

A consistent gravitational model is required to use these features as a test of GR. In particular, the model should accurately capture the EMRI evolution on and off-resonances. Work in this direction is currently underway. For instance, in Ref.~\cite{Pan:2023wau}, using the action-angle formalism, an effective resonant Hamiltonian was derived that describes the dynamics of the resonant degree of freedom, for the case that the EMRI motion across the resonance regime. They studied the dynamics of an EMRI  system near orbital resonances, assuming the background spacetime is weakly perturbed from Kerr described by dynamically modified Chern--Simons gravity~\cite{Alexander:2009tp}. Under this approach, they concluded that in the non-adiabatic regime, the transient resonance crossing gives rise to a GW dephasing given by
\begin{equation}
\delta \Psi_{\rm resonance~crossing} \sim \frac{ \mathcal{O}\left(\delta g\right)}{q^{3/2}} \,,
\end{equation}
which, depending on the perturbation, can have a significant observational impact. This study, however, did not included gravitational radiation reaction, and therefore more work is required to fully understand the behaviour around resonances. Without these effects under control, an EMRI waveform model for a generic Kerr perturbation will not be complete.

Based on the assumptions mentioned, several studies have concluded that the non-integrability nature of a modified theory of gravity is not likely to have a major effect on EMRIs. These studies have also demonstrated that a thorough understanding of dynamical systems can aid in comprehending the characteristics of gravity theories. However, it is still unknown how a system will react to realistic disturbances when all the physics are accounted for (see Ref.~\cite{Lukes-Gerakopoulos:2021ybx} for a discussion of non-linear effects in the dynamics of EMRIs and the possibility of detecting them in the emitted GWs).

\subsection{Dissipative changes to the EMRI evolution}
\label{Sec:Diss}

Conservative effects represent modifications of the stationary field of the primary, which change the orbital trajectory, and since the gravitational radiation emission strongly depends on the details of the orbit, it also affects the rate at which the inspiral indirectly takes place (see previous subsection). This subsection deals with direct changes to the gravitational radiation emission mechanism. That is, modifications of the theory affecting the radiative field. Looking for these effects has been frequently proposed as a method to test GR and alternative theories of gravity using GWs.

The radiation reaction effects are also properly accounted by the self-force, and the programme to compute it and the ''correct'' EMRI waveforms has already been introduced before in this review. In brief, the self-force encodes the effect of the gravitational field of the secondary on its own trajectory. It turns out that when we treat the secondary as a point particle, the \emph{bare} self-force (as computed from the retarded perturbative field) is ill-defined as it diverges at the location of the secondary. In a sense, we can say that the Coulombian gravitational field diverges at the location of the mass that creates it, as it happens with the gravitational field of particles in Newtonian theory. This is what generates what we referred before as the singular piece of the self-force, and which Detweiler and Whiting~\cite{Detweiler:2002mi} showed that does not contribute to the EMRI dynamics. Then, once we regularize the self-force (for instance, using the so-called \emph{mode sum} regularization scheme~\cite{Barack:1999wf,Barack:2000eh,Barack:2001bw,Mino:2001mq,Barack:2001gx,Barack:2002mha,Detweiler:2002gi,Barack:2002bt}), it can be included in the equation of motion, modifying in this way the geodesic equations, Eqs.~\eqref{geodesicNRR}, in the following way:
\begin{equation}
    \label{geodesic}
\frac{d^2 x^\mu}{d\tau^2} + \Gamma^\mu_{\nu\rho}  \frac{d x^\nu}{d\tau}
\frac{d x^\rho}{d\tau} = \mathcal{F}^{\mu}_{\rm SF} \,,
\end{equation}
where $\mathcal{F}^{\mu}_{\rm SF}$ denotes the first-order regularized self-force and the Christoffel symbols, $\Gamma^\mu_{\nu\rho}$, refer to the ones associated with the primary metric (the background metric). From this equation, and considering that modulus of the secondary 4-velocity~\eqref{4vel} is constant, we derive the following important relation:
\begin{equation}\label{fourforce}
g^{}_{\mu\nu}\frac{d x^\mu}{d\tau} \mathcal{F}^{\nu}_{\rm SF} = 0 \,. 
\end{equation}

Equation~\eqref{geodesic} is usually referred to as the MiSaTaQuWa equation~\cite{Mino:1996nk,Quinn:1996am}, derived in the framework of first-order black hole perturbation theory. One can also introduce higher perturbative orders of the self-force, which are necessary for a consistent computation of the EMRI waveforms (see, e.g., Ref.~\cite{Pound:2015tma}). There are different ways in which one can use this equation to follow the EMRI dynamics~\cite{Pound:2015tma}. One of them is the method of osculating orbits~\cite{Pound:2007th}, where at each instant of time one can use the self-force associated with the geodesic tangent to the secondary world-line at that time, or in other words, we can identify the unique geodesic in the phase space of the secondary. From the phase space perspective, each geodesic can be identified by a set of geodesic invariants. Two frequent and important sets of invariants are: (i) \emph{Orbital elements}. For instance, semi-latus rectum, eccentricity, and inclination. (ii) \emph{Physical constants of motion}. For instance: energy, the angular momentum component along the primary symmetry/spin axis, and the Carter constant. These values fully determine a given geodesic, and to complete the phase space information we need three positional elements identifying a point on the given geodesic. For instance, we can choose the phases of Kerr geodesic motion associated with the radial, polar and azimuthal motions. Then, we can write a formal expansion of the equations of motion in first-order form with respect to the secondary proper time (see, e.g., Ref.~\cite{Hinderer:2008dm} for details):
\begin{eqnarray}
\frac{d\mathcal{Q}^I}{d\tau} & = &  \omega^I(\mathcal{P}^J) + q g^I_1(\mathcal{Q}^J,\mathcal{P}^K) + q^2 g^I_2(\mathcal{Q}^J,\mathcal{P}^K) + \mathcal{O}(q^3) \,, \label{sf-equation-1} \\
\frac{d\mathcal{P}^I}{d\tau} & = &  q\, G^I_1(\mathcal{Q}^J,\mathcal{P}^K)  + q^2 G^I_2(\mathcal{Q}^J,\mathcal{P}^K) + \mathcal{O}(q^3) \,,
\label{sf-equation-2}
\end{eqnarray}
where $Q^I$ and $P^I$ denote the positional and geodesic elements, respectively. The functions $g^I_1$, $g^I_2$, $G^I_1$, and $G^I_2$ can be obtained from the self-force and hence, they contain the information on the deviations from geodesic motion. This formulation is suitable to study the different aspects of EMRI dynamics, in particular dissipative effects. It also provides a useful framework for introducing adiabatic approximations to the EMRI dynamics.  

The development of the self-force programme in the standard scenario is quite advanced~\cite{Pound:2015tma,vandeMeent:2017zgy,Barack:2018yvs}. From the superficial description we have given here, we can already see that it involves sophisticated GR and perturbation theory techniques. Moreover, the developments in the standard scenario use symmetries and the special structure of GR. It is clear that significant efforts have to be made in order to carry on this programme in alternative theories of gravity. Another alternative framework to study the full EMRI dynamics is the use of quasilocal conservation laws to formulate the equations of motion of a two-body system in the extreme-mass ratio regime~\cite{Oltean:2019jws,Oltean:2019ihp}.
 
From a PN perspective, one can classify the dissipative effects into three groups based on the order at which changes are made to the dissipative sector compared to the conservative sector~\cite{Cardenas-Avendano:2019zxd}: (i) dissipative corrections are made at a lower PN order than conservative modifications, (ii) they are made at the same PN order, or (iii) dissipative modifications are made at a higher PN order than conservative modifications. One has to study in a theory-by-theory basis how relevant such modifications can be.  

Within GR, we can express the different fluxes, $dE/dt$, $dL_z/dt$, and $dQ/dt$, in terms of the components of the self-force, for instance, by taking derivatives with respect to proper time, $\tau$, of Eqs.~\eqref{Eq:E},~\eqref{Eq:L} and~\eqref{Eq:CarterConstant}, respectively~\cite{Kennefick:1995za}:
\begin{eqnarray} \label{threeforce}
\frac{dE}{d \tau} & = & - g^{}_{tt}\mathcal{F}^{t}_{\rm SF} - g^{}_{t\phi}\mathcal{F}^{\phi}_{\rm SF}\,, \\
\frac{dL_{z}}{d\tau} & = & -g^{}_{t\phi}\mathcal{F}^{t}_{\rm SF} + g^{}_{\phi\phi}\mathcal{F}^{\phi}_{\rm SF} \,, \\
\frac{1}{2} \frac{dQ}{d\tau} &=& g_{\theta\theta}^{2}\frac{d x^\theta}{d\tau} \mathcal{F}^{\theta}_{\rm SF} 
%\nonumber \\ &&
-\cos^{2}\theta\left(a^{2}E\frac{dE}{d\tau}-\csc^{2}\theta L_{z}\frac{dL_{z}}{d\tau}\right)\,.
\label{Qdot}
\end{eqnarray}
The system is closed by considering the orthogonality of the 4-velocity of the secondary and the self-force (see Eq.~\eqref{fourforce}).

The adiabatic approximation is equivalent to taking only the leading term in Eq.~\eqref{sf-equation-1} and the orbit-averaged changes in the first term in Eq.~\eqref{sf-equation-2}, that is, $<G^I_1>(\mathcal{P}^J)$ (see Ref.~\cite{Pound:2005fs} for a study of the limitations of this approximation). These averaged terms are obtained from the GW fluxes to infinity and down to the horizon in the case of a BH primary (see, for details, Refs.~\cite{Mino:2003yg,Mino:2005yw,Mino:2005an}). 

In practice, and within the standard scenario (for a non-spinning secondary), these fluxes can be computed in the framework of the Teukolsky formalism~\cite{Teukolsky:1972my,Teukolsky:1973ha} for perturbations of Kerr BHs. Relevant developments in this line of research can be found in Refs.~\cite{Hughes:2005qb,Drasco:2005kz,Sundararajan:2008zm,Isoyama:2018sib,Fujita:2020zxe,Isoyama:2021jjd}. There are efforts to extend these computations to other scenarios. For instance, there are computations of the GW balance laws in the case of a spinning secondary~\cite{Akcay:2019bvk,Skoupy:2021asz}.

When the adiabatic fluxes are integrated, one obtains the phase space trajectory, i.e., the values of $E$, $L_z$ and $Q$ as functions of time. This procedure gives time dependent expressions to compute the trajectory of the inspiralling particle as a sequence of geodesic with adiabatically changing constants of motion.  For details on the computation of the fluxes of energy and angular momentum see Ref.~\cite{Teukolsky:1974yv}, and Refs.~\cite{Sago:2005fn,Ganz:2007rf} for the case of the Carter constant.

For instance, most of the non-Kerr inspirals, aimed towards studying the effect of modifications of the standard scenario, consider the radiation reaction effect written in terms of PN expansions and fits to perturbative calculations within GR~\cite{Gair:2007kr} for the fluxes ($dE/dt\,,\,dL_z/dt$ and $dQ/dt$) with the following two assumptions (see, e.g., Ref.~\cite{Canizares:2012is} in the case of dynamical Chern--Simons theory). First, the mass quadrupole moment has been modified by the theory (the Kerr metric is not a solution in theories like dynamical Chern--Simons theory) to accommodate the modification or ``bumpiness"\footnote{The use of this term has a historic reason, which will get clear once we discuss the multipole moments of the primary in Sec.~\ref{Sec:NatureofBHs}.} (see, e.g., Ref.~\cite{Barack:2006pq}) of the non-GR background, and second, these fluxes have been linearized~\cite{Canizares:2012is}. Then, the (approximate) evolution of the energy and angular momentum can be written as:
\begin{eqnarray} 
\label{linearizedfluxes}
E(t) & = & E(0) + \left.\frac{dE}{dt}\right|^{}_{0} N^{}_{r} T^{}_{r} \,, \\
L_{z}(t) & = & L_{z}(0) + \left.\frac{dL_z}{dt}\right|^{}_{0} N^{}_{r} T^{}_{r} \,,
\end{eqnarray}
where $T_r$ is the time that the orbit takes to travel from the periastron to apoastron and back, and $N_r$ denotes the number of cycles at which these equations are updated. The above equations are clearly a dissipative change, but it is \emph{not} a beyond-GR modification, and that is why we considered these types of analysis as changes of only the \emph{conservative} sector. It is important to mention that the adiabatic approximation is expected to be enough to produce EMRI waveforms that are good enough to perform EMRI detections, although not good enough to track the whole EMRI evolution in the band of a space-based detector with the accuracy required to make very precise estimation of the physical parameters (see~\cite{vandeMeent:2017zgy,Wardell:2021fyy}). Therefore, self-force computations are required. They can be used also to assess the validity of the adiabatic and post-adiabatic approximations.

During the EMRI evolution, due to gravitational radiation emission, the characteristic frequencies of the system evolve, and the dynamics gets more complicated. For bound Kerr geodesics, any function $\mathcal{G}$ describing the motion of the system, will depend, in general, on the three fundamental orbital frequencies $(\omega_{r},\omega_{\theta},\omega_{\phi})$, and one can write it in the frequency domain as a multi-frequency Fourier series~\cite{Drasco:2003ky,Ruangsri:2013hra}:
\begin{equation}
\label{eq:freqevol}
\mathcal{G}\left[z^{\mu}\right] = \sum_{j,k,l~\in~\mathbb{Z}} C^{}_{jkl} \, e^{i\left( j\omega_{r} + k\omega_{\theta} + l\omega_{\phi}\right)t}\,,
\end{equation}
where $z^\mu$ denotes the geodesic worldline and $C_{jkl}$ denotes a Fourier coefficient. For most orbits, the extreme-mass ratios involved make the evolution of the system to happen \emph{slowly}, and therefore, as the orbit evolves, there is a slow change in the Fourier components and frequencies. If this is the case, then the evolution is governed by a near-constant piece, when $(j=k=l=0)$, as the rapidly oscillating terms average to zero over multiple orbits. This is the fundamental assumption of the semi-relativistic approximation, which is the one heavily pursued for studies of EMRIs in modified theories of gravity.

Both the conservative quantities and the radiated flux are modified by $\mathcal{O}\left(\delta g\right)$ when radiation reaction is included, and therefore the resulting overall GW phase shift is~\cite{Pan:2023wau}
\begin{equation}
\delta \Psi_{\rm adiabatic} \sim \frac{ \mathcal{O}\left(\delta g\right)}{q} \,,
\end{equation}
which can, when $\left| \delta g \right| > q$, be larger than unity and therefore detected with future instruments. It is important to note that the nature of $\delta g$ may not necessarily originate from a beyond GR theory.

On the other hand, from Eq.~\eqref{eq:freqevol} one can directly see that under a resonance, $m\omega_{r}+n\omega_{\theta}+k\omega_{\phi}=0$, now all the terms will stop being be oscillatory, and won't average away over multiple orbits~\cite{Drasco:2003ky}. As explained above, this is the condition where the conservative dynamics changes the most and the dissipative contribution also has potential to significantly change the EMRI evolution. In other words, resonances also play an important role and require proper modelling. 

The results discussed in the previous subsection were derived using the adiabatic approximation of EMRIs around resonances. Therefore, these types of calculations should be replaced by full self-force computations. It is anticipated that the features mentioned earlier would be less pronounced with an instantaneous self-force evolution, similar to what occurs during the inspiral of spinning BHs of comparable mass~\cite{Cornish:2003uq}. This is due to GW dissipation. Nonetheless, this is a challenging task, and research is currently ongoing.

In the context of GR, see Ref.~\cite{Shen:2023pje}, the influence of mass-ratio corrections on the orbital motion (orbit, frequency, and phase) and gravitational radiation was investigated using EOB orbits and the Teukolsky formalism. It has been found that the mismatch between the EOB and test particle waveforms can be ignored for the mass-ratio $q\lesssim10^{-5}$. However, for the case of $q\gtrsim10^{-5}$, there is a risk of making the incorrect judgment that we have detected a deviation from GR. 

On the other hand, for a particular parametric deviation of the Kerr geometry that admits a rank-2 Killing tensor and is part of the family of spacetimes presented in Ref.~\cite{Benenti:1979erw}, the leading order PN corrections to the average loss of energy and angular momentum fluxes for eccentric equatorial motion were derived in Ref.~\cite{Kumar:2023bdf}. They found that the changes in the considered deviation parameters, which preserve the integrability of the spacetime, induce dephasings proportional to $\mathcal{O}\left(1/q\right)$, and therefore may be relevant for future observations. 

A remarkable exception to using linearized Einsteinian fluxes or studying pure geodesics, is the work presented in Ref.~\cite{Zimmerman:2015hua}. In this work, the equations of motion for a small body due to the coupled self-force in generic massive scalar-tensor theories in the Einstein frame to first-order in the mass-ratio were computed. In particular, the equations of motion can be decomposed as~\cite{Zimmerman:2015hua}
\begin{equation}
ma^{\mu} = F_{0}^{\mu} + F_{L}^{\mu} + F_{\textrm{tail}}^{\mu} \,,
\end{equation}
where $F_{0}^{\mu}$ is the force resulting from the gradient of the background potential, $F_{L}^{\mu}$ is the local contribution to the self-force which is built from background quantities evaluated on the world line, and $F_{\textrm{tail}}^{\mu}$ is the non-local contribution to the self-force which takes the form  of a tail time integral over the particle past history. 

For the theories considered in Ref.~\cite{Zimmerman:2015hua}, the resulting equations of motion are substantially different from those in vacuum GR due to the presence of the additional scalar field. These changes are mainly a consequence of the violation of the strong equivalence principle in scalar tensor theories, which makes the motion sensitive to the internal constitution of the body. In Ref.~\cite{Zimmerman:2015hua}, it was estimated that self-force effects become important for a large region of the parameter space, i.e., whenever $M/m\lesssim M^2/\mu^2$, where $\mu^2$ is the mass of the scalar field. 

\subsection{The effects of the spin to the motion}
\label{Sec:Spin}

Finding rotating solutions in modified theories of gravity is a challenging task. Consequently, the work carried out so far has been conducted either in solutions that are derived perturbatively or numerically~\cite{Canizares:2012is,Cardenas-Avendano:2018ocb}, or in an exact parametric BH spacetime~\cite{Collins:2004na,LukesGerakopoulos:2010rc,Destounis:2020kss}, that is not necessarily a solution of a theory, or can contain closed timelike curves or naked singularities. 

On the other hand, the force on the secondary is not only the gravitational self-force due to the perturbation caused to the background but also due to finite-size effects. In other words, there are additional contributions because the secondary is not a point particle. In particular, there is a finite size effects at the dipole level if the secondary is spinning. 

Within GR, the effects of the spin on an EMRI system were recently investigated by numerically integrating the motion of a spinning secondary in the field of a non-spinning primary and analyzing it using various methods~\cite{Zelenka:2019nyp}. They showed that resonances and chaos can be found even for astrophysically realistic spin values. However, the resonances observed were only caused by terms quadratic in the spin and are in general small in the small mass-ratio limit. It was also shown that the time series of the GW strain could be used to discern regular from chaotic motion of the source. While this study provides evidence that spin-induced chaos and resonances will not play a significant role in EMRIs with non-spinning MBHs (or primary), further research is needed to determine whether these effects are significant in other scenarios. A similar study in a modified theory of gravity is lacking. 

When considering a primary that is spinning, the role of the purely general relativistic effect of frame-dragging for tests of GR has eluded a definite answer. In Ref.~\cite{Gutierrez-Ruiz:2018tre}, it was shown that, when considering a neutral secondary around a family of stationary axially-symmetric analytical exact solutions to the Einstein-Maxwell field equations, frame-dragging is capable of reconstructing KAM-tori from initially highly chaotic configurations. In other words, Ref.~\cite{Gutierrez-Ruiz:2018tre} claimed that chaos suppression by frame-dragging can make the tests of GR with EMRIs even more challenging. This picture is, however, different when considering a charged secondary, and  previous works have not found any clear and unique indication of the spin dependence with chaos (see for instance, Refs.~\cite{Takahashi:2008zh,kopaeck2010}) when considering solutions with electromagnetic charge. 

For a primary described by a BH solution with synchronised scalar hair~\cite{Herdeiro:2014goa}, Ref.~\cite{Collodel:2021jwi} studied orbits on the equatorial plane using the quadrupole formula approximation. They showed that the frequencies of the emitted signals behave non-monotonically falling below LISA’s sensitivity range. They found that signals can chirp backward, and for some particular cases, they can become arbitrarily small. When comparing these BHs with Kerr BHs of the same mass and horizon radius, the signals emitted by the secondary around Kerr present overall larger frequencies than around the hairy ones, but those can be considerably smaller for the early stages of the evolution. They presented two sets of waveforms produced by a noncircular EMRI in which the secondary follows a type of geodesic motion typically present in spacetimes with a static ring, in which the compact object is periodically momentarily at rest. 

More recently, Ref.~\cite{Delgado:2023wnj} focused on how EMRI features evolve as one transitions from a highly spinning Kerr BH into a hairy one. Reference~\cite{Delgado:2023wnj} performed a comparison with a highly spinning Kerr BH, and found that EMRIs do not change significantly from the ones around extremal Kerr BHs. The only different feature happens at the endpoint of the evolution, as the perimetral radius can be more than twice as large as for the extremal Kerr BH, whereas, the angular frequency endpoint, and consequently the cut-off frequency of the produced GWs, can be around one-fifth of the extremal Kerr BH value.

In Ref.~\cite{Guo:2023mhq}, using the formalism we will review in Sec.~\ref{sec:secondary} developed mainly to detect scalar fields, was argued that the spin plays a relatively secondary role in the EMRI system, and therefore a limited influence on detecting the scalar charge. They argued that the spin of the secondary is more feasible to be detected in systems with less massive primaries, and therefore TianQin, which has a greater sensitivity in the high-frequency region, may be more suitable for these measurements. 

\subsection{Waveform generation}
\label{Sec:WaveformGeneration}

To detect and characterize sources, one needs the gravitational emission waveform corresponding to the different candidate sources expected to be detected. The quality of the waveform has a strong impact in the science output from the detection and also in the number of sources that will be detected. A review of current efforts to generate waveforms in the context of the LISA mission can be found in the whitepaper of the Waveform Working Group of the LISA Consortium~\cite{LISAConsortiumWaveformWorkingGroup:2023arg}. An important tool to be considered for the case of EMRIs is the Black Hole Perturbation Toolkit~\cite{BHPToolkit}, where different codes for the development of EMRI waveforms can be found.

Low-frequency sources as MBH binaries and EMRIs cannot be modelled in terms of a few number of parameters. The dimensionality of the parameter space for a given GW source is very important as it determines the computational cost of the searches for these sources. The case of EMRIs in GR is a clear case where we can in principle have waveforms for all the physically relevant cases. The parameter space of standard EMRIs in GR, the standard scenario, has dimension $14$ (that include source-intrinsic degrees of freedom, e.g., mass-ratio, spin of the primary, or eccentricity; and observer-dependent ones, such as the angular configuration of the EMRI). Typical modifications of the standard scenario will only increase this number, for instance, the spin of the secondary, possible hair of the primary or secondary, or environmental effects. See the following sections for effects that we may need to include, with their respective parametrization, in the EMRI waveforms. The relatively high dimensionality of an EMRI waveform, together with its expected duration (in the band of a LISA-like detector, which is of the order of $\sim 1\,$yr), tells us plenty about the complexity and cost of the search process (which has to be part of the global fit algorithm to analyze of the source at once for a given period).  Possible chaotic orbital motion (see Sec.~\ref{Sec:Conse} for more details) indicates the possibility of the loss of predictability, although in this case, we may focus instead on finding clear signatures of the non-integrable motion. 

The computation of EMRI waveforms requires a good understanding of the EMRI dynamics to be able to model the EMRI emission according to the accuracy requirements set by each detector or the particular application. In the ideal case, we expect to obtain this understanding from the developments in the self-force programme described before. But apart from the efforts to obtain precise EMRI waveforms, producing fast (in terms of computational time) EMRI waveform families is also essential. These waveforms are needed for different purposes, for instance, to carry out parameter estimation studies to make forecasts of the precision with which different detectors can detect and extract the EMRI parameters. This is very important to assess such detectors' scientific potential. 

Approximate (or \textit{kludge}, a frequently used term in the literature) waveforms have been shown to provide reliable forecasts (in the case of LISA they have been consolidated by using different families of kludge waveforms).
Another important use of these fast, although not accurate, EMRI waveforms is to use them for the development of data analysis strategies to detect EMRIs in the context of the \textit{global fit} (see Sec.~\ref{emri-detectability}). To that end, it is important to have algorithms that can generate EMRI waveforms quite fast, in a timescale of the order of a small fraction of a second, to allow for a feasible accurate data analysis. It is important that these waveforms models capture the main features of a real EMRI waveform. For instance, the different precessional effects (periastron precession, precession of the orbital plane, and others when we take into account the spin of the secondary or other potential physical effects) or the radiation-reaction timescales. 

Given that the modelling of EMRIs is a complex task, even when approximations and simplifications are made, it requires the use of different tools in GR (which at the same time can be applied to other theories of gravity).  Among them, it is worth mentioning: action-angle variables for EMRIs~\cite{Kerachian:2023oiw}; two-timescale evolution of EMRIs~\cite{Miller:2020bft}; PN and Post-Minkowskian (PM) techniques~\cite{Munna:2020som}; or hyperboloidal slicings~\cite{PanossoMacedo:2022fdi}.

The simplest approximation to the construction of EMRI waveforms is to assume that the orbital dynamics is Newtonian and that the gravitational radiation emission mechanism is well described by the quadrupole formula~\cite{Einstein:1918qf,Landau:1975pou}. This amounts to a Newtonian two-body system where the energy and angular momentum evolve \emph{adiabatically} as dictated by the quadrupole formula. More specifically, the evolution of the orbital elements is obtained by orbital averaging of the radiation emission. In this way, the rates of change of the Keplerian semimajor axis $a$ and eccentricity $e$ are given by:
\begin{eqnarray}
\frac{da}{dt} & = & - \frac{64}{5}\frac{G^3 \mu M^2}{c^5 a^3\left(1-e^2\right)^{7/2}}\left( 1 + \frac{73}{24}e^2 + \frac{37}{96}e^4\right) \,, \label{pandm-1} \\[2mm]
\frac{de}{dt} & = & -\frac{304}{15}\frac{G^3\mu M^2\, e}{c^5 a^4\left(1-e^2\right)^{5/2}}\left(1 + \frac{121}{304}e^2 \right)\,, \label{pandm-2}
\end{eqnarray}
where $\mu$ is the reduced mass of the EMRI: $\mu = mM/(m+M)$. The computations for this approximation were done first in Refs.~\cite{Peters:1963pm,Peters:1964PhRv..136.1224P}, and is usually known as the \textit{Newtonian approximation} (see also Ref.~\cite{Turner:1977mt}).  The EMRI dynamics can be seen as a sequence of Keplerian eccentric orbits that evolve according to Eqs.~\eqref{pandm-1} and~\eqref{pandm-2}, and the waveforms that we obtain from the quadrupole formula is a superposition of harmonics of the slowly varying radial frequency---the variation is given by Kepler's third law and Eqs.~\eqref{pandm-1} and~\eqref{pandm-2}.

%ACR-AK
One can go beyond the Newtonian approximation to improve the quality of the EMRI waveforms in many different ways.  One of the most popular kludge models of waveforms for EMRIs, and the most used one to date, was provided by Barack and Cutler~\cite{Barack:2003fp}, who added, to the Peters and Mathews waveform model described above, all the remaining ingredients in EMRI dynamics that were not present in the Newtonian approximation: pericenter precession, Lense-Thirring precession, and inspiral from radiation reaction.  All these ingredients were introduced in a very approximate way, using mainly (low order) PN approximations. One of the advantages of the Barack and Cutler waveform model is that the final waveforms are given in a nearly analytical form (up to integration of ODEs), and thus they have been named as Analytical Kludge (AK) EMRI waveforms. As we have already said, this model and its improvements, are still the most used ones, mainly for parameter estimation forecasts.  

For the purposes of this review, it is important to mention that there are extensions of the AK waveform model that allow for tests of the Kerr geometry or tests of alternative theories of gravity. In particular, Barack and Cutler themselves produced an extension to add an arbitrary (mass) quadrupole of the primary~\cite{Barack:2006pq}, which has also been extensively used to provide forecasts for the accuracy in this quantity that the LISA mission can provide (see, e.g., Ref.~\cite{eLISA:2013xep,LISA:2017pwj}). Another very interesting extension of the AK model was provided in Ref.~\cite{Gair:2011ym}, where the authors added parameterized post-Einsteinian
corrections that allow for generic, model-independent deformations of the primary away from the Kerr metric (by using a \emph{bumpy} metric). These deformations represent modified gravity effects and have been analytically mapped to several modified gravity BH solutions. Therefore, this modification allows for model-independent tests of GR with EMRIs.

%ACR-AAK
Regarding the improvements of the AK model, it has extended~\cite{Chua:2015mua,Chua:2017ujo} to include information from the exact Kerr geodesics, in particular from the three frequencies of the Kerr bound geodesic motion discussed above. The idea is to use a frequency map to the correct Kerr frequencies, to improve the faithfulness of AK waveforms without significantly increasing their computational cost. These waveforms can stay in phase for months. The resulting model has been named
\textit{Augmented Analytic Kludge} (AAK) waveforms. An extension of the AAK model to include an arbitrary  (mass) quadrupole moment of the EMRI primary has been provided in Ref.~\cite{Liu:2020ghq}.  

A different avenue to go beyond the Newtonian approximation is to change the description of the orbital motion from Newtonian to general relativistic kinematics, in particular Kerr geodesics in the standard scenario. The simplest such approximation is the so-called \emph{semi-relativistic} approximation~\cite{Ruffini:1981rs}, where the inspiral is described by a sequence of Kerr geodesics dictated by the quadrupole formula (see also Ref.~\cite{Gair:2005is}). In this way, the waveform computed is composed of harmonics of the three fundamental frequencies of Kerr geodesic motion [see Eq.~\eqref{eq:freqevol}].  This is the simplest model where locally in time, the orbital motion are full relativistic geodesics. An improvement of this model is the so-called Numerical Kludge (NK) waveform model~\cite{Babak:2006uv}. The orbital motion in the NK model is also a sequence of Kerr geodesics, but the inspiral is governed by a {\em kludge} adiabatic radiation reaction prescription~\cite{Gair:2005ih} based on PN approximations. The waveforms are constructed using a quadrupole-octupole approximation.  In this line, another model based on Kerr geodesics is the \emph{Chimera} scheme~\cite{Sopuerta:2011te,Sopuerta:2012de}.  In this waveform model the radiation-reaction prescription that drives the inspiral uses an approximate self-force obtained from a multipolar, PM expansion for the far-zone metric perturbations.  The waveform is computed using a multipolar expansion up to the mass hexadecapole and current octopole order. Another ingredient of this model is a map between the Boyer--Lindquist coordinates of the orbits to the harmonic coordinates in which the different multipolar PM quantities are defined. 

The next step is to improve the ingredients of these relativistic models up to what we have known as the \emph{adiabatic} approximation (see discussion above).  The inspiral can still be seen as a sequence of Kerr geodesics but the GW fluxes associated with energy, angular momentum and Carter constant, which dictate the transition between geodesics, are computed using the Teukolsky formalism~\cite{Teukolsky:1972my,Teukolsky:1973ha}. Also the EMRI waveforms are computed from the Teukolsky curvature variables (see Refs.~\cite{Hughes:2005qb,Drasco:2005kz,Sundararajan:2008zm,Isoyama:2018sib,Fujita:2020zxe,Isoyama:2021jjd}).  The computation of these waveforms is much more time consuming than previous more simple models. Recently, fast EMRI waveforms for millihertz GW data analysis~\cite{Katz:2021yft} have been produced. Techniques for fast generation of ERMI templates have been presented in Ref.~\cite{Chua:2020stf}. These techniques can be applied to the adiabatic approximation described above, making them a more attractive model to be used also for data analysis and forecast purposes.  On the other hand, extensions of the Teukolsky formalism to alternative theories of gravity, where the primary may not even be of Petrov type D\footnote{The Petrov classification distinguishes the different possible algebraic structures of the Weyl tensor. This can be done in practice in terms of the possible four \emph{principal} null directions of the Weyl tensor. The Petrov type D corresponds to the case where there are two repeated null principal directions. See Ref.~\cite{Stephani:2003tm} for details.} as the Kerr geometry is, have been recently appear~\cite{Li:2022pcy}. This opens the door to extend the adiabatic approximation to other theories of gravity. 

Other techniques have been used for the modelling of EMRIs. For instance, the EOB formalism~\cite{Buonanno:1998gg,Buonanno:2000ef}, already used for data analysis of ground-based detectors, is another candidate to produce EMRI waveforms. In the EOB approach, one maps the relativistic two-body dynamics to an EOB dynamics, which is inspired by what is done in classical and quantum-mechanical problems.  There are several developments in the EOB formalism to extend it towards high mass ratios~\cite{Yunes:2009ef,Yunes:2010zj,Albanesi:2021rby,Nagar:2022fep,Albanesi:2023bgi, Zhang:2020rxy,Zhang:2021fgy}, and has even been compared with second-order self-force computations~\cite{Albertini:2022rfe,Albertini:2022dmc,vandeMeent:2023ols}.

The modelling of EMRIs has to take into account the resonances, which have been described above mainly in Sec.~\ref{Sec:Conse}, since the inspiral will drive the system through several of them, specially in the last stages of the evolution. This will require to adapt the techniques we have been described until now to include them. There are some studies of resonances in the case of the scalar self-force, i.e., when the radiation reaction is due to an scalar field instead of the metric tensor, in~\cite{Nasipak:2021qfu,Nasipak:2022xjh}.
The impact of transient resonances has been discussed in Ref.~\cite{Speri:2021psr} in terms of an extension of the NK model~\cite{Babak:2006uv}. For one-year long inspirals it is found that $3:2$ resonance (the ratio of the polar and radial frequencies is $1.5$), which causes a dephasing of a few cycles, can induce systematic errors in the estimation of EMRI parameters that are up to several times the typical measurement precision of the parameters, specially when the resonance takes place close the horizon of the primary. This study also shows that EMRI observations can also be used to estimate the size of the changes of orbital parameters across the resonance. For instance, for a $1\,$yr inspiral duration with a SNR of $20$, it is found that these changes can determined with a relative precision of $10\%$. Actually, such a measurement can be seen as test of GR, by comparing with the predicted value, and hence it constitutes another fundamental physics test that we can carry out with EMRIs. 

On the other hand, in Ref.~\cite{Gupta:2021cno} tidal resonances, induced by the tidal field of a nearby astrophysical object, were studied and it was shown that they can significantly alter the orbital evolution, leading to a significant dephasing across observable parameter space. This studied was improved in Ref.~\cite{Gupta:2022fbe}, where a more general model of the tidal perturber was considered, which it is shown that could be applied to the study of self-force resonances.

In contrast to comparable-mass binaries, in EMRIs, the final merger and ringdown constitute a small part of the coalescence time, contributing to a small fraction of the accumulated SNR. This may change in non-standard scenarios.  For instance, in the case when the primary is a boson star, the secondary can penetrate inside the star, in which case the inspiral may last much longer and in a BBH case, where the signal will shut done after plunge, and we will not have a proper ringdown (see Ref.~\cite{Kesden:2004qx}). Despite in the standard scenario the merger-plunge seems negligible, it is worth to have a look at it to understand whether or not we can extract any information from it. In this sense, it is important to understand the transition from inspiral to plunge in EMRIs:~\cite{Ori:2000zn,Compere:2021zfj} (see also Ref.~\cite{Jaramillo:2023day} for connections of the equations of EMRI dynamics with integrable systems). 

From the discussion of this section, we can see that most efforts in the EMRI waveform modelling have been so far devoted either to the standard scenario or to scenarios within the framework of GR. For instance, taking into account the spin of the secondary can be quite relevant~\cite{Warburton:2017sxk,Drummond:2022efc,Piovano:2020zin}. 
There are some exceptions to this in the case of certain families of EMRI waveforms, in particular those that originate as extensions of the AK model.  This model, as some of the other mentioned above, share the property that are very modular in the sense that have been built from the union of different ingredients (such as the orbital motion, radiation reaction mechanism or the waveform construction) that can be modified independently and, in this way, improve the model, which is clearly exemplified by the AK model extensions.  This is a line of research that needs to be pursue in order to understand better the capability of space-based detectors like LISA to make fundamental physics tests that involve as many theories and models for the EMRI primary and secondary as possible. 

Nevertheless, as the time when these space-based detectors will become a reality is approaching, we also need to extend the high-accuracy waveforms (either improved versions of the adiabatic-approximation waveforms or directly, self-force based waveforms) in order to be able to carry out reliable fundamental physics tests with EMRIs. This is a challenging task but the advances in the self-force programme and related efforts show that it is a realistic task.  An interesting avenue when discussing fundamental physics tests is to design model-independent tests. An example of this would be to test the Kerr bound ($|a|\leq M$) with EMRIs (see Ref.~\cite{Piovano:2020ooe}). A detailed review of tests of the nature of BH and other dark compact objects can be found in Ref.~\cite{Cardoso:2019rvt}, including projected constraints with EMRIs.

\subsubsection{Towards a data analysis framework for testing General Relativity}

By now, we hope we have convinced (but not scared!) the reader that precision science with EMRIs will be very challenging \emph{both} in terms of developing accurate waveforms and the data analysis of the signals. The intrinsic complexity of the signal leads to costly model development and evaluations that will translate into expensive studies of the (highly) multimodal likelihood surface.  Thus, the development of efficient and faithful phenomenological EMRI waveform models will be crucial for the science output we can get from these observations.

A general framework for carrying out tests with LISA data was initiated in Ref.~\cite{Chua:2018yng}, where the authors used an extended version of the AK model to include (multipole) deviations from the EMRI standard scenario. They implemented the corrections to the AK waveforms by considering geodesics in a parametrically modified background, i.e., a bumpy spacetime, as defined in Sec.~\ref{multipole-moments-of-the-primary}. The chosen bumpy metric, derived in Refs.~\cite{Benenti:1979erw,Vigeland:2011ji}, ensures the existence of approximately conserved energy, angular momentum, a second-order Killing tensor, and a Carter constant without requiring that the Einstein equations be satisfied, which is due with the fact that symmetries are associated only with the metric form itself. As we discussed in Sec.~\ref{Sec:Conse}, the latter assumption may be too restrictive, as solutions for the primary in modified theories of gravity do not necessarily have to have a fourth constant of the motion. 

In the framework proposed in Ref.~\cite{Chua:2018yng}, both the long-timescale radiation reaction fluxes and the short-timescale precession rates of the EMRI system are altered by the bumpy parameters since the three first integrals of motion along a timelike geodesic are perturbed/modified. The framework of  Ref.~\cite{Chua:2018yng} consists in using these modified waveforms and various sophisticated state-of-the-art Bayesian statistical tools based on nested sampling techniques. 
In particular, the authors of Ref.~\cite{Chua:2018yng} assess the viability of a product-space method (a technique used to analyze multivariate data or data with multiple dimensions) in accelerating the convergence of nested sampling algorithms on EMRI likelihoods and applied a rethreading technique that provides error estimates on the Bayes factors (a measure that quantifies the strength of evidence for one model over another) obtained through this method.

Their resulting \emph{bumpy AK} waveform model is 19-dimensional!: the 14 GR parameters; four deformation parameters; and a submodel index (to distinguish between the 16 they considered). Since a full exploration of this space is currently forbidden (note that the likelihood surface over the GR parameter space is already highly multimodal~\cite{Gair:2004iv},) in Ref.~\cite{Chua:2018yng} they fix all but seven of the hypermodel parameters, allowing only the component masses, deformation parameters, and submodel index to vary.
They find that the product-space method attains better precision than regular nested sampling at the same computational cost level and indicates an even more significant cost improvement to reach $5\%$ error~\cite{Chua:2018yng}. We refer the interested reader to their work for details of developing a usable infrastructure for testing the EMRI standard scenario with future EMRI GW observations.

%%%%%%%%%%  Environmental Effects %%%%%%%%%%
\section{Environmental Effects}
\label{Sec:Enviroment}

When testing GR, it is crucial to accurately consider environmental influences to avoid misinterpreting them as signals of a GR violation. This is particularly true for EMRIs, as these systems are inherently intertwined with astrophysical surroundings; thus, their evolution inevitably should deviate from an uncontaminated vacuum scenario. How significant are these effects? Can environmental effects prevent precision tests with GWs from EMRIs? To be detectable, beyond-GR corrections should be more important than environmental effects, and that may not be always the case. 

If a test shows a ``smoking gun'' discrepancy from GR, one should consider how environmental contributions could affect that particular test. For instance, electromagnetic charges, magnetic fields, an accretion structure, a third body, the cosmological constant, a galactic halo, a DM distribution, or dynamical friction could have an impact in the EMRI waveform and the effect may be potentially degenerate with a modification of GR~\cite{Barausse:2014tra}.

While it is expected that for the majority of massive binary $m$Hz GW sources, environmental phase contributions are likely to be dominant over vacuum corrections of order higher than the current benchmark value of $5$PN~\cite{Zwick:2022dih}, for some EMRI systems, environmental perturbations can be completely negligible. For EMRIs formed in the standard dynamical (dry) astrophysical formation channel, their high initial eccentricity and rapid circularization implies that they might enter the band of the future space-based detectors directly in a regime where environmental perturbations are completely negligible, i.e.,  they ``skip'' over low-frequency GW emission~\cite{Zwick:2022dih}.

On the other hand, the so-called active galactic nucleus (wet) EMRI formation channel might produce low-eccentricity EMRIs, which, therefore, can likely enter the detector band at lower frequencies, damping out high-order PN effects and reducing the overall SNR~\cite{Zwick:2022dih}. These are the relevant cases where precision GR tests can be jeopardized. That is why Ref.~\cite{Zwick:2022dih} advocated for prioritizing the systematic inclusion of environmental effects in waveform templates. 

Within GR, Ref.~\cite{Barausse:2014tra}, after a comprehensive analysis, concluded that an EMRI \emph{detection} should not be significantly affected by environmental effects, as the considered environments (magnetic field, charge, gravitational effects due to matter and gas in binary systems on the emission of GWs, accretion onto the binary object, thick accretion disks, DM accretion onto the central BH, galactic DM halos, certain DM distributions, and cosmological constant) produce minor corrections to the periastron shift and GW phase. Their conclusions, however, were derived by considering the lowest-order effects of these environments using the SPA, the adiabatic approximation and assuming a mission with accuracy to phase measurements of $\left|\delta\Psi\right|>10/\textrm{SNR}$, i.e., a sensitivity to (absolute) dephasings larger than $1$ rad over the entire mission lifetime~\cite{Barausse:2014tra}. 

In the following, we will briefly mention some examples of such impacts to show how they can affect the GW emission. 

\subsection{An accretion structure}

Obtaining realistic estimates of the influence of accreting gas on a binary's orbital evolution and phase is challenging. This is due to the complexity of accretion dynamics, a largely unexplored 3D magneto hydrodynamics problem spanning a vast dynamic range~\cite{LISA:2022yao}. However, in broad terms, an accretion disk can affect the EMRI evolution due to a resulting ``hydrodynamic" drag force. This drag force consists of two contributions: energy and momentum transfer through matter accretion onto the secondary (short-range interaction), and the gravitational deflection of non-accreted material (long-range interaction), which still transfers momentum. This interaction is known as ``dynamical friction," and arises from the gravitational pull of the secondary and its gravitationally-induced wake (i.e., the density perturbations that the secondary excites, by gravitational interaction, in the medium). These two contributions to the motion of the secondary can be expressed as~\cite{Barausse:2007dy} 
\begin{equation}
    \label{Eq:AccInteract}
    \frac{dp^{\mu}}{d\tau} =\underbrace{\left.\frac{dp^{\mu}}{d\tau}\right|_{\textrm{accr}}}_{h\frac{dm_{0}}{d\tau}u_{\textrm{fluid}}^{\mu}}+\overbrace{\left.\frac{dp^{\mu}}{d\tau}\right|_{\textrm{defl}}}^{\left.\frac{dp}{d\tau}\right|_{\textrm{defl}}^{\textrm{tang}}\sigma^{\mu}+\left.\frac{dp}{d\tau}\right|_{\textrm{defl}}^{\textrm{rad}}\chi^{\mu}},
\end{equation}
where $m_0$ is rest-mass accretion onto the secondary; $h$ is the specific enthalpy of the fluid; $u_{\textrm{fluid}}$ is the perfect fluid 4-velocity of the considered material around the primary; $\sigma^\mu$ is a unit spacelike vector orthogonal to the secondary's 4-velocity and pointing in the direction of the motion of the fluid; while $\chi^\mu$ is a unit spacelike vector, orthogonal to both secondary's 4-velocity and $\sigma^\mu$, and pointing in the radial direction.

We now present three cases of accretion-disk geometries where their impact has been studied in the EMRI scenario. 

\subsubsection{Thin disks}

If the EMRI happens within a thin-disk environment, accretion, dynamical friction, or planetary-style migration could play a dominant role~\cite{Yunes:2011ws,Kocsis:2011dr,Barausse:2014tra}. In these cases, the modification of the orbital trajectory of an EMRI will come in the form of torques as the disk exchanges energy and angular momentum with the system. Such scenarios provide an excellent opportunity to measure accretion-disk properties, such as the accretion rate and disk viscosity~\cite{Speri:2022upm}, but imply a powerful source of degeneracies, making tests of GR very hard. 

One way to see how these effects can play a role can be found in Ref.~\cite{Cardoso:2019rou}. If one considers a BBH evolving in a medium of density
\begin{equation}
    \label{eq:DMProfile}
    \rho = \rho^{}_0 \left( \frac{R}{r}\right)^\beta \,,
\end{equation}
one can study the environmental effects produced by a constant magnetic field, constant-density fluid, or thick accretion disks at an order-of-magnitude level and agnostically~\cite{Cardoso:2019rou}. This profile, with a power law index $\beta$, is designed to create a high-density region, typically called the ``spike'' at a radius $R$. Note that, although the drag created by thin accretion disks technically falls outside the approximations made for the above density profile, it can still be mapped by taking $\beta = 15/8$~\cite{Barausse:2014tra,Cardoso:2019rou}.

Using the profile just given above, Ref.~\cite{Cardoso:2019rou} computed the corrections to the GW phase due to environmental effects, $\delta\psi_{\rm{env}}$, within the SPA for a quasicircular binary and obtained
\begin{equation}
    \delta\psi^{}_{\rm{env}} = \kappa^{}_\beta M^{2-\beta} R^{\beta} \rho_0 \left( M f \right)^{-\gamma}\,, 
\end{equation}
up to factors of order unity. In this expression, the exponent $\gamma$ and the coefficient $\kappa_\beta$ depend on the specific mechanism affecting the dynamics. For example, when $\beta=1$, for a gravitational pull $\gamma=4/3$ and $\kappa_1=1$. 
In Ref.~\cite{Cardoso:2019rou}, the GR waveform was computed using the PhenomB model~\cite{Ajith:2009bn} in the frequency domain for non-precessing spinning BHs. From a Fisher Matrix analysis, and considering an EMRI with $q=10^{-4}$ with spin parameters $(\chi_1, \chi_2) = (0.8, 0.5)$ located at 1 Gpc from the detector, with SNR of $\sim22$, they found that it could be the case that for a typical density or smaller than that of accretion disks, an observation will be unable to distinguish between a vacuum and a nontrivial environmental density. In other words, their results provide further evidence that such scenarios might hindering parameter estimation or a test of GR. 

One can, alternatively assume a profile of the form~\cite{Barausse:2014tra}
\begin{equation}
    \label{Eq:density}
    \rho \sim 169  \frac{f^{11/20}_{\rm Edd}}{\tilde{r}^{15/8}}  \left( \frac{0.1}{\beta}\right)^{7/10} \left( \frac{10^6 M_{\odot}}{M}\right)^{7/10} \,\, [\rm{kg/m^3}]\,, 
\end{equation}
where $\tilde{r}=r/\left( GM/c^2\right)$ and $f_{\rm Edd}=\dot{M}/M$ is Eddington ratio for mass accretion. With that density one can compute the energy lost by the secondary due to dynamical friction and compute the dephasing of a Shakura--Sunyaev thin disk in the SPA. Assuming the inspiral occurs on the equatorial plane where the disk is located at the innermost stable circular orbit (ISCO) the dephasing is~\cite{Barausse:2014tra}
\begin{equation}
    \label{thinBarausse}
    \delta \psi \sim 285 f^{}_{\rm Edd} \left( \frac{0.1}{\beta} \right) \left( \frac{q}{10^{-5}} \right) ^{1/2} \left( \frac{10^6 M^{}_{\odot}}{M}\right)^{-0.3}\,.
\end{equation}

\subsubsection{Thick disks}

Following Ref.~\cite{Barausse:2014tra} and for the same density profile used for the thin disk, Eq.~\eqref{Eq:density}, one gets a dephasing of~\cite{Barausse:2014tra} 
\begin{equation}
    \label{thickBarausse}
    \delta \psi \sim 3\times 10^{-9} \frac{f^{}_{\rm Edd}}{10^{-4}} \left( \frac{0.1}{\beta} \right) \left( \frac{q}{10^{-5}} \right) ^{0.48} \left( \frac{10^6 M^{}_{\odot}}{M}\right)^{-0.58} \,, 
\end{equation}
for a thick disk, assuming quasicircular geodesic of the central object and that the orbital radius changes adiabatically. This dephasing is many orders of magnitude smaller than in the case of thin disks and, under the above assumptions, can be safely neglected~\cite{Barausse:2014tra}, even when testing GR. 

One can, however, also consider a rather extreme scenario~\cite{Barausse:2006vt}: a configuration containing a massive and compact tori close to the event horizon of the central BH. Under this scenario, which should be considered as a phenomenological approach, however,  Ref.~\cite{Barausse:2006vt} found that for a large portion of the space of parameters the waveforms produced by EMRIs in these BH-torus systems are indistinguishable from pure-Kerr waveforms, leading to another type of the aforementioned ``confusion'' problem. In this case, however, it arises from the interplay between mass and spin in comparison to the semilatus rectum and eccentricity and, therefore, it poses a potentially more significant challenge. In particular, for a gravitational waveform produced by an EMRI within a BH-torus system below the dephasing timescale, discerning it from one generated in a pure Kerr spacetime becomes virtually impossible~\cite{Barausse:2006vt}. This can lead to a fundamental bias, i.e., one could incorrectly attribute the primary component of the EMRI to a Kerr MBH with the wrong values of mass and spin. 

The torus considered in Ref.~\cite{Barausse:2006vt} is non self-gravitating, stationary, axisymmetric and plane-symmetric. For the long-range drag, the second term in Eq.~\eqref{Eq:AccInteract}, Ref.~\cite{Barausse:2007dy} assumed that the secondary is moving on a circular planar orbit that experiences a drag in the tangential and radial directions.

With the above assumptions and under the adiabatic approximation, they reported that the dissipative effect caused by the hydrodynamic drag from the torus on the secondary is significantly smaller than the radiation reaction effect. This is the case when considering a system composed of a primary with mass $M=10^{6}M_{\odot}$ and a torus with mass $M_{t}\lesssim M$ and outer radius $r_{\rm out} = 10^{5}M$. However, under certain (very special) conditions, the hydrodynamic drag can have a comparable impact to radiation reaction during detectable phases of the inspiral that can be observed. This could be the case if, for instance, the radius of the torus is decreased to $r_{\rm out} = 10^{3}M-10^{4}M$ or, even for $r_{\rm out} = 10^{5}M$ and $M_{t}\lesssim M$, if $M=10^{5}M_{\odot}$. Note however, that the above analysis considered the metric of the primary to be pure Kerr and neglected the self-gravity of the fluid, even when its mass is comparable to the mass of the primary. 

The above result, as much of the ones mentioned in this review, relies on approximations that may (drastically) change the conclusions if relaxed. In particular, as listed in Ref.~\cite{Barausse:2006vt}: i) they employed \emph{kludge} waveforms, ii) they imposed a cut-off at the dephasing time where radiation-reaction effects become negligible, iii) they restricted their analysis to purely equatorial orbits, and iv) they used highly compact tori in close proximity to the BH. Future work needs to relax one or more of these approximations and revisit the potential confusion problem.

\subsection{Towards a generic fully-relativistic formalism for environmental effects}
\label{Sec:GeneralEnviroment}

Most of the results presented in this section on environments have adopted at least an approximation: a slow-motion quadrupole formula to estimate GW emission and the dynamics;  Newtonian dynamical friction; or considering vacuum backgrounds. More recently, Ref.~\cite{Cardoso:2022whc}, based on classical works on BH perturbation theory, developed a generic, fully-relativistic formalism to handle environmental effects in EMRIs in spherically-symmetric, but otherwise generic, backgrounds. Within their framework, the authors claim that treating GW generation and propagation is possible, including matter perturbations. Therefore, it is expected to be able to capture other environmental effects. 

As a proof of concept, Ref.~\cite{Cardoso:2022whc} studied a BH surrounded by a halo of matter of mass $M_{\rm halo}$. The (unperturbed) metric solution to Einstein’s gravity minimally coupled to an anisotropic fluid they used, was derived in Ref.~\cite{Cardoso:2021wlq} for the Hernquist profile
\begin{equation}
    \rho=\frac{M^{}_{\rm halo} a^{}_0}{2\pi r (r+a^{}_0)^3}\,,
\end{equation}
where $a_0$ denotes a scale or characteristic length. The secondary is considered to orbit the above primary BH and cause perturbations to the geometry and matter stress tensor of the form~\cite{Cardoso:2022whc}
\begin{equation}
g^{}_{\mu\nu} = g^{(0)}_{\mu\nu}+g^{(1)}_{\mu\nu}\,,\qquad 
T^{\rm env}_{\mu\nu} = T^{{\rm env} (0)}_{\mu\nu}+T^{{\rm env} (1)}_{\mu\nu}\,,
\end{equation}
respectively, where a superscript ``(1)'' denotes the perturbation order. They exploited the background's spherical symmetry and separated the variables in first-order quantities, expanding into tensor spherical harmonics, classified as axial and polar, and assumed that the point-like secondary is on a geodesic of the background spacetime. The \emph{Supplemental Material} of Ref.~\cite{Cardoso:2022whc} presents the explicit perturbed equations. 

They found that, for example, at a (dimensionless) gravitational frequency $\omega M = 0.02$, the relative flux difference between a vacuum and a primary immersed in a halo with $M_{\rm halo} = 0.1 M$ and $a_0 = 10^2 M,\,10^3 M$ is $\sim10\%$, $1\%$, respectively~\cite{Cardoso:2022whc}. These results show the ability of GW astronomy to strongly constrain smaller scale matter distributions around BHs. 

%%%%%%%%%% Dark Matter Detection  %%%%%%%%%%
\subsection{A Dark Matter distribution}

Constituting around $85\%$ of known matter, DM plays a vital role in comprehending the Universe and fundamental physics. Most GR tests with EMRIs are often investigated in the context of high-density regions as DM spikes—dense concentrations encompassing  MBHs in galactic cores—whose profiles, potentially measured through gravitational interactions, could constrain DM properties. Detecting a single DM spike from EMRIs would notably restrict popular DM candidates, including ultralight bosons, keV fermions, self-annihilating MeV–TeV DM, and sub-solar mass primordial BHs. All these scenarios (detection of DM spikes and the aforementioned specific DM models) were studied in detail in Ref.~\cite{Hannuksela:2019vip}.  

In the previous sections, we have considered fluids made of baryonic matter. However, if one assumes that the profile in Eq.~\eqref{eq:DMProfile} is due a DM effect, one similarly can consider a dephasing due to dynamical friction from the DM distribution assuming both thin and thick disks. Consequently, the dephasing associated is given by Eqs.~\eqref{thinBarausse} and~\eqref{thickBarausse} for thin and thick disks, respectively, made of DM~\cite{Barausse:2014tra}. As discussed above, the net effect of a thick disk contribution is many orders of magnitude smaller than in the case of thin disks, and therefore, under the above assumptions, a DM distribution like that can always be neglected and it will be hard to obtain constraints. 

The formalism and Hernquist-type density distribution reviewed in Sec.~\ref{Sec:GeneralEnviroment}, can also describe a DM halo. The halo will induce a correction to the ringdown stage that causes differences between the number of GW cycles (of order hundreds) accumulated by EMRIs with and without DM halos~\cite{Cardoso:2021wlq}. In Refs.~\cite{Dai:2023cft,Rahman:2023sof}, eccentric orbits were studied around a Schwarzschild BH immersed in a DM halo. In particular, Ref.~\cite{Dai:2023cft}, derived analytic formulae for the orbital period and orbital precession for eccentric EMRIs, to show that the presence of local DM halos slows down the decrease of the semi-latus rectum and the eccentricity during the EMRI evolution. Thus, when comparing the number of orbital cycles with and without DM halos over one-year evolution before the merger, they argued that DM halos with a compactness ($M_{\text {halo}}/a_0$) as small as $10^{-4}$ can be detected. More recently, Ref.~\cite{Rahman:2023sof} used the Regge-Wheeler-Zerilli formalism to calculate the gravitational wave flux and analyzed the system's evolution using the adiabatic approximation. Agreeing with previous results, they found that the inspiral time lengthens in the presence of dark matter and with higher eccentric orbits. They noted that the halo compactness parameter $M_{\text {halo}}/a_0$ can significantly affect the location of the last stable orbit, thereby extending the parameter space in the $p-e$ plane that allows for bound orbits.

After performing a Fisher-matrix parameter estimation analysis, Ref.~\cite{Zhang:2024ugv} concluded that the values of $a_0$ and $M_{\rm{halo}}$ are highly correlated, making it difficult for the detector to accurately measure DM, even though there is a significant difference between the signals with and without DM due to dephasing and mismatch. They also argued that by taking into account the effects of dynamical friction and accretion, it may be possible to reduce uncertainty by up to one order of magnitude and to break degeneracies between these parameters~\cite{Zhang:2024ugv}.

One can also consider a scalar cloud around the EMRI system, which may be formed through superradiant instabilities of massive scalar fields. This scenario was recently studied in Ref.~\cite{Brito:2023pyl} using a fully relativistic framework that treated the effect of the cloud on the geometry perturbatively. Under the point-particle approximation, for a circular, equatorial orbit around a non-spinning primary surrounded either by a spherically symmetric or a dipolar non-axisymmetric scalar cloud, they computed the leading-order power lost by the point particle due to scalar radiation and show that, in some regimes, it can dominate over gravitational radiation emission~\cite{Brito:2023pyl}. The presence of a scalar cloud can produce resonances that can give rise to sinking (a torque is exerted that increases the orbital frequency of the binary and, therefore, temporarily accelerates its shrinking) and floating orbits (the orbit stays unaffected during the resonance). These effects were present even with Newtonian approximations of the impact of the cloud~\cite{Macedo:2013qea}. 

In Ref.~\cite{Duque:2023cac}, axionic environments of ultra-light (scalar) fields were studied, which is a fully relativistic analysis complementing and extending the results presented in Ref.~\cite{Brito:2023pyl} on emission from scalar structures. In particular, in the context of a fuzzy DM soliton (a self-gravitating structure made of particles with masses $m_\Phi \sim 10^{-22}-10^{-19}$ eV) and a spherical boson cloud. Considering an EMRI immersed in such ultra-light DM structures, they computed the flux of scalar particles generated and the relativistic dynamical friction acting on an EMRI. Their results show how sensitive the dissipation is via dipolar scalar depletion to the scalar field profile of the background. Consequently, a GW detector like LISA may be able to probe such types of ultra-light DM distributions. 

\subsection{A matter distribution around a beyond GR primary}

Beyond GR, Ref.~\cite{Barausse:2014tra} parametrized possible deviations in \emph{both} the conservative and dissipative sector in a theory-agnostic way. In what follows, we will just give a brief summary of their analysis, present their results, and refer the reader to Ref.~\cite{Barausse:2014tra} for details. 

In the conservative sector, they considered weak-field deformations around the most general static and spherically symmetric geometry, by assuming a metric of the form
\begin{equation}
    ds^{2} = -A\left(r\right)dt^{2}+B^{-1}\left(r\right)dr^{2}+r^2d\Omega^{2}\,,
\end{equation}
where $d\Omega^{2}$ denotes the standard metric on the unit 2-sphere ($S^2$), $d \Omega^2=d\theta^2+\sin^2\theta\,d\phi^2$, with the metric coefficients given by
\begin{align}
    A(r)=& \left( 1 -\frac{r^{}_+}{r}\right) \left[1+ \sum_{i=1}^{N^{}_\alpha} \alpha^{}_i \frac{M^i}{r^i} \right]\,, \\
    B(r)=&  \left( 1 -\frac{r^{}_+}{r}\right) \left[1+ \sum_{i=1}^{N^{}_\beta} \beta^{}_i \frac{M^i}{r^i} \right]^{-1}\,,
\end{align}
where $r_+$ denotes the location of the event horizon. The dimensionless coefficients $\alpha_i$ and $\beta_i$ are considered to be small numbers capable to produce contributions to GW observables. In principle, one should be able to map such metric to a known theory, and therefore the expansion will depend on the contributions (theories) one wants to study. In other words, the number of coefficients ($N_\alpha$ and $N_\beta$) depend on each theory. 

On the other hand, the dissipative sector was parametrized as~\cite{Barausse:2014tra}
\begin{equation}
    \dot{E} = \dot{E}^{}_{\textrm{GR}}\left\{ 1+\mathcal{B}\left(\frac{M}{r}\right)^{\tilde{q}}+\mathcal{C}\left(\frac{M}{r}\right)^{\tilde{s}}\log\left[\tilde{\gamma}\left(\frac{r}{M}\right)^{\tilde{t}}\right]\right\}, 
\end{equation}
where $\mathcal{A}$ and $\mathcal{B}$ depend on the fundamental couplings of the considered theory, while $\mathcal{C}$ and the tilded variables are dimensionless quantities (which also depend on the specific theory). The GR terms were written to lowest PN order and $\dot{E}_{\textrm{GR}}$ is given by the quadrupole formula. 

Under the above assumptions, the authors of Ref.~\cite{Barausse:2014tra} compared the beyond-GR effects with those due to a matter distribution with total mass $\delta M$, to get a lower bound on possible constraints on modified theories from EMRI GW detections. This bound was estimated to be
\begin{equation}
    \left(\frac{a_{i}}{M^{x_{i}}}\right)^{n}\gtrsim10^{-6}\left(\frac{\delta M}{10^{-6}M}\right) \,,
\end{equation}
where $a_i$, which is proportional to $\alpha_i$, denotes a coupling constant that corresponds to a non-minimal interaction between the field and the metric, and is assumed to have a physical dimension of $\left[ a_i\right]= \rm{length} ^{x_i}$, with $x_i\in \mathbb{R}$. The values of all these quantities, as well as the exponent $n$, depend on the gravitational theory considered. For instance, for massive gravity $n=2$, $a_i=\mu^2_{\rm graviton}$ and $x_i=-1$~\cite{Barausse:2014tra}. 

Therefore, under the above assumptions, modified theories whose couplings are much smaller than the bound above would produce phase deviations than the ones made by a matter configuration around the system. Since such effects might be degenerate, the above bound is only an intrinsic lower bound on a theory~\cite{Barausse:2014tra}. 

After the seminal and comprehensive work presented in Ref.~\cite{Barausse:2014tra}, more studies have appeared, where several other possible environments surrounding the primary compact object were analyzed. For instance, Ref.~\cite{Cole:2022yzw} recently revisited this problem using a Bayesian approach to assess the detectability and measurability of three particular environments: accretion disks, DM spikes, and clouds of ultra-light scalar fields (also known as gravitational atoms). They confirmed, for a system with a mass-ratio of $q=10^{-4}$, that the characteristic imprint each environment leaves on the EMRI gravitational waveform would allow us to identify the environment that generated the signal and accurately reconstruct its model parameters. 
Most of the studies we have reviewed focused on analyzing one effect at a time to determine the constraints on its inference. However, to obtain a more realistic analysis, it is important to infer \emph{multiple} beyond vacuum-GR effects \emph{simultaneously}. This is because the parameters that describe these effects are usually correlated with each other and the vacuum EMRI parameters. In fact, as shown in Ref.~\cite{Kejriwal:2023djc}, even if any of these effects are absent in the actual signal, these correlations remain and can cause inference biases. As a result, the overall measurability of the entire parameter set may not be as optimistic as previously reported, leading to a potential decrease in accuracy.

The work presented in Ref.~\cite{Becker:2022wlo} has made progress towards distinguishing between various environmental effects for IMRI systems. In this work, eccentric Keplerian orbits with gravitational GW emission, dynamical friction with the DM spike, and gas interaction with an accretion disk were studied. They proposed using the dephasing index ($n_d$) and the braking index ($n_b$) as complementary observational signatures to distinguish the impact of the accretion disk and DM spike effects on the IMRI system. The dephasing index was defined as $n_{\textrm{d}}\equiv\left(d\log\Delta N\right)/\left(d\log\mathcal{F}_{\textrm{tot}}\right)$, where $\Delta N$ is the difference of the number of GW cycles completed in the cases with and without the considered environment (induced by dissipative force), $\mathcal{F}$ is the orbital frequency, and $\mathcal{F}_{\textrm{tot}}$ is the total frequency evolution, which takes into account the contribution from both the vacuum GW emission and the environmental effect. The braking index is defined as $n_{\rm{b}}\equiv\mathcal{F}\ddot{\mathcal{F}}/\dot{\mathcal{F}}^{2}$. Therefore, differences in braking and dephasing indexes, as well as differences in the evolution of eccentricity, may help distinguish between various environmental models. 

\section{Testing the nature of Black Holes}
\label{Sec:NatureofBHs}

As previously mentioned, GWs emitted from EMRIs occur for multiple cycles within the range of space-based detectors, and within the strong-field region of the primary. This makes EMRIs an important tool for testing the geometry of the primary object, which is reflected in the precise pattern of GWs emitted.

However, modifications to GR that are motivated by curvature couplings tend to be weaker for larger BHs. This may seem ironic, as we would expect to observe with future space-borne instruments the BHs that are least likely to have modifications with great detail. In GR, the structure of the secondary typically does not affect the dynamics, even at second order in the self-force~\cite{Pound:2019lzj}, which is where most of our intuition starts when testing it. However, a scalar charge on the secondary can significantly affect the waveform and be measurable~\cite{Yunes:2011aa,Maselli:2020zgv}, and it has even been argued that modelling the primary as a Kerr BH may be an adequate approximation in most cases. Therefore, when testing GR, it is possible to focus on testing both the primary and secondary structures. In what follows, we discuss both. 

\subsection{Multipole Moments}
\label{multipole-moments-of-the-primary}

Vacuum, stationary and axisymmetric BH solutions in GR are uniquely described by the Kerr metric (see, e.g. Ref.~\cite{Hawking:1973uf}).  
The description is tight as the multipole moments of a Kerr BH satisfy  closed form relations~\cite{Hansen:1974ro}. Due to the Kerr symmetries (axisymmetry and of equatorial symmetry) the azimuthal number of the multipolar decomposition $m\neq0$ vanish. Consequently, one is left with these (rather simple) expressions
\begin{align}
    \mathcal{Q}_{\ell}+i\mathcal{S}_{\ell}= M (ia)^{\ell}
\end{align}
where $\mathcal{Q}_{\ell}=\mathcal{Q}_{\ell,m=0}$ and $\mathcal{S}_{\ell}=\mathcal{S}_{\ell,m=0}$ are the mass and current moments respectively, and $\ell$ is the multipole harmonic number ($\ell=0,1,\ldots$). These precise functional forms are the consequence of the no-hair theorems~\cite{Israel:1966rt,Carter:1971zc,Hawking:1972}. In a modified theory of gravity, such relations may not hold, and in the presence of more degrees of freedom, may not even be determined in terms of only the BH’s mass and spin. 

Extensive research has been devoted to find a way to parametrize the characteristics of the geometry of the primary into the EMRI waveform models in such a way that we can test a large variety of models of the primary's geometry. One of the main approaches is to consider that the primary is, with good approximation, an stationary axisymmetric object that can be characterized in terms of its multipole moments, as computed at spatial infinity~\cite{Geroch:1970rg,Hansen:1974ro,Thorne:1980rm}. 
An alternative is to use multipole moments defined near the primary, for instance, using the framework of isolated horizons in the case of BHs~\cite{Ashtekar:2004gp}. Deviations away from the  Kerr mass quadrupole $\mathcal{Q}_{2}$,
\begin{equation}
\Delta \mathcal{Q}^{}_{2}\equiv\frac{\mathcal{Q}^{}_{2}-\mathcal{Q}_{2}^{\textrm{Kerr}}}{M^{3}} \,,
\end{equation}
may be measured with an accuracy of the order of $\sim 10^{-4}-10^{-2}$ using LISA~\cite{Babak:2017tow}. 

This programme was initiated by Ryan~\cite{Ryan:1995wh,Ryan:1997hg} and has been continued by other authors (see, e.g., Refs.~\cite{Sotiriou:2004bm,Brink:2008xx,Sopuerta:2010fte, Kastha:2018bcr,Datta:2019euh,Datta:2019epe}). One can also construct particular models (metrics) for the primary that introduce a specific deformation (\emph{bumpiness}) with respect to a Kerr BH (see, e.g., Refs.~\cite{Manko:1992mn,Collins:2004na,Vigeland:2009pr,Vigeland:2010xe,Vigeland:2011ji,Moore:2017lxy,Xin:2018urr} for examples), and search for these deformations. 

Initially, the idea started as a way to construct compact objects that are almost BHs but have some multipoles with the ``wrong'' value, and therefore generically named ``bumpy'' BHs~\cite{Collins:2004na}. At present, some parametric metrics are too general and may not solve the field equations of a theory. Additionally, many of the parameters used in these bumpy metrics do not have physical significance. 

In this type of test, one searches for the impact of the multipolar structure of compact objects on the GW signal. Most studies had only considered axial and equatorial symmetry. Such calculations have been recently extended in Ref.~\cite{Fransen:2022jtw} to include leading order corrections with broken equatorial symmetry, but still preserving axisymmetry, proportional to $\mathcal{S}_2$ and $\mathcal{Q}_3$. Using modified AK waveforms, they estimated how accurately LISA EMRIs can measure or constrain equatorial symmetry breaking. They reported that LISA could constrain deviations away from equatorial symmetry with an accuracy of $\mathcal{O}\left(1\right)\%$. 

These results were extended in Ref.~\cite{Loutrel:2022ant}, where analytic waveforms that model the effects of mass quadrupole moments, allowing for non-axisymmetric configurations were developed. In particular, Ref.~\cite{Loutrel:2022ant} solved for the dynamics of binary systems at relative Newtonian order, considering the perturbation caused by mass quadrupole effects at the $2$PN order for quasi-circular binaries. They found that the phase corrections enter at relative 2PN order, while the amplitude modulations enter at $-0.5$PN order. These findings suggest that constraints on generic quadrupole moments can be achieved through phase differences in GW signals, particularly for small quadrupole parameters corresponding to axial and polar corrections from oblateness/prolateness. Their study also computed leading-order corrections to binary systems' dynamics and GW properties with generic deformations, specifically focusing on the mass quadrupole tensor. By investigating the dephasing due to generic quadrupole moments, they found that a phase difference $\gtrsim0.1$ radians is achievable for $\epsilon_m\gtrsim10^{-3}$, where this dimensionless modulus parameter is defined as
\begin{equation}
    \epsilon^{}_{m} \equiv \sqrt{\frac{2}{3}\left[\left(\frac{Q_{+m}^{R}}{Q_{0}}\right)^{2}+\left(\frac{Q_{+m}^{I}}{Q_{0}}\right)^{2}\right]}\,,
\end{equation}
and $Q^R$ and $Q^I$ denote the real and imaginary parts, respectively, of the mass symmetric trace-free multipole tensors $Q_m$.

We end this section by mentioning that the field multipole moments of the Kerr spacetime are not enough to determine the nature of an object conclusively. For example, Ref.~\cite{Bonga:2021ouq} constructed examples in the Newtonian theory that have the \emph{same} multipole moments as a Kerr BH. Constructing such examples is possible because the uniqueness results depend on the ellipticity of the field equations. While the differential equation for the Newtonian potential is elliptic, Einstein's equations are only elliptic when they describe stationary vacuum spacetimes and are formulated on the manifold of trajectories of the time-like Killing vector field (written in suitable coordinates). When the ellipticity is lost, as is the case with the presence of matter, uniqueness typically fails~\cite{Bonga:2021ouq}. 

The study presented in Ref.~\cite{Bonga:2021ouq} has some limitations that are important to keep in mind. Firstly, PN corrections are necessary to accurately analyze compact objects with small sizes relative to their masses. Secondly, materials with non-uniform stress are required to create star profiles resembling Kerr's multipole moments, which can be challenging to fabricate naturally. Finally, the study only addresses whether such objects are allowed by the laws of nature, and the stability of the constructed objects under perturbations still needs to be explored due to factors such as non-uniform stress and non-homogeneous equations of state. It is worth noting that the constructed objects have non-negative mass density and velocity smaller than the speed of light~\cite{Bonga:2021ouq}. However, these limitations do not affect the overall conclusion drawn from the study: Kerr hypothesis tests based on multipole moments measurements should be considered as null tests, and conclusions derived from these future measurements have to be derived with great care. 

Capability of tests of the no-hair conjecture\footnote{The ``no-hair conjecture'' states that the external properties of a BH (which in GR are mass, charge, and angular momentum) completely determine its observable characteristics, effectively erasing detailed information about the matter that formed it.} with the planned space-based detector TianQin~\cite{TianQin:2015yph,TianQin:2020hid} has been described in Ref.~\cite{Zi:2021pdp}. In particular, Ref.~\cite{Zi:2021pdp} used AK waveforms with quadrupole moment corrections. They forecasted that TianQin can measure the dimensionless quadrupole moment parameter with accuracy to the level of $10^{-5}$ under suitable scenarios. In their work, the importance of the choice of the cutoff (when one assumes that the secondary plunges into the primary) when using the AK waveform model was reported to have a significant effect on the results: if the Schwarzschild cutoff is used, the accuracy depends strongly on the mass of the MBH, while the spin has negligible impact; if the Kerr cutoff is used, however, the dependence on the spin is more significant~\cite{Zi:2021pdp}. Further work is required in this specific area. 

\subsection{Tidal effects}

As we have discussed before, EMRI systems can be immersed in environments that can modify their dynamics and leave an imprint in the resulting waveform. In particular, studying EMRIs in clusters around MBHs can help understand cluster properties and the growth history of MBHs~\cite{Berry:2019wgg}. 

One of the consequences of the presence of, for example, a companion or material around, is tidal effects. At the conservative level, tidal effects can modify the geodesic motion of both null and timelike orbits~\cite{Cardoso:2021qqu}, and therefore the orbital frequency and therefore the GWs emitted by EMRIs~\cite{Camilloni:2023rra}. One can, for instance, understand the perturbation shown in Eq.~
(\ref{eq:modmetric}) as an external tidal field, and express its contribution using two symmetric and trace-free tensors: the polar tidal field $\mathcal{E}_{\rm L}$ and the axial tidal field $\mathcal{B}_{\rm L}$. While Ref.~\cite{Camilloni:2023rra}, in the context of EMRIs, specialized to a non-rotating primary, it provides explicit derivations for the electric and magnetic tidal moments at the quadrupole order for a Kerr BH.

Generating waveforms for these types of perturbed systems is paramount. In particular, studying tidal resonances, which are induced by the tidal field of nearby stars or stellar-mass BHs, a DM distribution, or the rest of the galactic potential, encode information about the EMRI system and its environment. However, including these effects makes the waveform model more complicated. Generically, these tidal resonances will cause a secular (long-timescale) shift to the orbital angular momentum, that, if properly modelled, would provide information about the distribution of mass near a galactic center MBH~\cite{Bonga:2019ycj}. Due to the intrinsic complexity and sources of resonant effects, a generic description is still lacking, and assessing the importance of resonant effects on future gravitational detections with future detectors has proven difficult~\cite{Speri:2021psr}. Current efforts are being made to make accurate EMRI models by iterating in perturbation theory to second order in the mass ratio and including effects such as the impact of the smaller body's spin~\cite{Bonga:2019ycj}. 

A Newtonian analog was used to study tidal effects in EMRIs in Ref.~\cite{Bronicki:2022eqa} by only changing the conservative part of the dynamics. The studied Newtonian system is the Euler’s 18th-century problem of two fixed gravitating centers, which is the unique stationary and axisymmetric Newtonian potential that shares several key properties of the Kerr metric~\cite{Eleni:2019wav}. By comparing the timescale of perturbation with radiation reaction at resonance, they estimated that regime in which tidal effects might be included when modelling EMRI gravitational scales as $\epsilon \gtrsim C q$, where $\epsilon$ measures the strength of the external tidal field, and $C$ is a real number that depends on the resonance and the shape of the tide. These types of studies show that tidal resonances, if not carefully modelled for, may ultimately limit the precision of waveform models. However, a thoughtful analysis, including dissipative effects, is required to obtain precise estimates of these timescales and the magnitude of the effects. 

\subsubsection{Tidal Love numbers}

The tidal deformability of a self-gravitating object is measured by its Love numbers $k_\ell$, which vanish\footnote{This is certainly the case for the static Love numbers. However, the corresponding dynamical tidal coefficients are generically non-zero~\cite{Perry:2023wmm}.} for BHs in GR~\cite{Damour:2009vw,Binnington:2009bb,Poisson:2021yau}, and depend on the equation of state for a NS~\cite{Chatziioannou:2020pqz}. The concept of tidal Love numbers has a long tradition in the literature~\cite{Bildsten:1992my,Flanagan:2007ix, Hinderer:2008dm} and has extensively been studied primarily in the context of binary NS and BH-NS systems~\cite{Damour:2009vw,Binnington:2009bb}. 

For massive (enough to definitely not be a NS) compact objects, if a non-vanishing value is accurately measured for the tidal Love numbers, it could indicate the presence of new types of ultracompact massive objects~\cite{Cardoso:2017cfl,Pani:2019cyc}. In Ref.~\cite{Cardoso:2017cfl}, the tidal Love numbers of various exotic compact objects (ECOs), such as different classes of boson stars, gravastars, and wormholes, were computed. For all these ECOs, which were described for spherically symmetric and static background geometries, it has been found that their tidal Love numbers vanish logarithmically~\cite{Cardoso:2017cfl} in the BH limit.

In PN theory, the tidal love numbers have an impact on both the conservative and dissipative components. However, the correction from the conservative term is not as significant as the one from the dissipative part~\cite{Pani:2019cyc}. In the EMRI limit and at leading PN order, the tidal correction to the instantaneous GW phase is of the form~\cite{Piovano:2022ojl}
\begin{equation}
\label{eq:tidalfreq}
    \phi^{}_{\rm tidal} \propto \frac{k_{1}}{q}v^5 \,, 
\end{equation}
where $k_1$ is the quadrupolar, electric tidal love number of the primary, $f$ is the GW frequency, and  $v=(\pi M f)^{1/3}$. The contribution of the tidal love number of the secondary is subdominant, as it scales as $q^4$~\cite{Pani:2019cyc}. 

Assuming the SPA to obtain approximate semi-analytical EMRI waveforms in the frequency domain, Ref.~\cite{Piovano:2022ojl} claimed that the tidal love number of the primary can be measured at the level of $10^{-3}$ if the primary's spin is high ($a\gtrsim0.9$). In particular, for a spinning primary with dimensionless spin $a= 0.9$, the bounds are approximately four orders of magnitude more stringent than what is achievable with current ground-based detectors for stellar-mass binaries. The constraining power, however, increases significantly with the spin, as for a spin of $a= 0.99$ the constraint gets better by two orders of magnitude. 

The inclusion of the spin to the secondary, however, did not affect the bounds on the tidal Love number of the primary. However, the study presented in Ref.~\cite{Piovano:2022ojl} only focused on circular, equatorial, and nonprecessing orbits. 

Using osculating orbits and an AK waveform, Ref.~\cite{Zi:2023pvl} studied eccentric orbits around a spinning primary and found similar results. In particular, for $q=10^{-5}$, $a = 0.8$, and an initial eccentricity of $e = 0.1$, at $D=1$Gpc, LISA can reach a level of accuracy of $10^{-4}$ with one year of observation. Therefore, using EMRIs to measure the tidal love number of the primary can be up to $6$ orders of magnitude better than what can be achieved for NSs with ground-based detectors~\cite{LIGOScientific:2018cki,Silva:2020acr}.

The above studies were carried out using the Fisher information matrix to compute numerical derivatives of the waveform parameters, which are then used to construct a matrix that quantifies the information content of the signal for each parameter. The Fisher matrix measures how well each parameter can be measured and how these measurements correlate. Computing the derivatives of the gravitational waveform signal can be challenging, especially for EMRIs~\cite{Vallisneri:2007ev}, as we need to solve the equations of motion for the smaller object in the strong gravitational field of the larger one, which can be computationally expensive. To address these difficulties, in Ref.~\cite{Piovano:2022ojl}, a semi-analytical approximation of the waveform using the SPA was implemented, providing an accurate description in the frequency domain. This same approach can be used for other gravity tests with EMRIs.

\subsection{Horizon absence} 

Within GR, BHs have spacetime curvature singularities that are conjectured to be always hidden by event horizons. At microscopical scales, however, it is theoretically expected that BHs should behave as thermodynamical systems with, respectively, the area of the event horizon and its surface gravity playing the role of the entropy and temperature~\cite{Hawking:1976de}. As a consequence, BHs can evaporate\footnote{An event horizon is sufficient but not necessary to have the evaporation process (see, for instance, Refs.~\cite{Barcelo:2006uw,Barcelo:2010pj})} and many paradoxes arise. Therefore, the existence, or not, of BH (event) horizons has fundamental implications in theoretical physics. 

Interestingly, and to add more fuel to the already enough complexities these paradoxes bring, there are even examples in which one can identify horizonless geometries that can mimic some observational features of BHs~\cite{Cardoso:2019nis}. For instance, a popular model of such objects are “fuzzballs,” in which a classical BH is replaced with a dense, star-like object from string theory, with no singularity and no event horizon. These objects therefore will have a nontrivial structure at the putative horizon scale. Fuzzballs will have differences, with respect to classical BHs, in the multipolar structure. In particular, Ref.~\cite{Bianchi:2020bxa} found numerical evidence that all multipole moments are typically larger (in absolute value) than those of a Kerr BH with the same mass and spin.

Horizonless objects with a reflecting surface, which can be parametrized  by a reflectivity coefficient $\mathcal{R} \left(\omega \right)$, where $\omega$ is the frequency of the absorbing radiation, can be unstable above a critical value of the spin due to the ergoregion instability~\cite{Maggio:2018ivz}: an instability that can occur in an asymptotically flat spacetime due to the presence of physical negative-energy states within the ergoregion, which can cascade towards even more negative-energy states. In Ref.~\cite{Maggio:2018ivz}, this instability was studied for electromagnetic perturbations of ultracompact Kerr-like objects with a reflecting surface, and derived an analytical result for the frequency and the instability timescale
\begin{equation}
    \tau^{}_{\rm{inst}} \sim 2 \left( \frac{M}{10^6 M^{}_\odot}\right) \, \rm{months} \,,
\end{equation}
of unstable modes which is valid at small frequencies. To date, the absence of any detectable GW stochastic background sets the most stringent constraints on these models~\cite{Barausse:2018vdb}.

Focusing on stable configurations, Ref.~\cite{Maggio:2021uge} studied EMRIs around a spinning horizonless compact object. In particular, they considered models of Kerr-like compact objects, from Ref.~\cite{Abedi:2016hgu}, approximated by the Kerr metric in the exterior spacetime and the properties of the interior modelled in terms of a reflectivity coefficient. The considered compact object has therefore a radius located at
\begin{equation}
    r_0 = r_+ \left( 1+\epsilon\right)  \,,
\end{equation}
where $r_+=M+\sqrt{M^2-a^2}$ is the value of the location of the horizon of a Kerr BH (in Boyer-Lindquist coordinates), and $\epsilon$ is a small parameter. 

These horizonless partially absorbing compact objects, i.e., $\left|\mathcal{R}\right|^2\leq1$, have quasi-trapped states in the photon sphere that are associated to low-frequency quasi-normal models. During the inspiral of a point particle, these modes can be excited when the orbital frequency equals the quasi-normal mode frequency~\cite{Cardoso:2019rvt}. In other words, when the orbital frequency resonates with the real part of the quasi-normal frequency interesting excitations in the energy fluxes happen. These excitations, however, are unimportant for non-spinning objects~\cite{Cardoso:2019nis}.  

Under this scenario, evolving the orbital parameters in an adiabatic expansion, Ref.~\cite{Maggio:2021uge} analyzed the motion of a point particle in a circular equatorial orbiting around a Kerr-like horizonless object, and found that EMRIs can potentially constrain the reflectivity at the level of
\begin{equation}
    \left|\mathcal{R}\right|^{2}\lesssim10^{-8} \,, 
\end{equation}
and for specific models such as~\cite{Wang:2019rcf} 
\begin{equation}
\mathcal{R}\left(\omega\right) = e^{\frac{\left|\omega-m\Omega_{\rm{H}}\right|}{2 T_{\rm{H}}}} \,,
\end{equation}
can be confirmed or ruled out. Here $\omega$ denotes the frequency of the GW wave, $m$ is the azimuthal number (not the mass of the secondary!), $\Omega_{\rm{H}}$ is the angular velocity at the horizon of the BH, and $T_{H}$ is the Hawking temperature. The work of Ref.~\cite{Maggio:2021uge} complements and builds upon the results of Refs.~\cite{Datta:2019euh,Datta:2019epe}, which found that tidal heating can be discriminator for horizons in EMRIs. 

We have seen that EMRIs have the potential to provide very robust limits on the reflectivity of ultracompact objects. In fact, these (projected) limits are orders of magnitude more stringent compared to those that could potentially be obtained from ``echo'' (secondary pulses that appear in the late-time ringdown waveform) searches in the post-merger phase of comparable mass coalescences~\cite{Cardoso:2016oxy,Cardoso:2016rao, Price:2017cjr, Cardoso:2017cqb}. Ironically, most echo studies have been done perturbatively on the test-particle limit, and therefore, in a formal sense~\cite{Cardoso:2016oxy}, only valid for EMRIs, but EMRIs are not a good source to detect them. This is because the SNR for echoes is expected to be (very) low, in comparison to the overall signal, even though the resulting energy in echoes can surpass the standard ringdown if the ECO's reflectivity is high~\cite{Cardoso:2019rvt}.

Future work, for example, could be devoted to generalizing these constraints to eccentric and inclined orbits. However, and perhaps more interesting, one could relax the stationary and adiabatic assumption and study these types of constraints in the presence of a ``greenhouse'' effect~\cite{Cardoso:2022fbq} that happens when dealing with such compact objects. The main point of Ref.~\cite{Cardoso:2022fbq} was that very compact objects can behave as a cavity, and therefore can have a very large “build-up time,” i.e., the time taken for arbitrary initial conditions to reach a stationary state. Therefore, assuming that the field is stationary, and then superposes an adiabatic evolution to evolve the particle in its motion, driven by GW emission, as in Ref.~\cite{Maggio:2021uge}, the time taken for arbitrary initial conditions to reach a stationary state can be very large. In particular, the build up time is of the order of the inverse of the resonance width itself,  which is typically much larger than the timescale for evolution via GWs. In other words, account for the energy piling up within the photon sphere of these objects. 

One can also further study the generality of parametric models, given that translating theoretical ideas into practical gravitational models is a complex process with potential limitations. Consequently, it is crucial to exercise caution when establishing connections between theoretical concepts, such as fuzzballs, and theory-agnostic parametrizations of reflective surfaces. Theory-agnostic models also have inherent limitations and may not necessarily be as general as one may expect, given the significant impact of the system's interior on boundary conditions. Many current parametric models assume spherical symmetry and employ the Schwarzschild metric in the exterior. However, the introduction of rotation can pose challenges, as there is no Birkhoff theorem in such cases. Assuming that GR remains unaltered up to the reflective surface is a robust assumption, considering that deviations at length scales comparable to the surface can indeed influence spacetime.

\subsection{Other types of compact objects}

If macroscopic ECOs exist, they could in principle create EMRI systems (either as the primary or the secondary, or both), providing a new target species for space-based GW detectors. ECOs with sub-solar mass can accumulate a large SNR in the LISA band by staying around the MBH's ISCO for a long time, in contrast to binary systems detected by LVKC-like sources. It is expected that for ECOs that are disrupted before reaching the ISCO, their radius, and therefore their compactness, can be estimated, helping to break the degeneracy with other types of objects. There are many possible self-gravitating compact objects (e.g., anisotropic stars, gravastars, AdS bubbles, wormholes, fuzzballs, or 2$-$2 holes, to name a few), and we refer the reader to specific works (e.g., Refs.~\cite{Carballo-Rubio:2018jzw}) or reviews (e.g., Refs.~\cite{Berti:2015itd,Cardoso:2019rvt}) for details. Below, we only highlight some examples where there has been recent progress. Consequently, this is \emph{not} an exhaustive list of ECOs, and the selection does not reflect their historical significance, as some of the objects considered below have a longer and richer history than the previously mentioned ECOs.

Before entering into the details of specific examples, it is important to mention that despite there being a substantial variety of ECOs proposed in the literature (some of them with many variants, see, for instance,  Ref.~\cite{Cardoso:2019rvt} for several examples), it is far from obvious how to incorporate them into the study of EMRIs that we carry out in the standard scenario, where at least the physical model for the primary and secondary, as well as the equations of motion are known. The main difficulty is solving the problem with different mathematical and numerical techniques.  

In many cases, we only know static/stationary models for ECOs. However, it is not clear what the precise equations that govern the dynamics of these objects are, since they are made of exotic matter whose mathematical description is, in the best of the cases, challenging to treat and, in the worst-case scenario, it is unknown. Therefore, since it is quite common that ECOs need a precise description of their dynamics out of the equilibrium state, we do not even know in many cases whether these objects are stable or not. All these difficulties are particularly relevant when it is the primary that is an ECO since we need to perturb its geometry and understand well how its geometry behaves out of the stationary state. In the case of the secondary, we may try to find an effective description and perhaps model its internal dynamics in terms of a set of parameters, like the Love numbers (see Ref.~\cite{Hinderer:2007mb} for a study focused on NSs), and in particular the tidal deformability used frequently in the case of GW searches of binaries involving NSs~\cite{Chatziioannou:2020pqz}.

In any case, a possible way out is to assume that no matter how exotic the components of an EMRI are, we can always describe them as deviations of known objects, with the hope that, in this way, we can figure out a proper treatment of the internal degrees of freedom of the object.  All this clearly shows the EMRI problem's dimension in a general context, beyond the standard scenario, and calls for a (unifying) framework where we can test as many EMRI configurations with future space-based GW detectors as possible.

\subsubsection{Hairy Black Holes}
\label{Sec:Hairy}

The ``no-hair'' theorems~\cite{Israel:1966rt,Carter:1971zc,Hawking:1972} are essential in understanding and restricting the possible characteristics of BHs in diverse gravitational scenarios. These theorems impose significant limitations on the BH characteristics beyond their mass, charge, and angular momentum. A remarkable exception has been identified for a linear coupling between a shift-symmetric, massless scalar field and gravity involving the Gauss--Bonnet invariant~\cite{Sotiriou:2013qea, Saravani:2019xwx}. This unique coupling evades the constraints imposed by the no-hair theorem without introducing additional degrees of freedom or compromising local Lorentz symmetry. Evading these theorems is a nontrivial task, and it is expected that realistic, astrophysical BHs in scalar-tensor theories will differ from their GR counterparts~\cite{Sotiriou:2015pka}.

A general framework to calculate analytical waveforms for EMRIs in scalar-tensor theories under the assumption that the primary features a scalar hair was developed in Ref.~\cite{Kuntz:2020yow}. They adopted an effective field theory (EFT) approach for perturbations in the presence of a scalar hair as described in Ref.~\cite{Franciolini:2018uyq}, to study departures from the BH quasi-normal frequencies predicted by GR. In that work, they wrote the most general action for the perturbations of a spherically symmetric solution up to some given order in derivatives or number of fields. They showed that at any PN order, the background metric and the relevant operators can be encoded in a finite number of parameters, which represent deviations of the metric and of the action from GR in the PN regime, that can be used to obtain an analytic expression for the dissipated power in the odd sector, up to $3.5$PN order, by solving perturbatively the Regge--Wheeler equation in the presence of a point-particle source. 

This formalism can provide a one-parameter family of templates that can be used to provide tests of GR, and, therefore, serve as a method that can provide modelled searches in a large class of modified gravity theories. Thus, it could be viewed as a parametrized PN framework applied to hairy BHs. However, it has not been used to forecast constraints. 

Another relevant example of hairy BHs are the ones endowed with the bumblebee field, known as a single vector or axial-vector field ($B_\mu$). This is a model with Lorentz violation that displays sharp features such as rotation, boost, and CPT violation, and yet allows for simple analytical solutions~\cite{Kostelecky:2003fs}. It can be seen as an extension of the Einstein--Maxwell theory, with the vector field non-minimally coupled to gravity. The detectability of this vector hair was studied in Ref.~\cite{Liang:2022gdk} for adiabatically decaying circular orbits and only considered the tensor modes. The extra modes of GWs, that couple to the perturbations of the bumblebee field, remain unstudied and are necessary for more accurately calculating the orbital evolution and building the waveform. Under this context, Ref.~\cite{Liang:2022gdk} considered the bumblebee field to be $b_\mu= (b_t, 0, 0, 0)$ that yields a vector bumblebee charge~\cite{Liang:2022gdk}
\begin{equation}
Q\propto -\lim_{r\rightarrow \infty} r^2 b^\prime_t \,,
\end{equation}
where the prime denotes the derivative with respect to $r$. For example, this model recovers the GR Reissner--Nordström solution when $Q=\sqrt{4\pi} r b_t $. For a non-rotating primary, they reported that combining the information from the orbital phase difference and match between the waveforms, one-year observation of an EMRI, can probe the vector charge as small as $Q\sim \mathcal{O} \left ( 10^{-3}\right)$~\cite{Liang:2022gdk}. 

\subsubsection{Boson stars}

New fields like axions and axion-like particles are crucial for explaining DM through cosmological models. The simplest version involves massive scalar fields that create self-gravitating objects known as boson stars. These macroscopic quantum states form naturally from gravitational collapse, are dark unless weakly interacting with regular particles, and can cluster into ultracompact configurations via ``gravitational cooling''~\cite{Cardoso:2019rvt}. When the secondary is a white dwarf, NS, or a primordial BH, the object is compact enough to avoid being tidally disrupted at the ISCO. However, this may not necessarily be true for boson stars, as they can be tidally disrupted outside the ISCO. Consequently, the GW signal will be cut off at the tidal radius, and the maximal frequency recorded by the detector will be smaller than what can be achieved at the ISCO, which could be a sharp feature to look for. 

A consistent description of EMRIs considering a boson star is missing. The only relevant study regarding this model of ECO, which has received considerable attention in the literature over the past years, is presented in Ref.~\cite{Guo:2019sns}. In this work, assuming circular orbit in the equatorial plane, using the Teuksolsky--Sasaki--Nakamura formalism~\cite{Teukolsky:1973ha,Sasaki:1981sx}, it was suggested that observing these compact objects for over five years with an SNR of $20$ with LISA may be able to probe them down to very sub-solar mass and compactness. 

\subsubsection{Braneworld model: extra dimensions}

According to the braneworld model (see, e.g., Ref.~\cite{Maartens:2010ar} for a review), all matter fields are confined to a four-dimensional slice (the brane) of a higher-dimensional bulk spacetime. In these models, gravity can propagate in all extra dimensions, which can be large or infinite. Higher-dimensional models typically introduce correction terms to the standard Einstein field equations that account for the local bulk effects on the matter on the brane and the gravitational field outside the brane. A particular vacuum, static, and spherically symmetric solution was derived in Ref.~\cite{Dadhich:2000am}, which has the same functional form of the GR solution Reissner--Nordström but instead of an electric charge $e$, there is a new parameter $\beta$ that does not necessarily have to be positive. Since it is a vacuum solution with no Maxwell fields outside the BH, the resulting spacetime is then considered to carry a ``tidal'' charge, encapsulated by $\beta$. 

In Ref.~\cite{Rahman:2022fay}, by considering an EMRI system where the primary is a spherically symmetric static braneworld BH that carries a tidal charge $Q=-\beta/M^2$, they estimated the effect of the tidal charge on the total GW flux and orbital phase due to a non-spinning secondary. Under this particular theory, a non-vanishing value of $Q$ signals the presence of an extra dimension. Studying equatorial and quasi-circular orbits, they found that even values of $Q\sim 10^{-6}$ would cause a detectable mismatch when compared to evolutions on a Schwarzschild ($Q=0$) background.

\subsection{Tests of the existence of new fundamental fields: when the secondary matters}
\label{sec:secondary}

Over the past few years, the conventional wisdom that EMRIs will trace the spacetime of the primary and probe whether it is a Kerr BH or not (as shown, for example, in Sec.~\ref{multipole-moments-of-the-primary}) has been expanded. Recent studies have shown that a scalar (and potentially other) charge on the secondary will also modify the waveform significantly to be measurable by LISA, and that modelling the primary as a Kerr MBH can be an adequate approximation for some cases when testing GR\footnote{There is also the possibility of using Pulsar Timing to follow an extreme-mass-ratio binary where the secondary is a NS~\cite{Kimpson:2019hwn}.}. In other words, the structure of the secondary also plays an important role. 

Measuring the properties of the EMRI secondary was brought back to the table a few years ago, in the context of alternative theories of gravity in which the central supermassive BH is described by a GR background, and the secondary carries a scalar charge. In Ref.~\cite{Yunes:2011aa}, it was shown, for a broad range of scalar-tensor theories in an asymptotically flat and stationary background, that, when expanding perturbatively in terms of the mass ratio, the equations governing the emission of gravitational and scalar waves depend solely on two parameters, namely the coupling constant and the mass of the scalar field. The specific scalar-tensor model would uniquely determine these parameters. 

For massless scalar-tensor theories, in Ref.~\cite{Yunes:2011aa}, the scalar energy flux carried to spatial infinity in the point-particle approximation and for quasi-circular orbits has been numerically calculated using the Teukolsky formalism. They concluded that EMRIs \emph{cannot} constrain Brans--Dicke's theory beyond the existing limits imposed by Solar System tests and also would be less stringent than those obtained through observations of comparable-mass binary inspirals when considering only the leading-order Brans--Dicke correction. This lack of constraining power of EMRIs is due to the fact that scalar-tensor theories do not necessarily entail substantial modifications to the curvature of GR. The scalar-field modification primarily influences the Ricci scalar in the Jordan frame at the action level without introducing higher curvature corrections. Consequently, the modified theory introduces a pre-Newtonian correction to the GW phase (a $-1\;$PN effect relative to GR) instead of a PN one~\cite{Yunes:2011aa}. In fact, they also showed that in the EMRI limit, BBHs in any scalar-tensor theory do \emph{not} lead to dipolar gravitational radiation to \emph{all} orders in PN theory, provided the background scalar field is constant at spatial infinity.

More recently, Ref.~\cite{Maselli:2020zgv} proposed a similar setup as the one studied in Ref.~\cite{Yunes:2011aa}, to study theories with higher-curvature corrections. In particular, Ref.~\cite{Maselli:2020zgv} considered that the primary is described by a GR background, and the secondary carries a scalar charge, and it is described by the action~\cite{Maselli:2020zgv}:
\begin{equation}
  S[\mathbf{g},\varphi,\Psi] = S^{}_0[\mathbf{g},\varphi] +
 \alpha S^{}_c[\mathbf{g},\varphi] + S^{}_{\rm m}[\mathbf{g},\varphi,\Psi]\,,\label{action}
\end{equation}
where
\begin{equation}
  \label{S_0}
S^{}_0= \int d^4x \frac{\sqrt{-g}}{16\pi}\left(R-\frac{1}{2}\partial_\mu\varphi\partial^\mu\varphi\right)\,,
\end{equation}
the Ricci scalar is denoted $R$, $\varphi$ is a scalar field, $\alpha S_c$ describes non-minimal couplings between the metric tensor $\mathbf{g}$ and $\varphi$, and $\alpha$ is a coupling constant with dimensions $[\alpha]=(\rm{mass})^n$, where $n$ is an integer. The action of the matter fields $\Psi$ is denoted with the last term, i.e., $S_{\rm m}=-\int m \left( \varphi\right) d\tau$, where $m\left( \varphi\right)$ is the mass of the point particle, which depends on the value of the scalar field at the location of the particle, and $\tau$ is the proper time. 

The above general action can enclose theories satisfying no-hair theorems~\cite{Israel:1966rt,Carter:1971zc,Hawking:1972}, for which stationary BHs will be described by the Kerr metric, as well as theories evading no-hair theorems. In particular, for the latter, some theories can predict BHs that carry a nontrivial scalar field, such as scalar Gauss--Bonnet gravity or dynamical Chern--Simons gravity, violating the strong equivalence principle. For these types of quadratic theories, if the scalar charge is fixed to be a constant and controlled by the coupling to the invariant, and the mass of the primary is the \emph{only} relevant scale, then the charge per unit mass carried by the primary is negligible~\cite{Saravani:2019xwx}. Therefore, for such theories,  ``large'' BHs, for which the curvature is low, are effectively Kerr BHs, and the effects of a massless scalar on MBHs is mostly negligible. 

However, this, at least for tests of gravity with EMRIs, ended up being a good property because the secondary then behaves as a scalar monopole that is accelerated gravitationally by the supermassive BH and thus emits scalar, mostly dipolar, radiation~\cite{Maselli:2020zgv}. Thus, if one assumes that: i) the primary is described by a GR solution, ii) that the secondary acts as a point particle moving on geodesics, and iii) that the scalar field is constant in the background spacetime, the resulting field equations are greatly simplified. In particular, one obtains that the tensor part is exactly as in GR, while the scalar sector is described simply as a scalar field propagating on a Kerr background with the source term depending on the orbital configuration of the secondary and the scalar charge~\cite{Maselli:2020zgv}
\begin{equation}
    \varphi=\varphi_{0}+\frac{\mu d}{r}+\mathcal{O}\left(\frac{\mu^{2}}{r^{2}}\right),
\end{equation}
where $\varphi_0$ is the asymptotic value of the scalar field, $\mu=m\left(\varphi_0 \right)$ is the mass of the point particle, and $d=4 m^\prime\left(\varphi_0 \right)/ m \left(\varphi_0 \right) $ denotes the dimensionless scalar charge. 

Under these assumptions, considering a non-spinning primary, using the quadrupole approximation for equatorial orbits, in Ref.~\cite{Maselli:2020zgv} was shown that the scalar field emission has a significant cumulative effect on the dynamics of the EMRI, and that one-year observation with LISA could detect scalar charges $d$ as small as $d\sim 5 \times 10^{-3}$. This particular analysis was extended in Ref.~\cite{Maselli:2021men} to a rotating primary described by the Kerr metric. As an example, for a mass ratio of $10^{-5}$ and dimensionless spin $0.9$, they forecast that LISA can put constraints on Gauss--Bonnet gravity as stringent as the already placed using ground based detectors~\cite{Nair:2019iur}, which is of the order of a few kilometers. Their results are independent of the origin of the scalar field and of the structure and other properties of the small compact object, so it can be seen as a generic assessment of LISA's capabilities to detect new fundamental fields. 

Since the energy fluxes and the GW dephasing accumulated is a function of the scalar charge, mass ratio, and spin of the central MBH, Ref.~\cite{Guo:2022euk} recently explored this parameters space. They showed that the dephasing changes monotonously and in proportion with these parameters, i.e., increasing either the mass ratio or the spin, strengthens both gravitational and scalar energy emissions. More recently, Refs.~\cite{Barsanti:2022ana,Zhang:2022rfr} included eccentricity and found that eccentric inspirals will improve these constraints, for $d  \gtrsim 0.01$.

The previous discussion assumed a massless scalar field and that any (self) interaction of the scalar respects shift-symmetry (the symmetry that protects the scalar from acquiring a mass). However, theoretical reasons support the existence of weakly interacting, massive scalar (spin $0$) or vector (spin $1$) particles~\cite{Berti:2015itd}. If these particles exist, they are expected to leave an observable imprint on compact objects only if their Compton wavelength, the inverse of their mass, is comparable to the size of the objects or the length scales of their system. Specifically, the condition is $\mu_{\text s} M \lesssim 1$ for a BH system, where $\mu_{\text s}$ is the mass of the scalar field~\cite{Barsanti:2022vvl}. The scalars that GW observations can currently probe would have masses smaller than $\sim 10^{-16}$eV, i.e., the estimate is ~\cite{Brito:2015oca}
\begin{equation}
    \mu_{\textrm{s}}\left[\textrm{eV}\right]\simeq \left(\frac{\mu_{\textrm{s}}M}{0.75}\right) \left(\frac{10^{6}M^{}_{\odot}}{M}\right)10^{-16}\textrm{eV} \,.
\end{equation}
Therefore, the search is aimed at ultralight scalar fields. In particular, if light bosons exist in nature, they will spontaneously form ``clouds'' by extracting rotational energy from rotating MBHs  through superradiance~\cite{Brito:2015oca}. Since the superradiant growth of the cloud sets the geometry of the final BH, and the BH geometry determines the shape of the cloud, 
\emph{both} the BH geometry and the cloud encode information about the light boson. Thus, GW measurements of the BH/cloud system are over-determined~\cite{Hannuksela:2018izj}.   

The idea, as proposed in Ref.~\cite{Hannuksela:2018izj}, is to obtain \emph{independent} measurements of the boson mass to confirm the existence of the cloud. This is in principle possible because the superradiant instability is expected to occur quickly ($\tau_{\text {instability}} \ll \tau_{\text {GW}}$) and that the BH/cloud system is in equilibrium over a typical LISA observation time, i.e.,  $T_{\text {observation}} \sim \mathcal{O}\left( \text{yr}\right) \ll \tau_{\text {instability}} \ll \tau_{\text {GW}}$~\cite{Hannuksela:2018izj}. Under the SPA, Ref.~\cite{Hannuksela:2018izj} constructed an EMRI gravitational-waveform template (for an equatorial orbit) in a BH/cloud potential given by $\Phi(r)=-M/r+\Phi_{\text b}$, where $\Phi_{\text b}$ is the gravitational potential produced by the boson cloud, as prescribed in Ref.~\cite{Ferreira:2017pth}. They found that EMRI observations with LISA can be used to confirm (or rule out) the formation of ultralight boson condensates around astrophysical BHs in the mass range $\mu_{\textrm{s}} \in \left[10^{-16.5},10^{-14} \right]$ eV. 

The dynamics when a massive scalar field is considered are also very rich and interesting. For instance, as shown in Ref.~\cite{Cardoso:2011xi}, the coupling of the field because of superradiance leads to a striking effect: matter can hover into ``floating orbits'' for which the BH’s rotational energy entirely provides the net gravitational energy loss at infinity. In other words, an extreme form of energy extraction from a Kerr BH is possible when considering massive scalars around rotating BHs. This phenomenon is produced by a resonance between orbital frequencies and proper oscillation frequencies of the BH, and should be general enough to apply to stationary background geometries different from the Kerr solution. 

Beyond this interesting effect, one can also consider the following modification to the action given in Eq.~\eqref{action}
\begin{equation}
  \label{S_0}
S^{}_0= \int d^4x \frac{\sqrt{-g}}{16\pi}\left(R-\frac{1}{2}\partial^{}_\mu\varphi\partial^\mu\varphi-\frac{1}{2} \mu^2_{\text s} \varphi^2 \right)\,,
\end{equation}
i.e., adding the last term to consider a scalar field with mass $\mu_{\text s}$, and study the EMRI dynamics. In particular, using this action, Ref.~\cite{Barsanti:2022vvl} argued that one can measure the scalar charge per unit mass of the secondary \emph{and} the mass of the scalar field simultaneously~\cite{Barsanti:2022vvl}, i.e., the combined effect of nonvanishing $d$ and $\mu_{\text s}$. By computing the FF, Eq.~\ref{eq:fitting-fact}, between signals from uncharged and charged secondaries, they showed that ultra-light scalar fields can leave a strong imprint on the GW emission, potentially detectable by LISA for a wide range of binary configurations. Specifically, they reported that for charges as small as $d \sim 0.05$, LISA could be able to distinguish fields with $\mu_{\text s} \gtrsim 0.01$ ($\mu_{\text s} \sim 10^{-18}$eV) from their massless counterpart~\cite{Barsanti:2022vvl}. This bound increases by almost an order of magnitude if, for example, $d \gtrsim 0.3$. In comparison with the massless case, the relative error on the scalar charge for the binary is typically larger due to correlations with $\mu_{\text s}$, which enters now as an additional parameter.

These studies need to be generalized to generic orbits and examine the effect of the small compact object’s spin, and include the self-force corrections. A recent development towards this direction was made in Ref.~\cite{Spiers:2023cva}. In that work, the perturbative self-force approach was generalized and they provide a new ansatz for the point-particle action~\cite{Spiers:2023cva}
\begin{equation}
    \label{eq:PPansatz}
    S = -\int_{\gamma} m[\tilde{\varphi}] d\tilde{s}= - \int_{\gamma} m[\tilde{\varphi}] \sqrt{\tilde{g}^{}_{\mu\nu} \tilde{u}^\mu\tilde{u}^\nu}d\tilde{\tau} \,,
\end{equation}
where the existence of a \textit{singular} (${\mathcal S}$) and \textit{regular} (${\mathcal R}$) split of 
the metric and scalar field perturbations~\cite{Zimmerman:2015hua} is assumed to exist and of the form $h^{(n)}_{\mu\nu}=h^{(n){\mathcal S}}_{\mu \nu} +h^{(n){\mathcal R}}_{\mu \nu}$, $\varphi^{(n)}=\varphi^{(n){\mathcal S}} + \varphi^{(n){\mathcal R}}$. Thus, they considered an effective metric and scalar field of the form
\begin{equation}
\tilde{g}^{}_{\mu \nu} = g^{}_{\mu \nu} + h^{{\mathcal R}}_{\mu \nu} \quad \tilde{\varphi}=\varphi^{{\mathcal R}}\,.
\end{equation} 
In Eq.~\eqref{eq:PPansatz}, the proper time in the effective spacetime is denoted by $\tilde\tau$, $\gamma$ is the worldline of the secondary, and $\tilde{u}^\alpha=dz^\alpha/d\tilde{\tau}$. 

This change to the matter action differs from the conventional ``skeletonized'' point particle action~\cite{Damour:1992we}, because near the worldline of the secondary, the singular nature of the metric and scalar field makes $g_{\mu \nu}$ and $m[\varphi]$ ill-defined. With this new point-particle action, for a non-spinning secondary with no scalar dipole, they derived the field equations for the first- and second-order metric and scalar perturbations, the first- and second-order equations of motion for the secondary object. It is expected that this formalism will produce sufficiently accurate waveform templates for precision measurements of the scalar charge of the secondary with LISA data on EMRIs~\cite{Spiers:2023cva}.

Beyond the introduction of new fundamental charges, compelling theoretical and observational evidence supports models where compact objects maintain global neutrality, thereby avoiding traditional dipole-emission limitations. A prime example of this are magnetars, which possess powerful magnetic dipole moments. Prompted by scenarios forecasting neutral compact objects with dipole moments, in Ref.~\cite{Lestingi:2023ovn}, a model-independent method for a fundamental scalar dipole was developed to assess GW emissions in EMRIs. Since the square of the mass ratio relative to the case of fundamental charges suppresses this effect, the additional flux from the dipole moment, compared to fundamental charges, was model for a fundamental scalar dipole significantly reduced, making its detection with EMRIs challenging. Notably, its impact on LISA is negligible for a dimensionless dipole moment $\lesssim 10$. 

\section{Tests beyond scales and the gravitational spectrum}
\label{sec:other}

\subsection{A tale of two specific theories}
\label{sec:testexample}

It is very challenging to test alternative theories of gravity with EMRIs due to the lack of suitable EMRI models. Given a concrete theory of gravity, there are some ingredients that we need in order to implement a reliable EMRI model. First, we must have definite models for the primary and the secondary in that theory. We need to ask questions like what the possible families of models for BHs are in that theory, as it may not be unique or parametrized just in terms of the mass and spin (assuming these are well-defined parameters). Then, we should understand the zero-order approximation to the EMRI dynamics, which are geodesics around a Kerr BH in the case of the standard scenario. Even more important is to figure out how we can describe the inspiral, and if the self-force formalism can be extended in a robust way. Here, the theory may contain extra fields mediating the gravitational interaction that need to be incorporated into the backreaction description. Finally, we have to understand the generation of the waveforms, where it is crucial to understand the number of independent polarizations in the theory and how these can be computed in practice.

Although this discussion may seem discouraging, we think that one can find ways of carrying out the description of EMRIs, at least in certain classes of alternative theories of gravity, that are necessary to carry out tests with space-based detectors. For instance, in the case of alternative theories that contain higher-curvature corrections (such as Gauss--Bonnet, Chern--Simons modified gravity, or $f(R)$ theories), which in principle would lead to evolution systems for the gravitational field with higher-order time derivatives (third order or higher), we can adopt the point of view that these corrections can be seen terms that come from some low-energy limit of a more fundamental unknown theory and that in this way can be treated in the spirit of corrections in an EFT. In that way, we can retain the main structure and properties of GR that make the description of an EMRI feasible. In the case of comparable-mass binaries, there are some studies that have adopted this point of view (see, for instance, Refs.~\cite{Okounkova:2019dfo,Okounkova:2019zjf}.  

In order to illustrate what to expect in the case of EMRIs, below we describe two different scenarios involving two different alternative theories of gravity, where EMRIs have been studied and some predictions have been made. Still, in both cases, several approximations have been made to make it possible, but nevertheless it is important to show these examples as a playground that illustrates the kind of difficulties that we may find and what type of fundamental physics tests may be possible with EMRIs. These two examples are: (i) EMRIs in dynamical Chern--Simons modified gravity; (ii) EMRIs in a particular five-dimensional braneworld model.

(i) Dynamical Chern--Simons gravitational theory is an alternative theory of gravity that includes a high-order gravitational correction to the well-known Einstein-Hilbert Lagrangian, which is not invariant under parity transformations. On the other hand, in 3+1 dimensions, it is a topological invariant, the Pontryagin invariant, and for this reason, in order to have an effect, it is introduced coupled to a scalar field, which has to be dynamical~\cite{Yunes:2007ss}, in contrast with the initial model by Jackiw and Pi~\cite{Jackiw:2003pm}. See~\cite{Alexander:2009tp} for details. This theory has features that are quite interesting for tests with a space-based GW detector: (a) The spinning BHs in this theory differ from the Kerr BH~\cite{Yunes:2009hc} (although non-spinning MBHs are still described by the Schwarzschild metric\footnote{It is still possible to study non-rotating solutions in this theory, as there are modifications to the dynamical equations of gravity caused by coupling the gravitational field with a scalar field. A study was conducted in this context and presented in Ref.~\cite{Pani:2011xj}, where it was shown that since the secondary spends many cycles in the bandwidth of LISA, the small effect of coupling to the dynamical Chern--Simons term accumulates, leading to measurable effects. In particular, this coupling decreases the number of cycles over a fixed frequency bandwidth.}); (b) The deviations from 
the Kerr geometry are controlled by a single parameter, $\xi_{\rm CS}$, that turns out to be a combination of universal coupling constants) approximation~\cite{Yunes:2009hc,Yagi:2012ya}; (c) These deviations change the multipole moments of the BH but, as in the Kerr solution in GR, they are still determined by only two numbers (mass and spin)--this indicates that the no-hair conjecture may survive in this theory; and (d) Despite the curvature corrections in this theory change the dynamics of the gravitational field, the energy-momentum tensor of the gravitational radiation is formally the same as in GR~\cite{Sopuerta:2011te}. An interesting point is that in this theory, GWs have three different polarizations, all transverse, but the non-GR one decays too fast with distance in order to be observable, and we can only see its indirect effects. In Refs.~\cite{Canizares:2012is,Canizares:2012he}, different EMRI evolutions were computed both in dynamical Chern--Simons gravity and GR, and it was shown that LISA can provide estimations of the Chern--Simons parameter that controls the deviations from GR, or to put strong constraints in the case GR is the correct theory of gravity. This theory is a good example of EMRIs being properly described and computations being carried out, which is not necessarily true for other alternative theories of gravity.

(ii) By the end of the last century, gravitational models with large (sub-millimeter) extra dimensions appeared, which were avoiding observational constraints by confining matter fields around a $3+1$ submanifold. Among other things, these extra dimensions were proposed as a possible solution to the hierarchy problem of the fundamental physical interactions and as a playground for holography. Very popular models were the ones proposed by Randall and Sundrum~\cite{Randall:1999ee,Randall:1999vf}. Of particular interest is the Randall--Sundrum II model~\cite{Randall:1999vf}, where standard model fields live on a $3+1$ brane moving in a five-dimensional spacetime where the direction out of the brane is not non-compact. It was proposed that in this scenario, an excess of degrees of freedom would accelerate the decay of a BH via Hawking radiation~\cite{Emparan:2002jp}. In an EMRI system, this may be relevant for the evolution of a BH secondary. Actually, in Ref.~\cite{McWilliams:2009ym}, it was found that a detector like LISA could constrain the size of the extra dimension to be below $5\,\mu$m. However, static solutions for BHs were later found~\cite{Figueras:2011gd}, invalidating the hypothesis on which the BH decay was based. In any case, the key lesson from this example is that EMRI GW observations have a tremendous potential to test fundamental physics scenarios linked with high-energy physics (see more details in Refs.~\cite{Cardoso:2012qm,Barack:2018yly}).

\subsection{Tests of the polarization components of the waves}
\label{sec:polarization}

Gravitational modes, or GW modes, refer to the different patterns or polarizations of gravitational radiation that can propagate through spacetime. These modes represent distinct ways the spacetime can be stretched and squeezed as GWs  pass through it. There are in general six different polarization modes associated with GWs: two tensor modes, plus ($h_+$), and cross ($h_\times$) and four vector and scalar modes~\cite{Will:2014kxa}. Understanding and detecting these modes is essential for studying astrophysical phenomena and testing the predictions of GR. 

Separating the different GW modes requires considering eight unknowns, including time series for six polarizations and two direction angles affecting the mode projection~\cite{Gair:2012nm}. However, only six observable components exist, making the problem indeterminate without additional information about the source's position or the limitation of gravitational radiation to transverse modes.

To measure the polarization of the source, therefore, more than a detector is needed. Luckily, space-based observatories typically have more than one \emph{independent} interferometric observable, depending on the laser links between the arms. Each interferometric component is therefore able to measure a different combination of polarizations. For example, a three-armed detector could distinguish non-GR polarization modes if the source direction is known or can be inferred from the signal through orbital motion or triangulation between signals from the components~\cite{Gair:2012nm}.

We have already seeing examples of theories that predict GWs with scalar polarization (e.g., Sec.~\ref{sec:secondary}), such as scalar-tensor theories, or vector modes (e.g., Sec.~\ref{Sec:Hairy}) such as the bumblebee gravity model. However, the above results have neglected the contribution of these modes and only the measurement of the extra ``hair'' has been studied.  

Laser frequency fluctuations are the primary noise source in a space-borne GW detector~\cite{Tinto:1999yr}. Time-delay interferometry (TDI) is proposed to mitigate this noise, which involves synthesizing virtual equal-arm-length interference through data post-processing using modulated laser beams exchanged between remote spacecraft~\cite{Tinto:2020fcc}. The detector's sensitivity depends on the polarization state of the incident GW, the detector's configuration, and the GW amplitudes of the GWs in different polarization channels. Additionally, the response function's angular dependence is averaged out for all-sky coverage due to the unknown location of the GW source~\cite{Wang:2021owg}. 

In the context of LISA, the averaged response functions for all six possible polarizations and different TDI combinations was calculated in Ref.~\cite{Tinto:2010hz}. In this work it was estimated the sensitivities to (general) vector- and scalar-type waves. They found that at frequencies larger than ($\sim6\times10^{-2}$ Hz) LISA is more than ten times more sensitive to scalar-longitudinal and vector signals than to tensor and scalar-transverse waves, and that in the low part of its frequency band is equally sensitive to tensor and vector waves and somewhat less sensitive to scalar signals~\cite{Tinto:2010hz}. These results, however, have not been applied to get constraints on specific modified theories of gravity. 

\subsection{Tests of the propagation of gravity}
\label{sec:propagation}

According to GR, gravitons have a rest mass of zero ($m_{\rm{g}}=0$), and their dispersion relation is given by $E=pc$, where $E$ and $p$ represent the total energy and momentum of the graviton, respectively. Conversely, some modified theories of gravity endow gravitons with mass, resulting in their speed depending on energy or wavelength~\cite{Will:2014kxa}. As GWs emitted by compact binary inspirals exhibit frequency chirping, gravitons emitted during the early inspiral phase will travel at a slower pace than those emitted closer to the merger. This discrepancy leads to a frequency-dependent GW dephasing, in contrast to the phasing of a massless graviton as predicted by GR. If such a dephasing is not observed, it could serve as a means to constrain the graviton's mass~\cite{Will:1997bb}.

A generic Lorentz-violating dispersion relation on the propagation of GWs was proposed in Ref.~\cite{Mirshekari:2011yq} that parametrically introduces Lorentz-violating deviations in a continuous way. It only has two parameters, 
\begin{equation}
   E^2 = p^2 c^2 + m_{\rm{g}}^2 c^4 + \mathbb{A}p^\alpha c^\alpha\,,
\end{equation}
where $\alpha$ is a dimensionless quantity while $\mathbb{A}$ has dimensions of $\left[ \rm{energy}\right]^{(2-\alpha)}$. Thus, if one considers such a dispersive case, two gravitons emitted at $t_e$ and $t^\prime_e$ with different frequencies, $f_e$ and $f^\prime_e$, respectively, will be received at corresponding arrival times $t_a$ and $t^\prime_a$. If during the difference of emitting time ($\Delta t_e=t_e-t^\prime e$) there is a change on the cosmological scale factor $a$, then there will be a delay of arrival times of two gravitons ($\Delta t_a=t_a-t^\prime a$). Assuming that $\Delta t_e=0$ Ref.~\cite{Yang:2018ioy} studied the arrival time delay
\begin{equation}
    \Delta t^{}_a = (1+Z)\frac{cD}{2\lambda^{}_{\rm g}}\left( \frac{1}{f^2_e}-\frac{1}{{f^\prime}^2_e}\right)\,,
\end{equation}
where $\lambda_{\rm g}$ is the Compton wavelength of the graviton, $Z$ is the cosmological redshift, and $D$ is the luminosity distance. Note in passing, that all these quantities depend on cosmological parameters, such as Hubble constant, matter density or dark energy density, which may also be modified with respect their GR values. Since EMRIs may be eccentric when observed, the GW modes (or ``voices,'' using the language of the harmonic numbers for the decomposition of the waveform~\cite{Hughes:2001jr}) with different frequencies will be emitted at the same time. Circular orbits can also be used to test dispersion, as performed with ground-based detectors. However, for EMRIs the frequency evolution is very slow, so without eccentricity one will need a very long observation to accurately perform this test. 

Comparing the differences between dispersive and nondispersive GWs, Ref.~\cite{Yang:2018ioy} by fixing the cosmological-parameters to current measured values, reported that the observations of EMRIs with LISA can constrain the $\lambda_{\rm g}$ around two orders better than the current LVKC-like observations. 

\subsection{Multimessenger tests}
\label{sec:multimessenger}

To the best of our knowledge, no model reliably provides an electromagnetic counterpart for an EMRI. It will be very unfortunate if merging a small compact object with a larger one only leaves gravitational radiation, as there are exciting tests that could be performed if there also were another type of radiation.  In particular, there are some proposals in the literature of how an electromagnetic counterpart to an EMRI event may be produced, e.g., an emission due to the presence of EMRIs embedded in thin accretion disks~\cite{Kocsis:2011dr}, or
quasi-periodic eruptions from impacts between the secondary and a rigidly precessing accretion disc~\cite{Linial:2023nqs,Franchini:2023bou}.

\subsubsection{Constrains on the mass of the graviton}

Gravitational radiation sources with an electromagnetic counterparts can provide constraints on the mass of the graviton. In GR the speed of the graviton, $v_{\rm{g}}$, is
\begin{equation}
\label{Eq:gravitonspeed}
v^{}_{\rm{g}} = c \sqrt{1-\frac{m^2_{\rm{g}}c^4}{E^2}}\,.
\end{equation}
Thus, a \emph{simultaneous} measurement of GWs and photons can put a constraint on the mass of the graviton. For instance, the multi-messenger observation of GW170817 (a binary NS merger) found a $1.7$s delay of the gamma-ray burst, which translates to a constraint of $m_{\rm{g}}\leq\times10^{-22}\rm{eV}/c^2$~\cite{Baker:2017hug}. If an EMRI can leave an electromagnetic signature, a constraint may be achieved with a joint measurement. 

Polarization measurements with estimates of the propagation speed of the observed GW signal can also determinate the mass of the graviton. However, LISA will not be able to distinguish the propagation speeds of scalar and vector polarizations from the speed of light~\cite{Tinto:2010hz}. Their argument goes as follows. If one writes Eq.~\eqref{Eq:gravitonspeed} in terms of the frequency of the wave, at $10^{-4}$ Hz the group velocity's fractional difference, $\Delta_v$, from the speed of light is $\Delta_v=\left|v-c\right|/c\approx10^{-16}$, assuming a very tight constraint of $m_{\rm{g}}\leq5\times10^{-27}\rm{eV}/c^2$. Taking the (weaker) constraint from GW170817~\cite{Baker:2017hug}, one gets $\Delta_v\approx 3\times 10^{-8}$, which is still far from the time-separations ($\mathcal{O}\left(s\right)$) of their imprints in the TDI combinations. Thus, LISA fundamentally cannot resolve the propagation speeds of the different polarization components~\cite{Tinto:2010hz}. The derivations presented in Ref.~\cite{Tinto:2010hz} for the expression of LISA TDI responses to vector and scalar GWs are valid for any wavelength, not just in the long wavelength limit as in other places (see also Ref.~\cite{Zhang:2020khm}).

%%%%%%%%%%
\subsubsection{Constrains on cosmological parameters}

If one assumes a theory of gravity, the measurement of GWs provides a way to obtain the absolute distance of the source~\cite{Schutz:1986gp}. The amplitude of the GW signal decreases as it travels towards Earth. This is because, like other forms of radiation, the energy density decays following an inverse square law, and the amplitude only as the inverse of the distance. The observed signal strength (amplitude) and the intrinsic properties of the source are related to the luminosity distance through the inverse square law. However, to obtain an accurate cosmological distance, the Universe's expansion must also be considered.

That is why one can use gravitational sources as standard sirens, in analogy to the traditional standard candles in the electromagnetic spectrum, when used with independently observed electromagnetic counterparts, to obtain a relationship between luminosity distance and redshift and, therefore, constrain cosmological parameters ~\cite{Berry:2019wgg}. 

In Ref.~\cite{MacLeod:2007jd} this idea was considered in the context of EMRIs observed by LISA. What is needed for these inferences of cosmological parameters is an estimation of direction on the sky, $\vec{\theta}$, and the luminosity distance, $D_L$, and a cosmological model relating $D_L$ and redshift $z$, which must be estimated from other techniques. A galaxy redshift survey in this $\vec{\theta},z$ space (a three-dimensional ``error box'') provides a statistical estimate of the host redshift. This relation maps out the expansion history of the Universe and therefore, a way to constrain cosmological parameters. Assuming a $\Lambda$CDM cosmology, mock LISA data, and data from the Sloan Digital Sky Survey,  Ref.~\cite{MacLeod:2007jd} argued for an accurate and unbiased estimator of the Hubble constant, $H_0$, with sub-percent precision~\cite{MacLeod:2007jd}.

Ref.~\cite{Laghi:2021pqk} revisited the this problem and considered an updated configuration of LISA, and relaxed the assumption of Ref.~\cite{MacLeod:2007jd} of a linear cosmic expansion, which neglects the acceleration of the Universe. They found that by using the loudest EMRIs (SNR$>100$) detected by LISA as dark standard sirens, statistically matching their sky localisation region with mock galaxy catalogs, constraints on $H_0$ can reach $\sim1.1\%$ ($\sim3.6\%$) accuracy, at the $90\%$ credible level, in their best (worst) case scenario. By considering a dynamical dark energy cosmological model, with $\Lambda$CDM parameters fixed by other observations, they also showed that their best (worst) case scenario $\sim5.9\%$ ($\sim12.3\%$) relative uncertainties at the $90\%$ credible level can be obtained on the dark energy equation of state parameter $w_0$. 

To summarize, in the case when the host galaxy cannot be identified, i.e., there is not an electromagnetic counterpart, one can still cross-correlate the sky localization region of the GW event with a galaxy catalog and assign to each galaxy within this region a probability of being the host galaxy. 

On the other hand, if an electromagnetic counterpart exists, then joint observations (i.e., an event measured \emph{both} in the gravitational and electromagnetic spectra) should also allow for these types of cosmological measurements~\cite{Dalal:2006qt,Nissanke:2013fka,Speri:2020hwc}. For flat cosmologies, for example, a measurement of $H_0$ at the percent level, combined with precision CMB measurements of the absolute distance to the last scattering surface, would constrain the dark energy equation of state parameter to around $10\%$~\cite{Nissanke:2013fka}. 

The results of this particular test of cosmological parameters can be significantly improved if there were EM precursors to EMRIs, i.e., if the localization can be better constrained through another messenger (photons, in this case). For example, Ref.~\cite{Wang:2019bbk} proposed a formation channel of EMRIs with tidal disruption flares as EM counterparts. In this formation scenario, flares can be produced from the tidal stripping of the helium envelope of a massive star by a MBH. The left compact core of the massive star will evolve into an EMRI, and depending on the initial eccentricity and semimajor axis, the GW frequency of the inspiral can enter, for example, LISA band within a few years, making the tidal disruption flare an EM precursor to EMRI~\cite{Wang:2019bbk}. 

\subsection{Multiband observations}

In theories with additional scalar fields, the enhanced energy emission during the inspiral leads to a cumulative dephasing of the gravitational waveform, as we briefly mentioned in Sec.~\ref{sec:secondary}. Ground-based observatories only see BH binaries near the merger, where the GW dipole term is subdominant and thus can only be weakly constrained~\cite{Yunes:2016jcc}. However, if, for example, a GW 150914-like BBH is first observed by LISA and then by ground-based detectors~\cite{Sesana:2016ljz}, dipole emission can be better constrained~\cite{Perkins:2020tra}. Joint observations of GW 150914-like systems with both ground-based and space-based detectors is expected to improve bounds on dipole emission from BBHs by six orders of magnitude relative to current constraints~\cite{Barausse:2016eii}.

Note that, as shown in, for example, Refs.~\cite{Carson:2019rda,Gnocchi:2019jzp,Carson:2019kkh}, for specific theories of gravity, multiband observations are the \emph{only} way to obtain valid constraints with GWs. This is the case for dynamical Chern--Simons~\cite{Alexander:2009tp}, where the constraints from measurements from comparable-mass BHs~\cite{Nair:2019iur}, and the forecasts with EMRIs fall outside the region of validity where the theory was derived. In fact, the most stringent constraint on this theory is from multimessenger observations~\cite{Silva:2020acr} of NSs.

One can also consider the possible constraints and tests with space-based detector networks, such as LISA and TianQin~\cite{Zi:2021pdp}. Such configurations can boost the accuracy of the measurement of source parameters by a couple of orders of magnitude, with the improvement in sky localization being the most significant~\cite{Huang:2020rjf}.

Another multiband test concerns the use of binary-EMRIs (b-EMRIs)~\cite{Chen:2018axp}, in which the secondary consists of a BBH. According to Ref.~\cite{Chen:2018axp} when the eccentricity of a b-EMRI drops to about $0.85$, the two stellar BHs will quickly merge due to the tidal perturbation by the primary. The high-frequency ($\sim10^2$Hz) GWs generated during the coalescence coincide with the low-frequency ($\sim10^{-3}$Hz) waves from the b-EMRI, making this system an ideal target for future multi-band GW observations~\cite{Chen:2018axp}.

This multiband test makes use of the fact that according to GR, after the coalescence, the resulting post-merger BH will recoil because GWs carry linear momentum, and this radiation is asymmetric~\cite{Fitchett1983}. The magnitude of the recoil velocity is a function of the mass ratio of the two merging BHs as well as their spin magnitudes and directions~\cite{Centrella:2010mx}. Thus, in this scenario when the (secondary) binary merges, the resulting recoil will alter its orbital elements and a glitch is induced in the low-frequency EMRI waveform. Under this scenario, the original detected EMRI will be ``lost'' due to the rapid change of the parameters, but a ``new'' one, from the same sky location, luminosity distance and with almost the same BH masses, should emerge immediately after the glitch because the recoiling BH remains bound to the MBH~\cite{Chen:2018axp}. Since b-EMRIs emit simultaneously milli- and hundred-Hertz GWs (when the BBHs in them coalesce), it can be used to test the dispersion relation and check if there is a frequency dependence (as the speed of the wave depends on the mass of graviton).

%%%%%%%%%%  Outlook  %%%%%%%%%%
\section{Outlook}\label{Outlook}

Do EMRIs provide a remarkable opportunity to test GR? This review is dedicated to furnishing the reader with a compelling rationale for a positive response to this query. EMRIs represent systems featuring highly asymmetric binaries, which we anticipate observing over numerous cycles with a reasonably high SNR. Consequently, they provide an avenue to encode properties of its components (i.e., the primary, secondary, and their environment), scrutinizing whether these properties align with the predictions of GR.

Despite the recent impressive calculations emanating from BH perturbation theory within the framework of GR, the task of modeling generic EMRIs still needs to be completed. Current methodologies employed to generate EMRI waveforms capture the fundamental aspects of the dynamics. Nevertheless, they might not suffice to attain the level of precision essential for discerning whether an inspiral corresponds to a GR BH. However, in regard to the EMRI standard scenario, the self-force  programme has made significant advances in recent years, solving many of the conceptual issues involved in the EMRI waveform computation. It is quite reasonable to think that having precise waveforms for standard scenario EMRIs is a matter of time and resources. It would be desirable to reach a similar situation in the case of EMRI systems outside the standard scenario.

The recent findings scrutinized in this review offer encouragement for future outstanding tests of GR and the nature of the most (astrophysical) compact objects. To that end, more work is needed to understand the different possible compact objects that we may find in an EMRI and the possible classes of alternative gravity theories that are physically reasonable. Finding out the precise impact of these modifications of the standard scenario on GW observations of EMRIs constitutes a great theoretical challenge. In particular, future analyses need to consider the post-adiabatic gravitational self-force terms, which are recognized even within the framework of GR as having a significant impact on the waveform. 

Further efforts should also focus on producing reliable gravitational waveforms and conducting extensive data analyses to measure and identify distinct features arising from the complex interaction between strong-field corrections and all other relevant relativistic effects. However, given the complexity of EMRI modeling, not all the models for compact objects can be equally developed (in particular, some of the ECO concepts proposed in the literature), because we do not have a complete description of their dynamics. In most cases, we only possess a depiction of their equilibrium state. In other words, it may not always possible to express the EMRI problem mathematically consistent. However, it might be possible if one considers an EFT approach, in which the dynamics inherits the GR structure, and the modifications to the gravitational dynamics can be seen as effective corrections to GR-type equations. In order to test models of ECOs and modified theories of gravity with EMRIs, it is essential to ensure that all the necessary theoretical components are in place, which may require significant theoretical advancements.

Numerous challenges arise in studying beyond-vacuum GR effects in EMRIs, primarily stemming from using simplistic models and the resulting parameter correlations. Advancement hinges on improving modeling, such as refining the description of the PN effects. While attempts at joint Bayesian inference have been made, the perturbative (short-amplitude) and secular (long-timescale) nature of these effects, disentangling them is challenging. Population-level studies can help differentiate between global and local effects, while cataloging detected sources can aid in addressing correlated effects in EMRIs. 

Choosing between a theory-specific or theory-agnostic approach remains a pivotal decision in pursuing GR testing. Opting for a theory-specific approach involves tailoring tests to the predictions of a particular theoretical framework, while a theory-agnostic strategy aims for broader examinations without preconceived commitments. The rationale behind this choice lies in balancing specificity and generality: a theory-specific approach allows for targeted assessments, but the risk of overlooking unexpected phenomena exists. Deviating from the paradigm involves exploring alternative theories or modifications to GR, and the seriousness attributed to these deviations should be informed by empirical evidence and mathematical consistency. While critically evaluating and refining our understanding of gravity is essential, the worth of alternative theories lies in their ability to provide new insights or resolve existing challenges within the framework of observational and experimental constraints. Future perspectives in the field hinge on striking a balance between theory-specific and theory-agnostic methodologies, fostering open-minded exploration, and embracing the potential paradigm shifts that could reshape our understanding of gravity and the fundamental nature of gravity. EMRIs have the potential to facilitate such exploration with flying colors.

%%%%%%%%%%  Acknowledgments  %%%%%%%%%%
\begin{acknowledgement}
We thank Cosimo Bambi, Vitor Cardoso, Anna Heffernan, Paolo Pani, and Thomas Sotiriou for their helpful feedback. A.C.-A.  is thankful to Frans Pretorius for valuable discussions. C.F.S thanks Michele Lenzi and Iv\'an Mart\'in~V\'ilchez for enlightening discussions. We also thank our colleagues in the LISA Science Group of the LISA consortium for pushing the LISA science programme and, in particular, the science with EMRIs described in this review.
A.C.-A. acknowledges support from the Simons Foundation.
C.F.S. is supported by contracts PID 2019-106515GB- I00/AEI/10.13039/501100011033 and PID 2022-137674NB- I00/AEI/10.13039/501100011033 (Spanish Ministry of Science and Innovation), and by the programme {\em Unidad de Excelencia Mar\'{\i}a de Maeztu} CEX2020-001058-M (Spanish Ministry of Science and Innovation). He also acknowledges the Consolidated Research Group seals 2017-SGR-1469 and 2021-SGR-01529 from the SGR programme (AGAUR, Generalitat de Catalunya).
\end{acknowledgement}

\vspace{1cm}

%%%%%%%%%%%%%%%%%%%%%%%%%%%%%%%%%%%%%%%%%%%%%%%%%%%%%%%%%

\bibliographystyle{myspringer}
\bibliography{refs}

%\printbibliography[title={References}]

%%%%%%%%%%%%%%%%%%%%%%%%%%%%%%%%%%%%%%%%%%%%%%%%%%%%%%%%%

\end{document}